\documentclass[12pt]{article}
\usepackage{amsmath}
\usepackage{multirow}
\usepackage{amsfonts}
\usepackage{amssymb,graphics,psfrag}
\usepackage{array,epsfig,multirow,stmaryrd,graphicx}
\usepackage{comment}
\usepackage{slashed}
\usepackage{hyperref}

\def\hybrid{\topmargin -20pt    \oddsidemargin 0pt
        \headheight 0pt \headsep 0pt 
        \textwidth 6.25in      
        \textheight 9 in      
        \marginparwidth .875in
        \parskip 5pt plus 1pt
          \jot = 1.5ex
  }
\hybrid
\numberwithin{equation}{section}
\numberwithin{table}{section}

\newcommand{\beq}{\begin{equation}}
\newcommand{\eeq}{\end{equation}}
\newcommand{\be}{\begin{equation}}
\newcommand{\ee}{\end{equation}}
\newcommand{\bea}{\begin{eqnarray}}
\newcommand{\eea}{\end{eqnarray}}
\newcommand{\ben}{\begin{eqnarray*}}
\newcommand{\een}{\end{eqnarray*}}               
\newcommand{\ba}{\begin{aligned}}
\newcommand{\ea}{\end{aligned}}
\newcommand{\bt}{\begin{tabular}}
\newcommand{\et}{\end{tabular}}
\newcommand{\bc}{\begin{center}}
\newcommand{\ec}{\end{center}}

\newcommand{\cT}{\mathcal{T}}
\newcommand{\cE}{\mathcal{E}}

\newcommand{\cC}{\mathcal{C}}

\newcommand{\cL}{\mathcal{L}}

\newcommand{\cN}{\mathcal{N}}

\newcommand{\cB}{\mathcal{B}}

\newcommand{\cR}{\mathcal{R}}

\newcommand{\fg}{\mathfrak{g}}

\DeclareMathOperator{\MW}{MW}
\DeclareMathOperator{\rk}{rank}

\DeclareMathOperator{\sign}{sign}

\newcommand{\rep}[1]{\mathbf{#1}}
\newcommand{\frep}[1]{\mathcal{#1}}

\newcommand{\bbP}{\mathbb{P}}

\newcommand{\nn}{\nonumber}

\newcommand{\cref}{{\bf [check ref]}}

\newcommand{\tr}{\mathrm{tr}}

\psfrag{n1}{$\nu_1$}
\psfrag{n2}{$\nu_2$}
\psfrag{n1'}{$\nu_1'$}
\psfrag{n2'}{$\nu_2'$}
\psfrag{n9}{$\nu_9$}
\psfrag{n10'}{$\nu_{10}'$}
\psfrag{t1}{${\nu}_1$}
\psfrag{t2}{${\nu}_2$}
\psfrag{t9}{${\nu}_9$}
\psfrag{t1'}{${\nu}_1'$}
\psfrag{t2'}{${\nu}_2'$}
\psfrag{t10'}{${\nu}_{10}'$}

\DeclareMathOperator{\ad}{ad}

\def\blfootnote{\xdef\@thefnmark{}\@footnotetext}
\long\def\symbolfootnote[#1]#2{\begingroup%
\def\thefootnote{\fnsymbol{footnote}}\footnote[#1]{#2}\endgroup}

\begin{document}

\baselineskip=15pt

\begin{titlepage}
\begin{flushright}
\parbox[t]{1.8in}{\begin{flushright} MPP-2013-125 \end{flushright}}
\end{flushright}

\begin{center}

\vspace*{ 1.2cm}

{\large \bf Effective action of 6D F-theory with U(1) factors:\\[.2cm]
                Rational sections make Chern-Simons terms jump}


\vskip 1.2cm

\begin{center}
 {Thomas W.~Grimm, Andreas Kapfer and Jan Keitel\ \footnote{grimm,\ kapfer,\ jkeitel@mpp.mpg.de}}
\end{center}
\vskip .2cm
\renewcommand{\thefootnote}{\arabic{footnote}}

{Max-Planck-Institut f\"ur Physik, \\
F\"ohringer Ring 6, 80805 Munich, Germany}

 \vspace*{1cm}

\end{center}

\vskip 0.2cm
 
\begin{center} {\bf ABSTRACT } \end{center}

We derive the six-dimensional $(1,0)$ effective action arising 
from F-theory on an elliptically fibered Calabi-Yau threefold with 
multiple sections. The considered theories admit both non-Abelian and
Abelian gauge symmetries. Our derivation employs the M-theory to 
F-theory duality in five-dimensions after circle reduction. Five-dimensional
gauge and gravitational Chern-Simons terms are shown to arise at one-loop by integrating out 
massive Coulomb branch and Kaluza-Klein modes. In the presence of a non-holomorphic zero 
section, we find an improved systematic for performing the F-theory limit by using the
concept of the extended relative Mori cone.
In this situation Kaluza-Klein modes can become lighter than Coulomb branch modes and 
a jump in the Chern-Simons levels occurs.
By determining  
Chern-Simons terms for various threefold examples we are able to compute the 
complete six-dimensional charged matter spectrum and show consistency with 
six-dimensional anomalies.

\hfill {May, 2013}
\end{titlepage}

\tableofcontents




\section{Introduction}

Compactifications of F-theory on elliptically fibered Calabi-Yau threefolds 
have long been known to yield a large class of six-dimensional (6D) supergravity theories with 
eight supercharges implying $\cN=(1,0)$ supersymmetry \cite{Vafa:1996xn,Morrison:1996na,Morrison:1996pp}. 
These $(1,0)$ theories are chiral, can admit non-Abelian gauge groups 
with a charged matter spectrum, and support a number of tensors 
with self-dual or anti-self-dual field strength. Since F-theory arises 
from the consistent Type IIB string theory, its 
effective theories are quantum consistent and meet all known low-energy 
constraints \cite{Taylor:2011wt,Sadov:1996zm,Kumar:2009ac,Seiberg:2011dr,Park:2011wv,Grimm:2012yq,Cvetic:2012xn}. 
In particular, since all fermions in the 6D spectrum are 
chiral, anomaly cancelation imposes strong constraints on the number of multiplets and their connection to
the couplings of the low-energy effective supergravity action. The effective action of F-theory 
reduced on an elliptically fibered Calabi-Yau threefold with purely non-Abelian gauge
symmetries was derived via the duality of F-theory to M-theory 
in \cite{Ferrara:1996wv,Bonetti:2011mw,Grimm:2013fua}. 
In this work we generalize the 
derivation of \cite{Bonetti:2011mw} to include Abelian gauge symmetries
and investigate their physics from the F-theory and M-theory point of view.

Recently, global F-theory compactifications with Abelian gauge symmetries
arising on seven-brane world-volumes have been
investigated intensively \cite{Grimm:2010ez,Park:2011wv,Park:2011ji,Morrison:2012ei,Cvetic:2012xn,
Mayrhofer:2012zy,Braun:2013yti,Borchmann:2013jwa,Cvetic:2013nia} \footnote{For a systematic survey
of $U(1)$s in local models see \cite{Dolan:2011iu}.}. At weak string coupling, $U(1)$ symmetries 
arise on stacks of D7-branes and can be analyzed from the global configuration of D7-branes and O7-planes \cite{Blumenhagen:2005mu,Blumenhagen:2006ci,Ibanez:2012zz}. 
Their origin in F-theory is purely geometrical, since geometrically massless $U(1)$ symmetries 
are counted by the number of sections of the elliptic fibration. More mathematically 
stated, the geometry of the Abelian gauge groups is captured by
the rank of the Mordell-Weil group of the elliptic fiber \cite{Morrison:1996pp,Park:2011ji,Morrison:2012ei}. As we
discuss in detail in this work, the properties of the sections parametrizing 
the $U(1)$s have a crucial impact on the physics of the low-energy effective
action. In particular, we consider non-holomorphic sections defined by rational instead of holomorphic equations.
These induce rich new physics in six- and four-dimensional F-theory reductions \cite{Morrison:2012ei,Cvetic:2012xn,
Mayrhofer:2012zy,Braun:2013yti,Borchmann:2013jwa,Cvetic:2013nia}.
Notably, there is no physical reason for the zero section of the Calabi-Yau manifold to be holomorphic and
abandoning this constraint leads to interesting new phenomena, which we
investigate in this work. The geometry of elliptic fibrations with a non-holomorphic
zero section has recently been investigated in \cite{Braun:2013yti,Cvetic:2013nia}.

To determine the effective action of F-theory on Calabi-Yau threefolds,
one has to consider these reductions as a limit of M-theory \cite{Denef:2008wq}. This can be traced back to 
the fact that there is no fundamental low-energy effective action of F-theory. We
thus proceed as follows. Determining the effective action 
of M-theory on a resolved elliptically fibered Calabi-Yau threefold at 
large volume allows us to use eleven-dimensional supergravity plus 
known higher curvature corrections to derive a five-dimensional (5D)
effective action \cite{Cadavid:1995bk,Antoniadis:1997eg}. 
This action is compared with the low-energy effective action 
obtained by compactifying a general six-dimensional $(1,0)$ action with 
Abelian and non-Abelian gauge groups on a circle. The M-theory reduction 
and the circle compactification have to be compared in the regime where 
all massive modes are integrated out. On the circle side this implies that 
both massive modes arising from moving to the five-dimensional Coulomb branch 
and the modes arising in the Kaluza-Klein tower have to be integrated out \cite{Bonetti:2011mw}. 
The classical comparison of the two five-dimensional theories allows deriving the 
various F-theory couplings specifying the six-dimensional action. Importantly, the 
loop corrections arising from integrating out the massive modes teaches us about 
the six-dimensional charged spectrum even in the phase where the Calabi-Yau threefold 
is smooth.

Before comparing the five-dimensional theories, one first has to integrate out 
massive modes arising in the Coulomb branch after circle reduction. In order to do that, we compute 
the one-loop Chern-Simons levels using the general results of \cite{Bonetti:2013ela} (see also 
references therein). The massive modes include fields that become massive 
when moving to the five-dimensional Coulomb branch of the gauge group. 
These modes can be charged under the non-Abelian as well as the Abelian gauge
groups of the six-dimensional theory. Accordingly, they introduce gauge and 
gravitational Chern-Simons terms involving the dimensionally reduced  
Cartan generators of the six-dimensional gauge group.
In addition, also all massive Kaluza-Klein modes have to 
be integrated out. These are charged under the Kaluza-Klein vector
arising in the reduction of the six-dimensional to the five-dimensional metric,
and may also admit charges under the six-dimensional gauge group.
Integrating out all massive spin--$1/2$, spin--$3/2$ and tensor modes, we 
derive the form of the Chern-Simons levels in the presence of
six-dimensional $U(1)$-symmetries. Since these Chern-Simons terms are 
independent of the mass scale, one-loop corrections have to be included irrespective of the size of the 
compactifying circle and the values of the VEVs parametrizing the Coulomb 
branch vacuum. 
Supersymmetry further relates classical and one-loop Chern-Simons terms to 
the kinetic terms. This allows to determine the characteristic data specifying  
the complete six-dimensional (1,0) vector and tensor sector. The hypermultiplet sector can 
only be studied for the neutral multiplets that reduce trivially from six to five dimensions.
While the number of charged hypermultiplets can be determined via the one-loop Chern-Simons terms,
their precise metric remains elusive.

Having derived the 6D (1,0) spectrum and the characteristic data of the F-theory 
effective action, one can verify that anomaly conditions are fulfilled. 
The 6D anomaly conditions were translated into intersection relations in 
\cite{Sadov:1996zm,Kumar:2009ac,Grassi:2011hq,Park:2011ji}, where, however, it was implicitly assumed that the 
zero section was holomorphic. We show that loop-induced 
Chern-Simons terms can be employed to analyze equivalent 
conditions even in the case of a non-holomorphic zero section.

Hence, one particular focus of this work lies on Calabi-Yau threefold examples with rational 
sections. In particular, if the zero section is not holomorphic, then 
certain topological identities used in \cite{Park:2011ji,Bonetti:2011mw,Grimm:2012yq} have to be refined. We show 
that this has a crucial impact on the F-theory limit and the structure of the five-dimensional 
one-loop Chern-Simons terms. For a non-holomorphic zero section, one can no longer independently 
shrink the generic torus fiber of the elliptic fibration and the exceptional divisors resolving
the non-Abelian singularity. The physical interpretation of this situation for the 
circle reduced 6D (1,0) theory is that some of the 
Kaluza-Klein modes are in fact lighter than the Coulomb branch modes. 
Let us denote by $m_{CB}$ the Coulomb branch mass of a given mode and by $m_{KK}=1/r$
the Kaluza-Klein scale of a circle with circumference $r$. Then 
the inequality  
\beq \label{normal_hierarchy}
    m_{CB} \ < \ m_{\rm KK}
\eeq
considered in \cite{Bonetti:2011mw,Cvetic:2012xn} is no longer satisfied for all states when dealing with 
a non-holomorphic zero section. 
The effect of this modified hierarchy 
is to induce a discrete shift in the one-loop Chern-Simons levels, as was already noted in \cite{Cvetic:2012xn}.
In particular, this implies that various cancelations among 
contributions of Kaluza-Klein modes, which were previously encountered for 
holomorphic zero sections, do not occur. 
On the resolved Calabi-Yau threefold, the mass hierarchy is dictated 
by the Mori cone of the manifold. We find it necessary to replace the relative Mori cone
by a more refined version, the extended relative Mori cone introduced in a different context in \cite{Grimm:2011fx},
which captures not only information about the mass hierarchy among the curves associated to the gauge groups,
but also contains data about the zero section.

The paper is organized as follows.
To begin with, we review in \autoref{On_geometry} the geometric properties of elliptically fibered 
Calabi-Yau threefolds with rational sections. In \autoref{sec:eff_action} we derive the effective action 
of 6D F-theory compactifications with Abelian gauge factors by comparing a circle reduced 
$\mathcal{N}=(1,0)$ supergravity on the Coulomb branch and M-theory on a resolved Calabi-Yau 
threefold. We proceed in \autoref{sec:one-loop_CS} by including one-loop induced Chern-Simons 
terms in the circle reduced theory and relating them to 6D anomalies. We formulate the matching of 
the one-loop Chern-Simons coefficients with the analogous expressions in M-theory. Finally, we verify 
the matching in some examples in \autoref{s:examples}. In all examples we find a connection between 
the existence of a holomorphic zero section in the geometry and a hierarchy of Coulomb branch masses 
$m_{CB}$ and Kaluza-Klein masses $m_{KK}$. This suggests an improved systematic 
of performing the F-theory limit.

\section{On the geometry of Calabi-Yau threefolds with Abelian gauge factors} \label{On_geometry}

 In this section we review some of the essential features of elliptically fibered 
 Calabi-Yau manifolds. In particular, we recall the connection between the 
 Mordell-Weil group of the elliptic fiber and aspects of Abelian gauge group factors in 
 the resulting effective theory. Furthermore, we comment on the implications of having 
 a non-holomorphic zero section, which has recently appeared in examples constructed in 
 \cite{Braun:2013yti,Cvetic:2013nia}. A more detailed geometric discussion of its features 
 can be found in \cite{Cvetic:2013nia}. 
 While much of the following discussion is independent of the dimension of the base manifold, 
 we deal with Calabi-Yau threefolds in this work and hence 
 take $\dim_{\mathbb{C}} \mathcal{B} = 2$.
  
  \subsection{General remarks on elliptic fibrations}
  
 A fibered Calabi-Yau threefold is defined by specifying its total space $Y_3$ 
 together with a holomorphic and surjective map
 \begin{equation}
  \pi: Y_3 \to \mathcal{B}
 \end{equation}
 called the projection map defining the base manifold $\mathcal{B}$. $Y_3$ is said 
 to be elliptically fibered if the preimages of points $p$ on $\mathcal{B}$ under $\pi$ 
 are elliptic curves
 \begin{equation}
    \mathcal{E} = \pi^{-1}(p) \quad \mathrm{for} \ p \in \mathcal{B}\ .
 \end{equation}
 Up to \emph{birational} equivalence\footnote{Two varieties are said to be birationally 
 equivalent if they are isomorphic to each other inside open subsets. Here, openness
 is to be understood in terms of the Zariski topology, in which every open subset is automatically dense. 
 Therefore two birationally equivalent varieties can only differ inside lower-dimensional subsets.}, 
 every elliptic curve can be described in terms of a Weierstrass equation 
  \begin{equation}
   y^2 = x^3 + f x z^4 + g z^6\ , \label{e:weierstrass}
  \end{equation}
  where $z$, $x$ and $y$ are the homogeneous coordinates of $\mathbb{P}_{1,2,3}$ and $f$ and $g$ 
  are elements of the field $K$ that $\mathcal{E}$ is defined over. In terms of the Weierstrass 
  coefficients $f$ and $g$, the discriminant of the torus is given as
  \begin{equation}
   \Delta = 4 f^3 + 27 g^2
  \end{equation}
  and the elliptic curve is singular when $\Delta = 0$.

  Given an elliptic curve $\cE$, its Mordell-Weil group $\MW(\cE)$ is formed by 
  the set\footnote{Since two rational points on an elliptic curve can be added to obtain a third rational point, 
  addition endows this set with a group structure. The neutral element is the zero point on $\cE$. For a 
  concrete example see the appendix of \cite{Morrison:2012ei}.} of rational points on $\mathcal{E}$. $\MW(\cE)$ 
  is a finitely generated Abelian group, and therefore takes the following form
  \begin{align}
   \MW(\cE) = \mathbb{Z}^{r} \oplus T\,,
  \end{align}
where $r$ is the rank of the Mordell-Weil group
and $T$ is the torsion subgroup, which 
is the finite Abelian group completing the decomposition.
  
In our specific case we consider fibrations of $\cE$ over the base manifold $\cB$. 
$K$ is then the fraction field of the coordinate ring of the base manifold and in order 
for $Y_3$ to be a Calabi-Yau threefold, $f$ and $g$ must be sections of $K_{\mathcal{B}}^{-4}$
and $K_{\mathcal{B}}^{-6}$, respectively, where $K_{\mathcal{B}}$ is the canonical 
bundle of the base manifold. Rational points, including the zero point on $\mathcal{E}$, 
then become rational sections of the Calabi-Yau threefold, which we 
denote by $\sigma_m$.
  
As explained for example in \cite{Morrison:2012ei,Mayrhofer:2012zy,Braun:2013yti,Cvetic:2013nia}, rational sections 
do not necessarily define injective maps from the base manifold into the Calabi-Yau threefold anymore. 
Instead, they are allowed to wrap entire fiber components over certain loci in the base manifold. 
Therefore they are only properly defined over the blow-up $\hat{\cB}$ along these loci
\begin{align}
  \sigma_m: \hat{\cB} \to \cB \hookrightarrow Y_3\,,
\end{align}
where the first arrow is given by the blow-down map.
For Weierstrass models defined by \eqref{e:weierstrass}, the zero section given
by $[x:y:z] = [1:1:0]$ is always holomorphic, as was noted in \cite{Cvetic:2013nia}. 
However, since an arbitrary elliptic curve can be mapped to such a model only \emph{birationally}, 
there exist elliptic fibrations for which none of the sections is holomorphic.
While rational zero sections still describe physically well-defined 
compactifications\footnote{See for example the manifolds in \cite{Braun:2013yti,Cvetic:2013nia}.}, in most 
examples in the existing literature the zero section is taken to be holomorphic. We therefore try to make 
sure to point out the implications of having a non-holomorphic zero section. In particular, we argue 
that in this case the mass hierarchy \eqref{normal_hierarchy} can be violated.
  
  \subsection{Topology of elliptically fibered Calabi-Yau threefolds}
  
  Having recalled the key features of an elliptic fibration, we now choose our notation 
  essentially as in \cite{Grimm:2011fx,Bonetti:2011mw,Cvetic:2012xn,Cvetic:2013nia} and select a 
  convenient basis of divisors.
  To be able to perform meaningful calculations, we take $\hat{Y_3} \to Y_3$ to be 
  the smooth blow-up of $Y_3$ along all singular loci. We then choose the following 
  basis of divisors $D_\Lambda$ and their respective dual two-forms $\omega_\Lambda \in H^{1,1}(\hat{Y_3}, \mathbb{Z})$:
  \begin{itemize}
   \item The divisor $D_0$ dual to the two-form along which $C_3$ is expanded 
   to give the Kaluza-Klein 
   vector field $A^0$. $D_0$ is obtained by shifting the zero section $D_{\hat{0}}$ 
   according to \eqref{e:base_shift}.
   \item Vertical divisors $D_{\alpha} = \pi^{*}(D_{\alpha}^b),\ \alpha=1,\ldots, h^{1,1}(\cB)$ 
   obtained as pullbacks of a basis of divisors $D_{\alpha}^b$ on $\cB$.
   \item Exceptional divisors $D_{I}$ obtained by resolving singularities of the elliptic 
   fibration along the divisor $S^b$ in the base manifold.
   \item $U(1)$ divisors $D_m$ obtained by applying the Shioda map given 
   in \eqref{e:shioda} to each of the rational sections $\sigma_m$, where 
   $m = 1,\ldots,\rk \MW(Y_3)$.
  \end{itemize}
  
  In order to define the shifts mentioned above, it is convenient to introduce 
  the intersection product on the base manifold as 
  \begin{align}
   D^b_{\alpha} \cdot D^b_{\beta} = ( D_{\alpha} \cdot D_{\beta} )_{\cB} 
                                                 = D_{\alpha} \cdot D_{\beta} \cdot D_{\hat{0}} \equiv \eta_{\alpha \beta}\,, \label{e:intersection_base}
  \end{align}
  so that we can lower and raise Greek indices using $\eta_{\alpha \beta}$ and 
  its inverse, $\eta^{\alpha \beta}$. Furthermore, we can project a two-cycle $X \subset Y_3$ to the base via
  \begin{equation}
   \pi(X) = \left(X \cdot D^{\alpha} \right) D_{\alpha}^b\,. \label{e:projection_two_cycle}
  \end{equation}
  As was noted in \cite{Grimm:2011sk,Park:2011ji,Cvetic:2012xn}, 
  $D_0$ is obtained by requiring that
  \begin{align}
   D_0 \cdot D_0 \cdot D_\alpha = 0\,,
  \end{align}
  which can be achieved by choosing
  \begin{equation}
    D_0 = D_{\hat{0}} - \frac{1}{2} ( D_{\hat{0}} \cdot D_{\hat{0}} \cdot D^\alpha ) D_{\alpha}\;. \label{e:base_shift}
  \end{equation}
  
  In a similar fashion, the Shioda map shifts the rational sections $\sigma_m$ 
  such that specific intersection numbers of $D_m$ with $D_0, D_I$ and $D_{\alpha}$ vanish, as we will see in \eqref{e:i_shioda}. This orthogonalization procedure turns out to be crucial for the matching of M-theory and F-theory later. First, however, 
  we must recall the intersection properties of the exceptional divisors obtained by 
  blowing up the singularity of the elliptic fibration. Given a base divisor $S^b$ over 
  which the elliptic fiber of $Y_3$ develops singularities,
  the blow-up divisors of $\hat{Y}_3$ intersect as
  \begin{align}
   D_{I} \cdot D_{J} \cdot D_{\alpha} = - \cC_{I J} \left( S^b \cdot D^b_{\alpha} \right)\ , \label{e:intersection_cartan}
  \end{align}
  where $\cC_{I J}$ denotes the coroot intersection matrix, which we define in the group theory conventions of appendix \ref{app:group_id}.
  
  Having 
  associated the exceptional blow-up divisors $D_{I}$ with the Cartan generators of $\fg$, 
  one can go a step further and define a rational curve localized over a single point in the base 
  manifold for each root of $\fg$.  For the simple roots $\alpha_I$ of $\fg$, one chooses a base divisor 
  $D^b$ intersecting $S^b$ exactly once and takes the intersection product between $D = \pi^{*} (D^b)$ and $D_I$:
  \begin{equation}
   \cC_{\alpha_{I}} = - D_{I} \cdot D \qquad \textrm{for} \quad D^b \cdot S^b = 1
  \end{equation}
  From \eqref{e:intersection_cartan}, one can see that the intersection $D_I \cdot \cC_{\alpha_J}$ reproduces the $I$th component
  of the simple root $\alpha_J$ in the Dynkin basis of the root system of $\fg$.
  With these definitions, we are ready to give an explicit formula for the Shioda 
  map relating rational sections $\sigma_m$ and their associated $U(1)$-divisors:
  \begin{equation}
   D_m = \sigma_m - D_{\hat{0}} - \left( (\sigma_m - D_{\hat{0}}) \cdot D_{\hat{0}} \cdot D^{\alpha} \right) D_{\alpha}
   - \left( \sigma_m \cdot \cC_{\alpha_{I}} \right) \left(\cC^{-1} \right)^{I J} D_{J} \label{e:shioda}
  \end{equation}
  
  Let us now discuss the intersection numbers in this basis and emphasize clearly 
  what impact a rational zero section has. We begin by examining the geometry 
  of the blow-up divisors $D_I$. A \emph{holomorphic} zero section marks a single 
  point in each fiber. In particular, when this point lies over $S^b$, it is on the original fiber component\footnote{Assuming 
  that the resolution locus in the base is $S^b$, one can associate the divisor $\pi^{*}(S^b) - \sum_I D_I$ 
  with the affine node of the Dynkin diagram of $\fg$. Intersecting this divisor with $\pi^{*}(D^b)$ 
  such that $D^b \cdot S^b=1$ gives the rational curve associated with the original fiber component.}
  and not on the resolution $\mathbb{P}^1$s of the rational curves $\cC_{\alpha_I}$. 
  Therefore the following equation holds as an identity in the Chow ring of $\hat{Y}_3$:\footnote{The Chow ring of 
  an algebraic variety $X$ is formed by equivalence classes of the subvarieties of $X$, where the equivalence 
  relation is given by rational equivalence. The multiplicative structure is defined by taking the intersection of two subvarieties.}
  \begin{equation}
   D_{\hat{0}} \cdot D_I = 0\ , \qquad \textrm{if } D_{\hat{0}} \textrm{ is holomorphic.} \label{e:i_0I}
  \end{equation}
  On the other hand, a rational zero section may wrap the entire fiber component over 
  lower-dimensional loci of the base. Since this fiber component intersects the resolution 
  divisors as the affine node in the extended Dynkin diagram of $\fg$, its intersection 
  \emph{can} be non-zero.
  However, since the locus over which a rational zero section can wrap the entire fiber 
  component has at least codimension one in the base and is generically different from 
  $S^b$, $D_{\hat{0}} \cdot D_I$ has at least codimension two in the base manifold. 
  The intersection with a vertical divisor therefore vanishes and we find that
  \begin{equation}
   D_{\alpha} \cdot D_{\hat{0}} \cdot D_I = 0
  \end{equation}
  even for a non-holomorphic zero section.
  
  The other peculiarity of having a non-holomorphic zero section is that one can no longer
  evaluate expressions involving $D_{\hat{0}}$ by using adjunction to the base manifold.
  Recall that
  \begin{equation}
   D_{\hat{0}} \cdot D_{\hat{0}} = D_{\hat{0}} \rvert_{\cB} = K_{\cB}\ , \qquad \textrm{if } D_{\hat{0}} \textrm{ is holomorphic.} \label{e:adjunction}
  \end{equation}
  However, for a rational zero section this needs no longer be the case, since the divisor $D_{\hat{0}}$ and $\cB$ are only rationally equivalent, but \emph{not isomorphic}.
  
  To put it in a nutshell, a rational zero section may intersect blow-up divisors over points 
  in the base and the divisor corresponding to that section is no longer isomorphic to the base manifold. With this in mind, we can now 
  list the intersection numbers both for a rational zero section and for its holomorphic 
  counterpart. We begin by stating intersections that hold \emph{both for a rational and 
  for a holomorphic} zero section:
  \begin{subequations} \label{e:intersections_rational}
  \begin{align}
    &D_{\alpha} \cdot D_{\beta} \cdot D_{\gamma}  = 0\, ,  && D_{0} \cdot D_{\alpha} \cdot D_{\beta} = \eta_{\alpha \beta}\, ,
    && D_{0} \cdot D_{0} \cdot D_{\alpha} = 0\, , & \label{e:i_base} \\  
    &D_{\alpha} \cdot D_{\beta} \cdot D_{I} = 0\, , &&D_{\alpha} \cdot  D_0 \cdot D_{I} = 0\, ,  && D_{\alpha} \cdot D_{I} \cdot D_{J} = - \cC_{I J} (S^b \cdot D_{\alpha}^b)\, , & \label{e:i_cartan} \\
    &D_{\alpha} \cdot D_{\beta} \cdot D_m = 0\, , & & D_{\alpha} \cdot D_{I} \cdot D_m  = 0 \, , & & D_0 \cdot D_{\alpha} \cdot D_m = 0\, ,\label{e:i_shioda} \\
    &&&&& D_{\alpha} \cdot D_m \cdot D_n = \pi (D_m \cdot D_n )_\alpha \, . & \label{e:i_pi}
  \end{align}
  \end{subequations}
  All three equations in \eqref{e:i_base} describe intersections on the base manifold. 
  The first one is a triple intersection product between codimension 1 objects in the 
  base and therefore vanishes. Using this fact, the second equation simply reduces 
  to the definition in \eqref{e:intersection_base} and the third equation can be 
  verified directly by inserting \eqref{e:base_shift}.
  Next of all, the three equations in \eqref{e:i_cartan} are a direct consequence of 
  the blow-up geometry and were discussed above.
  Equation \eqref{e:i_pi} is just a formal rewriting of the intersection number using \eqref{e:projection_two_cycle} and we stress that unlike in \cite{Cvetic:2012xn}, we do not require $D_m$ and $D_n$ be orthogonal to each other.
  Lastly, the remaining three 
  equations \eqref{e:i_shioda} follow from the orthogonalization 
  properties of the Shioda map. They can be verified by inserting the expression 
  in \eqref{e:shioda} and exploiting that all sections intersect the generic fiber 
  component precisely once, that is
  \begin{equation}
   \sigma_m \cdot \cE = D_{\hat{0}} \cdot \cE = D_0 \cdot \cE = 1\,, \label{e:i_section_fiber}
  \end{equation}
  where the class of the generic fiber $\cE$ is given as
  \begin{equation}
   D_{\alpha} \cdot D_{\beta} = \cE \eta_{\alpha \beta}\,.
  \end{equation}
  
In a second step, we now assume to have a \emph{holomorphic} zero 
section $D_{\hat{0}}$. Using the definition of the Shioda map we evaluate
  \begin{equation}
   D_{\hat{0}} \cdot D_m = 0\ , \qquad \textrm{if $D_{\hat{0}}$ is holomorphic.} \label{e:i_0m}
  \end{equation}
Exploiting \eqref{e:i_0I}, \eqref{e:i_0m} and \eqref{e:adjunction} 
one can then show that
\begin{subequations} \label{e:intersections_holomorphic}
\begin{align}
   &D_{0} \cdot D_{m} \cdot D_n 
    = -\frac{1}{2}\pi (D_m \cdot D_n)_\alpha K^\alpha\, ,  \qquad D_{0} \cdot D_{I} \cdot D_{J} 
  = \frac{1}{2} K^{\alpha} (D_{\alpha} \cdot D_{I} \cdot D_{J})\, ,     \label{e:i_0gauge} \\
    &D_{0} \cdot D_{0} \cdot D_I = 0\, , \qquad  \qquad 
      D_{0} \cdot D_{0} \cdot D_m = 0\, , \qquad  \qquad 
      D_0 \cdot D_{I} \cdot D_m = 0\, , \label{e:i_hol_0I} \\
    &D_{0} \cdot D_{0} \cdot D_{0} = \frac{1}{4} K^{\alpha} K_{\alpha}\, ,    \label{e:i_000}
\end{align}
\end{subequations}
where $K^{\alpha}$ are the expansion coefficients of the canonical class of $\cB$ 
in $K_{\cB} = K^{\alpha} D^b_{\alpha}$. All equations in \eqref{e:i_hol_0I} are a direct 
consequence of \eqref{e:i_0I} and \eqref{e:i_0m}. Equation \eqref{e:i_000} follows from applying the adjunction formula.
Finally, the two equations in \eqref{e:i_0gauge} both follow from applying \eqref{e:i_0I}, \eqref{e:i_0m} and the adjunction formula.
We stress that \eqref{e:intersections_holomorphic} are not valid for a non-holomorphic zero section.

\section{Effective action via M-theory}
\label{sec:eff_action}

In this section we perform the circle reduction of a 6D $(1,0)$ supergravity with Abelian gauge factors describing
the F-theory effective action. We push the reduced theory to the Coulomb branch and compare it with M-theory on a resolved
elliptically fibered Calabi-Yau threefold. We focus in particular on the Chern-Simons terms and the prepotential of the
resulting 5D $\mathcal{N}=2$ supergravity. On the M-theory side, we find classical terms surviving the F-theory limit and
one-loop induced terms that vanish in the limit. In contrast, the F-theory side only accounts for the classical terms in 
the reduction. By matching the classical terms, we find a geometric interpretation of the 6D F-theory data.

\subsection{Reducing $(1,0)$ supergravity on a circle}
\label{sec:circle_reduction}
The effective action of F-theory compactified on a 
singular Calabi-Yau threefold is a 6D $(1,0)$ supergravity theory.
Let us denote the 6D space-time manifold by $M_6$.
The $(1,0)$ supermultiplets with their individual bosonic and fermionic 
constituents are listed in the Table \ref{T:spectrum_(1,0)}.
In the following, we denote the number of vector multiplets by $V$, the number of tensor 
multiplets by $T$, and the number of hypermultiplets by $H$.

\begin{table}[h]
\begin{center}
\begin{tabular}{|c||l|}
\hline \rule[-.2cm]{.0cm}{.7cm} multiplet & field content \\
\hline \hline
\rule[-.2cm]{.0cm}{.7cm} gravity & 1 graviton, 1 left-handed Weyl gravitino, 1 self-dual 2-form \\ 
\hline 
\rule[-.2cm]{.0cm}{.7cm} vector & 1 vector, 1 left-handed Weyl gaugino \\ 
\hline 
\rule[-.2cm]{.0cm}{.7cm} tensor & 1 anti-self-dual 2-form, 1 right-handed Weyl tensorino, 1 real scalar \\ 
\hline 
\rule[-.2cm]{.0cm}{.7cm} hyper & 1 right-handed Weyl hyperino, 4 real scalars \\ 
\hline 
\end{tabular}
\caption{The spectrum of 6D $(1,0)$ supergravity. Note that one can substitute each Weyl spinor by two symplectic Majorana-Weyl spinors.} \label{T:spectrum_(1,0)}
\end{center}
\end{table}

We allow for a non-Abelian gauge group $G$, which splits into a simple non-Abelian part $G_{nA}$ and $n_{U(1)}$ $U(1)$-factors as
\begin{align}
G=G_{nA} \times U(1)^{n_{U(1)}}.
\end{align}
Our goal is to find the F-theory effective action of a $(1,0)$ theory with gauge group $G$. Since the tensors 
in the spectrum obey (anti-)self-duality constraints, we can only give a pseudo-action for this theory
for which the additional constraints have to be imposed manually at the level of the equations of motion.
For the sake of simplicity we only display the bosonic part of this pseudo-action. The fermionic couplings can then
be inferred by using the general supergravity actions found in \cite{Nishino:1997ff,Ferrara:1997gh,Riccioni:1997ik,Riccioni:1999xq}. 
Our conventions are summarized in appendix \ref{a:conventions} and follow largely the ones used in \cite{Bonetti:2011mw}.

Let us collectively denote the anti-self-dual tensors from the tensor multiplets and the self-dual tensor from the gravity multiplet 
by $\hat{B}^\alpha$, $\alpha = 1\dots T+1$. 
The real scalars in the tensor multiplets parametrize the manifold
\begin{align}
SO(1,T)/SO(T) \ .
\end{align}
For a convenient description of this coset space we introduce $T+1$ scalars $j^\alpha$ and a constant 
metric $\Omega_{\alpha \beta}$ with signature $(+,-,\ldots,-)$. Due to the constraint
\begin{align}
\Omega_{\alpha \beta}j^\alpha j^\beta \overset{!}{=} 1
\end{align} 
one scalar degree of freedom is redundant. Furthermore, it is useful to define another non-constant positive metric
\begin{align}
g_{\alpha \beta} = 2 j_\alpha j_\beta -\Omega_{\alpha \beta}\ .
\end{align}
Here and in the following indices are raised and lowered using $\Omega_{\alpha \beta}$.

The gauge connection for the simple non-Abelian group is denoted by $\hat{A}$ 
and the Abelian ones are denoted by $\hat{A}^m$, where $m=1\dots n_{U(1)}$. The field strength two-forms read
\begin{align}
\hat{F}=d\hat{A}+\hat{A}\wedge\hat{A}\ , \qquad \hat{F}^m = d \hat{A}^m
\end{align}
and the Chern-Simons forms are defined as
\begin{align}
\hat{\omega}^{CS}=\tr(\hat{A}\wedge d\hat{A}+\frac{2}{3}\hat{A} \wedge \hat{A} \wedge \hat{A})\ , \qquad \hat{\omega}^{CS,mn}=\hat{A}^m \wedge d\hat{A}^n\ .
\end{align}
In order to distinguish 6D and 5D fields, we use hats for fields of the 6D theory. 

Let us now turn to the gravity sector, which is described by the spin connection $\hat{\omega}$
on $M_6$, the curvature two-form
\begin{align}
\hat{\mathcal{R}}= d\hat{\omega}+\hat{\omega} \wedge \hat{\omega}
\end{align}
and the Ricci-Scalar $\hat{R}$. The gravitational Chern-Simons form is defined as
\begin{align}
\hat{\omega}_{grav}^{CS}=\tr(\hat{\omega}\wedge d \hat{\omega} + \frac{2}{3} \hat{\omega} \wedge \hat{\omega} \wedge \hat{\omega})\ .
\end{align}

Moreover, there are four real scalars in each hypermultiplet, which we collectively denote by $q^U$, $U=1\dots 4H$. 
These parametrize a quaternionic manifold with metric $h_{UV}$. Since the hypermultiplets may transform in some representation 
$\rep{R}$ of the simple non-Abelian gauge group and may also carry $U(1)$-charges, we introduce the covariant derivative
\begin{align}
\hat{D}q^U=dq^U + \hat{A}^{\rep{R}} q^U - i q_m \hat{A}^m q^U ,
\end{align}
where $\hat{A}^{\rep{R}}$ denotes the Lie-algebra valued gauge connection of $G_{nA}$ in the representation $\rep{R}$.

Since the 6D $(1,0)$ spectrum is chiral, the theory is potentially anomalous. For some spectra, one can employ
the Green-Schwarz mechanism \cite{Green:1984sg,Sagnotti:1992qw,Sadov:1996zm} to cancel these anomalies.
We therefore include the Green-Schwarz counterterm in the action, which reads
\begin{align} \label{e:GS_factor}
\hat{S}^{GS}=-\frac{1}{2} \int _{M_6}\Omega_{\alpha \beta}\hat{B}^\alpha \wedge \hat{X}_4^\beta\ ,
\end{align}
where
\begin{align}
\hat{X}_4^\alpha = \frac{1}{2}a^\alpha \tr\,\hat{\mathcal{R}}\wedge\hat{\mathcal{R}}
                             +2\frac{b^\alpha}{\lambda (\mathfrak{g})} \tr\,\hat{F}\wedge\hat{F} 
                             + 2b^\alpha_{mn}\hat{F}^m \wedge \hat{F}^n\ .
\end{align}
The constants $a^\alpha$, $b^\alpha$, $b^\alpha_{mn}$ will later be given in terms 
of geometrical data of the internal Calabi-Yau space. We have furthermore inserted a group theoretical 
factor $\lambda (\mathfrak{g})$ defined in \eqref{lambda} for later convenience.
The Green-Schwarz term can be used to cancel those anomalies whose anomaly polynomial factorizes as
\begin{align}\label{factorization}
\hat{I}_8 = -\frac{1}{2}\Omega_{\alpha \beta}\hat{X}_4^\alpha \wedge \hat{X}_4^\beta\ ,
\end{align}
provided that we assign an appropriate transformation to the tensors under gauge and local Lorentz transformations, which turns out to be
\begin{align}
 \delta \hat B^\alpha = d \hat \Lambda^\alpha - \frac{1}{2}a^\alpha \tr \, \hat l d \hat \omega -2 b^\alpha \tr \, \hat \lambda d \hat A -2 b^\alpha_{mn}  \hat \lambda^m d \hat A^n \ , 
\end{align}
where $\hat l$, $\hat \lambda$, $\hat \lambda^m$ are the parameters of local Lorentz and gauge transformations respectively
\begin{align}
 \delta \hat\omega = d \hat l + [\hat\omega , \hat l ]\ , \qquad
 \delta \hat A = d \hat \lambda + [\hat A , \hat \lambda ]\ , \qquad
 \delta \hat A^m = d \hat \lambda^m
\end{align}
and the one-forms $\hat \Lambda^\alpha$ encode the standard gauge transformations of two-forms.
The precise conditions the matter spectrum has to satisfy in order for the factorization \eqref{factorization} to take place
will be reviewed in \autoref{anom_rev}.
The gauge invariant field strength for the tensors then takes the form
\begin{align} \label{def-hatG}
\hat{G}^\alpha = d\hat{B}^\alpha + \frac{1}{2}a^\alpha\hat{\omega}^{CS}_{grav} + 2\frac{b^\alpha}{\lambda (\mathfrak{g})}\hat{\omega}^{CS} + 2b^\alpha_{mn}\hat{\omega}^{CS,mn}\ .
\end{align}
Note that the $\hat{G}^\alpha$ are subject to a duality constraint
\beq \label{self_duality}
g_{\alpha \beta} \hat{\ast}\hat{G}^\beta = \Omega_{\alpha \beta} \hat{G}^\beta\ ,
\eeq
which has to be enforced in addition to the equations of motion 
derived from the pseudo-action.
The bosonic part of the pseudo-action for 6D $(1,0)$ supergravity with gauge group $G$ reads
\begin{align}\begin{split}
\label{e:6D_action}
\hat{S}^{(6)}  = \int _{M_6}& + \frac{1}{2} \hat{R} \hat{\ast}1 -\frac{1}{4} g_{\alpha \beta} \hat{G}^\alpha \wedge \hat{\ast} \hat{G}^\beta 
                      - \frac{1}{2} g_{\alpha \beta} dj^\alpha \wedge \hat{\ast} dj^\beta - h_{UV} \hat{D}q^U \wedge \hat{\ast} \hat{D}q^V\\
&  -2 \Omega_{\alpha \beta} j^\alpha \frac{b^\beta}{\lambda (\mathfrak{g})} \tr\, \hat{F} \wedge \hat{\ast} \hat{F} -2 \Omega_{\alpha \beta} j^\alpha b^\beta _{mn} \hat{F}^m \wedge \hat{\ast} \hat{F}^n\\
&  - \Omega_{\alpha \beta} \frac{b^\alpha}{\lambda (\mathfrak{g})} \hat{B}^\beta \wedge \tr\, \hat{F}\wedge \hat{F} - \Omega_{\alpha \beta} b^\alpha _{mn} \hat{B}^\beta \wedge \hat{F}^m \wedge \hat{F}^n\\
&-\frac{1}{4}\Omega_{\alpha \beta}a^\alpha \hat{B}^\beta \wedge \tr\, \hat{\mathcal{R}} \wedge \hat{\mathcal{R}} -\hat{V} \hat{\ast}1\ ,
\end{split}\end{align}
where $\hat V$ is the scalar potential. In the following we do not need the precise 
form of $\hat V$ and refer for example to~\cite{Nishino:1997ff,Ferrara:1997gh,Suzuki:2005vu,Bergshoeff:2012ax,Grimm:2013fua} 
for more details.

In a next step we compactify this theory on a circle of radius $r$
and thus choose the 6D space-time to be of the form $M_6=S^1 \times M_5$. 
Let us briefly summarize the results of this reduction 
here and defer technical details and conventions to \autoref{app:circle_reduction}.
The coordinate along the circle is denoted by $y$.
We write $A^0$ for the Kaluza-Klein vector and call the corresponding field-strength $F^0 = dA^0$.
Let us also define $Dy=dy-A^0$.
Recall that expressions without hats are of 5D origin and are hence independent of $y$.
It is important to stress here that we only approach a two-derivative reduction for the moment. 
We therefore also neglect higher curvature contributions. This implies that we can omit the 
gravitational contribution in the Green-Schwarz terms \eqref{e:GS_factor}
and all other gravitational contributions from the tensors proportional to $a^\alpha$. Later on, we revisit
these terms and discuss them in more detail.

Hypermultiplets in six dimensions reduce trivially to 5D hypermultiplets.
The 6D vectors $\hat{A}$, $\hat{A}^m$ reduce to 5D vectors $A$, $A^m$ and scalars $\zeta$, $\zeta^m$.
Tensors $\hat{B}^\alpha$ in the six--dimensional theory reduce to 5D tensors $B^\alpha$ with 
field-strength $G^\alpha$ and vectors $A^\alpha$ with field-strength $F^\alpha = dA^\alpha$.
These reductions can be inserted into the 6D pseudo-action. 
One then has to integrate over the circle direction to obtain a 5D pseudo-action. 
Reducing the (anti-)self-duality constraint \eqref{self_duality} yields a relation 
between the tensor field-strength $G^\alpha$ and the vector
field-strength $\mathcal{F}^\alpha$ given by
\begin{align} \label{modified_cF}
\mathcal{F}^\alpha = F^\alpha - 4 \frac{b^\alpha}{\lambda (\mathfrak{g})} \tr (\zeta F) 
                               + 2 \frac{b^\alpha}{\lambda (\mathfrak{g})} \tr(\zeta \zeta)F^0 
                               -4 b^\alpha _{mn} \zeta ^m F^n +2 b^\alpha _{mn} \zeta ^m \zeta ^n F^0\ .
\end{align}
This condition can be used to obtain a proper 5D supergravity action depending 
only on $\mathcal{F}^\alpha$ by eliminating the dependence of the 5D pseudo-action 
on the tensors $B^\alpha$ in favor of the vectors $A^\alpha$. While this is always 
possible at the massless Kaluza-Klein level for the compactified 
tensors, doing so will no longer work at the massive level.
Furthermore, we also perform a Weyl rescaling to arrive at the canonical 
form of the Einstein-Hilbert term.

The last step is to push the theory onto the 5D Coulomb branch by switching on vacuum expectation values for the scalars 
in the vector multiplets. This results in giving mass terms to the W-bosons (and by supersymmetry also to their fermionic partners) 
and the charged hypermultiplets. The massive W-bosons break the simple non-Abelian gauge group to its 
maximal torus $U(1)^{\rk(G_{nA})}$. Below the mass scale characteristic of the gauge group breaking, all 
massive states have to be integrated out from the 5D effective action. We discuss the induced corrections
in \autoref{sec:one-loop_CS}.
On the massless level we are only left with the Cartan generators and the generators of the Abelian 
gauge symmetry, which generically stay massless. 
We thus find the residual gauge symmetry
\begin{align} \label{CB_group}
U(1)^{\rk(G_{nA})} \times U(1)^{n_{U(1)}}\ .
\end{align}
In the following, the $U(1)$s originating from the non-Abelian Cartan generators are labeled by 
$I=1,\dots, \rk(G_{nA})$.

Let us summarize the massless bosonic fields of the Coulomb branch effective theory 
and their completion into 5D $\mathcal{N}=2$ 
multiplets. We distinguish three types of 5D multiplets:
\begin{itemize}
\item The gravity multiplet consists of the 5D metric (graviton) and in general a linear 
        combination of $A^0$ and $A^\alpha$ (graviphoton).
\item We find $\rk(G_{nA})+ n_{U(1)} +T+1$ vector multiplets. 
        The vectors are $A^I$, $A^m$ and $T+1$ linear combinations of $A^0$ and $A^\alpha$. 
        The corresponding scalar degrees of freedom are provided by $\zeta^I$, $\zeta^m$, $r$ and $j^\alpha$ 
        supplemented by the constraint $\Omega_{\alpha \beta} j^\alpha j^\beta \overset {!}{=}1$ from the 6D theory.
        Recall that $\alpha = 1,\ldots , T+1$, $m=1,\ldots, n_{U(1)}$, and $I=1,\dots, \rk(G_{nA})$.
\item The only massless 5D hypermultiplets arise from $H^{\textrm{neutral}}$ 6D hypermultiplets that transform trivially under $G$. 
\end{itemize}

To specify the Coulomb branch action, we first need to introduce some additional 
notation.
The Cartan generators $\cT_I$ are chosen to be in the coroot basis, i.e.~we have the following relation to the Cartan generators
$T^M$ in the usual basis given around \eqref{norm_cartan}
\begin{align}
 \mathcal{T}_I = \alpha^\vee_I \cdot T \ .
\end{align}
According to the convention \eqref{norm_cartan}, the trace normalization for the Cartan generators in the coroot basis reads
\begin{align}
\tr \, (\mathcal{T}_I \mathcal{T}_J ) = \lambda (\mathfrak{g}) \mathcal{C}_{IJ}\ , 
\end{align}
where the coroot inner product matrix $\cC_{IJ}$ is defined in \eqref{def-cCIJ}.\footnote{Note that all roots and weights
appearing in this work are still associated to the Cartan generators $T^M$ and not to $\cT_{I}$.}

To simplify our expressions we introduce indices $\hat I=(I,m)$, $\hat J = (J,n)$, etc.~running over all $U(1)$s in the 
Coulomb branch group \eqref{CB_group}. In particular, we define
\beq \label{e:hat_notation}
  b^\alpha_{\hat I \hat J} = \left( \begin{array}{cc} b^\alpha \cC_{I J} & 0 \\ 
                                                      0 &  b^\alpha_{mn} 
                                    \end{array}\right)\ ,
\eeq
where $\hat I,\hat J = 1,\ldots, \rk(G)+n_{U(1)}$.

The 5D action on the Coulomb branch then reads
\begin{align} \label{circle_action}
S^{(5)F}  = \int _{M_5} &+ \frac{1}{2} R \ast 1 -\frac{2}{3}r^{-2} dr \wedge \ast dr 
         - \frac{1}{2} g_{\alpha \beta} dj^\alpha \wedge \ast dj^\beta -h_{uv} dq^u \wedge \ast dq^v \\
&  -2 r^{-2} \Omega_{\alpha \beta} j^\alpha b^\beta_{\hat I \hat J} \, d \zeta ^{\hat I} \wedge \ast d \zeta ^{\hat J}
    - \frac{1}{4} r^{8/3} F^0 \wedge \ast F^0   - \frac{1}{2} r^{-4/3} g_{\alpha \beta}\, \mathcal{F}^\alpha \wedge \ast \mathcal{F}^\beta  \nn  \\
&  
     -2 r^{2/3}  \Omega_{\alpha \beta} j^\alpha  b^\beta_{\hat I \hat J} \, (F^{\hat I} - \zeta ^{\hat I} F^0) \wedge \ast(F^{\hat J} - \zeta ^{\hat J} F^0) 
     + \cL_{\rm CS}^{\rm p} +  \cL_{\rm CS}^{\rm np} \nn \ ,
\end{align}
where  gauge-invariant Chern-Simons terms are given by
\beq \label{CSp}
  \cL_{\rm CS}^{\rm p} = - \frac{1}{2} \Omega _{\alpha \beta} \, A^0 \wedge F^\alpha \wedge F^\beta  
                                  +  2 \Omega_{\alpha \beta} b^\alpha_{\hat I \hat J} \, A^\beta \wedge F^{\hat I} \wedge F^{\hat J} \ ,
\eeq
and non-gauge-invariant Chern-Simons terms read
\begin{align} \label{CSnp}
\cL_{\rm CS}^{\rm np}  = 
& - 2\Omega_{\alpha \beta} b^\alpha_{\hat I \hat J} b^\beta_{\hat K \hat L} \zeta ^{\hat K} \zeta ^{\hat L} \zeta ^{\hat I} A^{\hat J} \wedge F^0 \wedge F^0 \\
&  + 2  \Omega_{\alpha \beta} (b^\alpha_{\hat I \hat J} b^\beta_{\hat K \hat L} + 2 b^\alpha_{\hat I \hat K} b^\beta_{\hat J \hat L}) \zeta ^{\hat K} \zeta ^{\hat L} A^{\hat I} \wedge F^{\hat J} \wedge F^0 \nn \\
&  - 2\Omega_{\alpha \beta} (2 b^\alpha_{\hat I \hat J} b^\beta_{\hat K \hat L} + b^\alpha_{\hat I \hat L} b^\beta_{\hat J \hat K}) \zeta ^{\hat L} A^{\hat I} \wedge F^{\hat J} \wedge F^{\hat K}\ . \nn
\end{align}
Note that the 5D expression $\cL_{\rm CS}^{\rm np}$ arises from the reduction of the 6D non-gauge-invariant 
Green-Schwarz term \eqref{e:GS_factor}. In contrast to six dimensions, $\cL_{\rm CS}^{\rm np}$ can be canceled by adding a 
one-loop counter-term in five-dimensions that renders the action gauge invariant \cite{Ferrara:1996wv,Bonetti:2011mw}.
In the vector field sector, we have only kept Cartan and Abelian gauge fields (and their respective scalar partners)
and, similarly, in the hyper sector also only the massless,
i.e.~uncharged scalars, denoted by $q^u$, $u=1\dots 4H^{\textrm{neutral}}$.

The information about the gravity and vector sector of 5D $\mathcal{N}=2$ supergravity is 
contained entirely in the real prepotential $\mathcal{N}$. In the canonical form of the supergravity, $\cN$
is a cubic polynomial in the scalar fields $M^\Lambda$. The $M^\Lambda$  are so-called very special coordinates and 
encode the scalar degrees of freedom in the 5D $\cN=2$ vector multiplets subjected to one 
normalization constraint
\begin{align}
\mathcal{N}\overset{!}{=}1\ ,
\end{align}
which reduces the degrees of freedom by one.
Generally, the prepotential can be written as 
\begin{align}
\mathcal{N}=\frac{1}{3!} k_{\Lambda\Sigma\Theta}M^\Lambda M^\Sigma M^\Theta ,
\end{align}
where $k_{\Lambda\Sigma\Theta}$ is constant and symmetric in all indices. 
The canonical form of the action then reads
\begin{align}\begin{split}\label{e:canonical_action}
S^{(5)}=\int_{M_5}&+\frac{1}{2}R\ast 1 -\frac{1}{2}G_{\Lambda \Sigma}dM^\Lambda\wedge\ast dM^\Sigma -h_{uv} dq^u\wedge\ast dq^v \\
&-\frac{1}{2}G_{\Lambda\Sigma}F^\Lambda \wedge \ast F^\Sigma - \frac{1}{12} k_{\Lambda\Sigma\Theta}A^\Lambda \wedge F^\Sigma \wedge F^\Theta .
\end{split}\end{align}
Note that the fields $A^\Lambda$ comprise the graviphoton and the vectors from the vector multiplet.
Here, we have also defined the metric 
\begin{align}\begin{split}\label{e:metric}
G_{\Lambda\Sigma}=-\frac{1}{2}\partial_{M^\Lambda}\partial_{M^\Sigma}\log \mathcal{N}\mid_{\mathcal{N}=1}\,.
\end{split}\end{align}

The effective action \eqref{circle_action} of the circle reduced 6D $(1,0)$ supergravity is
not yet in the canonical form \eqref{e:canonical_action} of 5D $\mathcal{N}=2$ supergravity
and we therefore have to perform a field redefinition. It turns out that the fields
\begin{align}\begin{split}
& M^0 = r^{-4/3} \\
& M^\alpha = r^{2/3} (j^\alpha +2 r^{-2} b^\alpha_{\hat I \hat J} \zeta ^{\hat I} \zeta ^{\hat J}) \\
& M^{\hat I} = r^{-4/3} \zeta ^{\hat I} 
\end{split}\end{align}
yield the right structure, which is analogous to the redefinition found in \cite{Bonetti:2011mw}.
Let us further define
\begin{align}
\mathcal{N}_{\rm p}^F = \Omega_{\alpha \beta} M^0 M^\alpha M^\beta -4 \Omega_{\alpha \beta} b^\alpha_{\hat I \hat J} M^\beta M^{\hat I} M^{\hat J}\,,
\end{align}
which is the polynomial part of the prepotential for our setting. As was already pointed out in \cite{Bonetti:2011mw}, 
this has to be supplemented by a non-polynomial part $\mathcal{N}^F_{\rm np}$, which is found by imposing the special geometry constraint
\begin{align}
\mathcal{N}_{\rm p}^F + \mathcal{N}_{\rm np}^F \overset{!}{=} \Omega_{\alpha \beta} j^\alpha j^\beta =1
\end{align}
to be
\begin{align}
\mathcal{N}_{\rm np}^F & = 4\Omega_{\alpha \beta} b^\alpha_{\hat I \hat J} b^\beta_{\hat K \hat L} \frac{M^{\hat I} M^{\hat J} M^{\hat K} M^{\hat L}}{M^0}\ .
\end{align}
Hence, the prepotential is not a cubic polynomial, but still a homogeneous function of degree three. 
The reason for deviating from the canonical case lies in the non-trivial transformation behavior of 
the six-dimensional tensors under gauge transformations. This required introducing
the redefined field strength \eqref{def-hatG}, which, when reduced to five dimensions, yields the
modified vector field strength \eqref{modified_cF}. In this way, all non-gauge-invariance of 
the classical 6D action is contained in the Green-Schwarz terms, while all non-gauge-invariance of the 5D action is 
encoded in the Chern-Simons terms \eqref{CSnp}.
Apart from the Chern-Simons terms \eqref{CSnp}, the action is therefore obtained 
in exactly the same way as the canonical supergravity action \eqref{e:canonical_action}. 
The metric $G_{\Lambda \Sigma}$ again has to be calculated using \eqref{e:metric}, this time taking into account both the 
polynomial and non-polynomial parts, i.e.~the sum $\mathcal{N}^F_{\rm p} + \mathcal{N}^F_{\rm np}$. 
More subtleties arise in the analysis of the Chern-Simons terms. In turns out that the two contributions \eqref{CSp} and \eqref{CSnp} 
can be brought into the form 
\begin{align}
S^{(5)F}_{CS}= -\frac{1}{12}\int_{M_5} (\mathcal{N}^F_{\rm p})_{\Lambda \Sigma \Theta} A^\Lambda \wedge F^\Sigma \wedge F^\Theta 
                - \frac{1}{16}\int_{M_5} (\mathcal{N}^F_{\rm np})_{\hat I \Sigma \Theta} A^{\hat I} \wedge F^\Sigma \wedge F^\Theta \ ,
\end{align}
where the indices on $\cN^{F}$ indicate that derivatives are taken 
with respect to the corresponding scalar fields. 
Note that the second part is not symmetric in the indices, since one cannot integrate by parts.

Finally, let us make a short remark on higher curvature terms. Their reduction proceeds along the same 
lines as in \cite{Bonetti:2011mw}. By including gravitational contributions in the Green-Schwarz 
terms and in the tensor transformations, one induces a 5D Chern-Simons term
\begin{align}\label{e:F_CS_hc}
S^{(5)F}_{A\mathcal{R}\mathcal{R}}=\frac{1}{2}\int_{M_5}\Omega_{\alpha \beta} a^\alpha A^\beta \wedge \tr\, \mathcal{R} \wedge \mathcal{R}\ .
\end{align}
We note that there are additional higher curvature corrections to the circle reduced action when including 
higher curvature terms in six dimensions. However, the new Chern-Simons term \eqref{e:F_CS_hc} 
turns out to be sufficient to extract the geometrical interpretation of $a^\alpha$ in F-theory
when the matching with M-theory is performed.

\subsection{M-theory on a Calabi-Yau threefold with $U(1)$s} \label{s:MtheorywithU(1)}

After having discussed 6D F-theory with Abelian gauge factors on a circle, 
we turn to the dual setting, M-theory on an elliptically fibered resolved 
Calabi-Yau threefold $\hat{Y}_3$ with rational sections. 
The geometry of these spaces was discussed in \autoref{On_geometry}.
To perform the dimensional reduction one expands the M-theory three-form $\hat{C}_3$
along the harmonic forms of $\hat{Y}_3$. Recall that the non-vanishing Hodge numbers are
$h^{0,0}(\hat{Y}_3)=h^{3,3}(\hat{Y}_3)=1$, $h^{1,1}(\hat{Y}_3)= h^{2,2}(\hat{Y}_3)$, $h^{2,1}(\hat{Y}_3)=h^{1,2}(\hat{Y}_3)$, 
and $h^{3,0}(\hat{Y}_3)=h^{0,3}(\hat{Y}_3)=1$.
The cohomology group $H^{1,1}(\hat{Y}_3)$ consists of the cohomology classes Poincar\'e-dual to the divisors of the 
Calabi-Yau threefold introduced in \autoref{On_geometry}. For $H^3(\hat{Y}_3)$ we introduce a real symplectic basis 
$(\alpha_K , \beta^K)$, $K=1\dots h^{2,1}+1$.
The reduction then reads  
\begin{align}
\hat{C}_3 = \xi^K \alpha_K - \tilde{\xi}_K \beta^K + A^0 \wedge \omega_0 + A^\alpha \wedge \omega_\alpha + A^I \wedge \omega_I + A^m \wedge \omega_m + C_3 \ ,
\end{align}
where we have introduced the vectors 
\begin{align}
(A^\Lambda) = (A^0 , A^\alpha , A^I , A^m) \ ,
\end{align}
a 5D three-form $C_3$ and real scalars $(\xi^K , \tilde{\xi}_K)$.
Similarly, one can expand the K\"ahler form of $\hat{Y}_3$ as
\begin{align}
\hat{J} = v^0  \omega_0 + v^\alpha  \omega_\alpha + v^I  \omega_I + v^m  \omega_m 
\end{align}
to obtain the 5D scalars $v^\Lambda$.
One of the vectors from the $\hat{C}_3$-reduction belongs to the gravity multiplet and comprises the graviphoton, while
the remaining vectors form $h^{1,1}(\hat{Y}_3)-1$ vector multiplets. The corresponding scalars are the $v^\Lambda$. 
Note that these $h^{1,1}(\hat{Y}_3)$ scalars are distributed among $h^{1,1}(\hat{Y}_3)-1$ vector multiplets and 
the universal hypermultiplet. The vector multiplets contain normalized scalars 
\beq
   L^\Lambda = \mathcal{V}^{-1/3}v^\Lambda\ , \qquad (L^\Lambda) \equiv (R,L^\alpha , \xi^I , \xi^m)\ ,
\eeq
while the total volume, given by 
\begin{align}
\mathcal{V}=\frac{1}{3!}\mathcal{V}_{\Lambda \Sigma \Theta}v^\Lambda v^\Sigma v^\Theta
\end{align}
is part of the universal hypermultiplet.
The 5D three-form $C_3$ is dualized into a real scalar $\Phi$ and also sits in the universal hypermultiplet.
Concerning the scalars $(\xi^K , \tilde{\xi}_K)$, we note that $2 h^{1,2}(\hat{Y}_3)$ degrees of freedom 
together with the complex structure moduli form $h^{1,2}(\hat{Y}_3)$ hypermultiplets. The remaining two 
degrees of freedom from these scalars enter the universal hypermultiplet.

Having obtained the above data of the massless modes, we can easily derive the gravity and vector sector in the
canonical form of 5D $\mathcal{N}=2$ supergravity. The prepotential is given by
\begin{align}
\mathcal{N}=\frac{1}{3!}\mathcal{V}_{\Lambda \Sigma \Theta}L^\Lambda L^\Sigma L^\Theta\ ,
\end{align}
where we have defined the intersection numbers
\begin{align}
\mathcal{V}_{\Lambda \Sigma \Theta}=D_\Lambda \cdot D_\Sigma \cdot D_\Theta\ .
\end{align}
Recall that these intersections were discussed in \autoref{On_geometry} and that
they take the special form \eqref{e:intersections_rational} in the case of an elliptic fibration. 
If the manifold admits a holomorphic zero section, then the additional relations in \eqref{e:intersections_holomorphic} hold.
We are now in a position to write down the prepotential. As discussed in more detail in
\cite{Witten:1996qb,Intriligator:1997pq,Bonetti:2011mw,Bonetti:2013ela}, the prepotential
of the resolved threefold contains both classical and one-loop terms when interpreted in the dual 
F-theory setup. To distinguish these contributions in M-theory, let us define an 
$\epsilon$-scaling for the 5D M-theory fields. The limit $\epsilon \rightarrow 0$ corresponds to the F-theory limit
and enforces that both the volume of the elliptic fiber and the blow-up divisors shrink to zero. 
For the scalar fields $v^\Lambda$ we set\footnote{For consistency checks on these scaling relations we refer to \cite{Bonetti:2011mw}.}
\begin{align}
v^0 \mapsto \epsilon v^0, \qquad v^\alpha \mapsto \epsilon^{-1/2}v^\alpha, \qquad v^I \mapsto \epsilon^{1/4} v^I, \qquad v^m \mapsto \epsilon^{1/4} v^m\ .
\end{align}
On the level of the redefined fields this reads
\begin{align}
R \mapsto \epsilon R, \qquad L^\alpha \mapsto \epsilon^{-1/2}L^\alpha, \qquad \xi^I \mapsto \epsilon^{1/4} \xi^I, \qquad \xi^m \mapsto \epsilon^{1/4} \xi^m\ .
\end{align}
 In this limit only classical terms are non-zero. Hence, we can divide the prepotential into a part surviving as $\epsilon \rightarrow 0$ and a part that vanishes in the limit.
Accordingly, the classical part of the prepotential is given by
\begin{align}\begin{split}
\mathcal{N}^M_{class}=&\frac{1}{2}\eta_{\alpha \beta}RL^\alpha L^\beta - \frac{1}{2}\mathcal{C}_{IJ} \eta_{\alpha \beta}S^{b,\alpha} L^\beta \xi^I \xi^J \\
& +\frac{1}{2}\pi (D_m \cdot D_n)^\alpha \eta_{\alpha \beta}L^\beta \xi^m \xi^n\ .
\end{split}\end{align}
The one-loop part of the prepotential cannot be given in such an explicit form. It reads 
\begin{align}
\mathcal{N}^M_{loop}=&\frac{1}{6}\mathcal{V}_{000} RRR + \frac{1}{2}\mathcal{V}_{00m} RR\xi^m + \frac{1}{2}\mathcal{V}_{00I}RR\xi^I +\frac{1}{2}\mathcal{V}_{0IJ} R \xi^I \xi^J \\
& +\frac{1}{2}\mathcal{V}_{0mn} R \xi^m \xi^n + \mathcal{V}_{0mI}R\xi^m \xi^I + \frac{1}{6} \mathcal{V}_{IJK}\xi^I \xi^J \xi^K \nn \\
& + \frac{1}{6} \mathcal{V}_{mnk}\xi^m \xi^n \xi^k + \frac{1}{2} \mathcal{V}_{mIJ}\xi^m \xi^I \xi^J + \frac{1}{2} \mathcal{V}_{Imn}\xi^I \xi^m \xi^n\ . \nn
\end{align}
In case there is a holomorphic zero section, one can use \eqref{e:intersections_holomorphic} to simplify the above expression to
\begin{align}
\mathcal{N}^M_{loop}=&\frac{1}{24}K^\alpha K^\beta \eta_{\alpha \beta} RRR + \frac{1}{4}\mathcal{C}_{IJ} K^\alpha S^{b,\beta} \eta_{\alpha \beta} R \xi^I \xi^J \\
& -\frac{1}{4}\pi (D_m \cdot D_n)^\alpha K^\beta \eta_{\alpha \beta} R \xi^m \xi^n \nn \\
& + \frac{1}{6} \mathcal{V}_{IJK}\xi^I \xi^J \xi^K + \frac{1}{6} \mathcal{V}_{mnk}\xi^m \xi^n \xi^k \nn \\
& + \frac{1}{2} \mathcal{V}_{mIJ}\xi^m \xi^I \xi^J + \frac{1}{2} \mathcal{V}_{Imn}\xi^I \xi^m \xi^n\ . \nn
\end{align}
In fact,by inserting the $\epsilon$-rescaled fields  one can check that $\mathcal{N}^M_{loop}$ vanishes
in the limit $\epsilon \rightarrow 0$, while $\mathcal{N}^M_{class}$ stays finite.

The above analysis leads to an effective action in which massive modes appearing 
in the M-theory reduction have been integrated out already. Let us remark on how 
these massive states arise in 5D M-theory. On the Coulomb branch of the dual circle 
reduced 6D $(1,0)$ theory, non-Cartan vector multiplets, charged hypermultiplets and 
KK-modes become massive. By taking the decompactification limit $r \rightarrow\infty$ 
and by moving to the origin of the Coulomb branch all these modes therefore 
become massless again. In the dual M-theory setting they arise from M2-branes wrapping 
rational curves in the fiber that shrink to zero volume in the F-theory limit. These modes, 
which are massive on the Coulomb branch, wrap the $\mathbb{P}^1$s resolving the singularities 
in the fibration. In fact, as we move towards the origin of the Coulomb branch, 
the $\mathbb{P}^1$s shrink in size and the M2-brane states become light. 
Similarly, the KK-modes arise from M2-branes with volume contribution depending 
on the volume of the generic elliptic fiber. The KK-mass also 
becomes zero as $r \rightarrow\infty$  in the decompactification limit 
and all such modes become massless.

Before we conclude this section, let us discuss the dimensional reduction of known 
higher curvature corrections in M-theory. Their lift to F-theory proceeds along the lines of \cite{Bonetti:2011mw,Grimm:2013gma}, but we focus here 
on the term quartic in the curvature two-form and linear in $\hat C_3$.
Concretely, this term in the 11D action is given by
\begin{align}
\hat{S}^{(11)}_{C\mathcal{R}^4}=-\frac{1}{96}\int_{X_{11}}\hat{C}_3 \wedge [\tr\hat{\mathcal{R}}^4 - \frac{1}{4}(\tr\hat{\mathcal{R}}^2)^2]\ .
\end{align}
Upon dimensional reduction on a general Calabi-Yau threefold, one finds, among other terms, the 5D Chern-Simons terms \cite{Antoniadis:1997eg}
\begin{align}\label{e:hc_action_M}
S^{(5)M}_{A\mathcal{R}\mathcal{R}}=\frac{1}{48}c_\Lambda \int_{M_5}A^\Lambda \wedge \tr\, \mathcal{R} \wedge \mathcal{R}\ ,
\end{align}
where
\begin{align}\label{e:hc_coeff_M}
c_\Lambda = \int_{\hat{Y}_3} \omega_\Lambda \wedge c_2 (\hat{Y}_3)\ .
\end{align}
The comparison with F-theory will show that the $c_\alpha$-term is a classical Chern-Simons term, 
while the other terms involving $c_0$, $c_I$, $c_m$ are induced at one-loop. 
We discuss this matter in more detail in the next sections.

On the M-theory side, one can use the geometry of $\hat{Y}_3$ 
to evaluate the various components $(c_\Lambda)=(c_\alpha,c_0, c_I, c_m)$. In the case of $c_{\alpha}$,
it is possible to perform this calculation without knowledge of the specific manifold. One finds that
\begin{equation}
 c_{\alpha} = - 12 K_{\alpha}\,, \label{calpha_geom}
\end{equation}
where $K_{\alpha} = \eta_{\alpha \beta} K^{\beta}$ and $K^{\beta}$ are the expansion coefficients of the canonical
class in terms of vertical divisors. Notably, the result is independent of
whether the zero section of $\hat{Y}_3$ is holomorphic or not. For details on the calculation, we refer to appendix \ref{a:c_2}.

If, on the other hand, we do have a holomorphic zero section, then we can explicitly evaluate another coefficient to find that
\begin{equation} \label{explicit_c0}
 c_0 = 52 - 4 h^{1,1}(\cB) \qquad \textrm{if } D_{\hat{0}} \textrm{ is holomorphic.}
\end{equation}
Again, we defer details to appendix \ref{a:c_2}.

\subsection{Classical matching of F-theory and M-theory}

In the last two subsections we have found the prepotentials for the 5D M-theory and F-theory reduction. 
The circle reduction of the general 6D $(1,0)$ theory results in a 5D action, which can 
be brought into standard form after dropping all massive Coulomb branch modes and 
Kaluza-Klein modes as well as the non-gauge invariant terms \eqref{CSnp}. However, 
the resulting 5D theory can only be matched with the parts of the M-theory reduction 
obtained from $\cN^M_{class}$ and the gravitational Chern-Simons term proportional 
to $c_\alpha$. In the next section, we show that the remaining terms in the 5D M-theory
vector sector are induced in the circle compactification of the 6D $(1,0)$ theory by 
integrating out certain massive modes.

Before doing so, let us first match the classical parts of the prepotentials. One obtains relations 
among the fields given by
\begin{align} \label{M-match}
 M^0 &= 2R &M^\alpha & = \frac{1}{2}L^\alpha \\
 M^I &= \frac{1}{2}\xi^I & M^m &= \frac{1}{2} \xi^m \,. \nn 
\end{align}
In addition, the constant couplings specifying the 6D (1,0) action are 
identified as 
\begin{align}  \label{bbOmega-match}
 b^\alpha &= S^{b,\alpha} &b^\alpha_{mn}&=-\pi (D_m \cdot D_n)^\alpha \\
 \Omega_{\alpha \beta}&=\eta_{\alpha \beta}\ . \nn
\end{align}
Furthermore, matching the classical higher curvature terms \eqref{e:F_CS_hc} and \eqref{e:hc_action_M}
gives
\begin{align}
\label{e:a_matching}
a^\alpha = K^\alpha ,
\end{align}
after identifying $c_\alpha = -12 \eta_{\alpha \beta} K^\beta$ as in \eqref{calpha_geom}.
The identifications \eqref{M-match},\eqref{bbOmega-match},\eqref{e:a_matching} and 
the discussion of the proceeding subsections imply that the Hodge numbers of the resolved 
Calabi-Yau threefold $\hat X$ and its base $\cB$ are related to the spectrum as
  \begin{align}
   h^{1,1}(\hat{X}) &= 1 + h^{1,1}(\cB) + \rk \fg + n_{U(1)} \label{e:h11_spectrum} \\
   h^{1,1}(\cB) &= T+1\,,\\
   h^{2,1}(\hat{X}) &= H^{\textrm{neutral}} - 1\;.
  \end{align}
In particular, inverting \eqref{e:h11_spectrum} provides an easy way of 
calculating the rank  $n_{U(1)}$ of the Mordell-Weil group of a given Calabi-Yau manifold. 
These identifications of geometrical quantities with the 
characteristic data of the effective action 
are in accordance with the matchings found 
in \cite{Morrison:1996na,Morrison:1996pp,Sadov:1996zm,Kumar:2009ac,Seiberg:2011dr,Park:2011wv,Bonetti:2011mw}.

Summing up, we have matched all classical Chern-Simons terms in M-theory and F-theory. 
In the next section we account for the one-loop induced terms by explicitly 
evaluating the corresponding amplitudes in the circle reduced (1,0) theory.

\section{One-loop Chern-Simons terms and anomaly cancelation}
\label{sec:one-loop_CS}
In this section we first review in \autoref{anom_rev} the anomaly conditions of 6D $(1,0)$
supergravity in the presence of $U(1)$ gauge factors. Our presentation is adapted
to conform with the later application to the 5D theories. Concretely, for the 5D effective theories 
obtained by circle compactification and M-theory reduction, we analyze the 5D gauge and 
gravitational Chern-Simons terms given by
\begin{align} \label{general_CS}
S^{(5)}_{CS} = -\frac{1}{12}\int_{M_5} k_{\Lambda \Sigma \Theta} \, A^\Lambda \wedge F^\Sigma \wedge F^\Theta 
+\frac{1}{48} \int_{M_5} k_{\Lambda}\, A^\Lambda \wedge \tr\,\mathcal{R} \wedge \mathcal{R}\ , 
\end{align}
where $k_{\Lambda \Sigma \Theta}$ and $k_{\Lambda}$ are the constant Chern-Simons coefficients of interest. 
Comparing the effective Chern-Simons terms in the circle reduced theory with the couplings from
$\cN^M_{loop}$ and $c_0,c_I,c_m$ in M-theory, two 
applications are possible: (1) Automatic anomaly cancelation in F-theory can be shown. (2) The 6D F-theory 
matter spectrum can be determined.

To derive the effective Chern-Simons terms of the form \eqref{general_CS} in 
the circle compactified 5D theory, we integrate out massive fermionic modes and massive self-dual tensors to 
derive a 5D quantum effective action.
We summarize the general formulas to perform this computation in \autoref{sec:summary_1loop}.
In \autoref{sec:mixed_an} we consider the couplings 
$k_{000}$, $k_{0mn}$, $k_{0IJ}$, and $k_0$, discuss their matching with the 6D anomaly contributions, and
study F-theory anomaly cancelation by comparing with M-theory. We argue that this can be done generally 
when the elliptic fibration has a holomorphic zero section. In this case, the hierarchy \eqref{normal_hierarchy}
is satisfied and the matching reproduces the cancelation conditions for gravitational and all mixed anomalies exactly. 
On the other hand, when the hierarchy \eqref{normal_hierarchy} is violated, the one-loop Chern-Simons terms get shifted 
and a direct match with 6D anomalies is no longer possible. In the examples of \autoref{s:examples} we nevertheless
show that 6D anomaly conditions are satisfied even though in a more intricate way.

The one-loop Chern-Simons terms associated to pure gauge anomalies are discussed 
in subsection \ref{sec:pure_anom}. While their one-loop expressions can be given generally, a 
direct match with 6D anomalies conditions is only possible on a case by case basis
and we show that these are satisfied for concrete examples in \autoref{s:examples}.
Finally, in \ref{sec:new_CS} we show that additional one-loop Chern-Simons terms 
are induced if condition \eqref{normal_hierarchy} is violated. This is 
in agreement with the fact that the intersection numbers of the Calabi-Yau threefold
are less restricted if the zero section is non-holomorphic.

\subsection{Review of 6D anomalies with $U(1)$s} \label{anom_rev}

Anomalies in quantum field theory describe the breakdown of a classical symmetry of the Lagrangian under quantization. 
Even if the classical action is invariant under some symmetry, the path integral measure need not be.
In those cases where it is not, the quantum effective action does not exhibit the classical symmetry anymore. For gauge symmetries, this spells a
disaster, because certain current conservation laws are violated at the quantum level.
For $2n$-dimensional theories a useful method of capturing anomalies in a gauge invariant way proceeds via the 
anomaly polynomial, a formal polynomial of degree $n+1$ in the curvature two-forms, where two auxiliary 
dimensions are introduced. These polynomials were worked out in \cite{AlvarezGaume:1983ig}.

Before writing down the anomaly polynomial for our setting, it is important to fix our notation.
\begin{itemize}
\item We write $\frep{R}$ for some representation of the whole gauge group $G$, while representations of $G_{nA}$ are referred to as $\rep{R}$.
        The $U(1)$ charges are denoted by $q_m$.
        
\item For a representation $\frep{R}$ we denote the weights of the whole representation (including $U(1)$-factors) by $w$. The roots of the whole group $G$
        are referred to as $\alpha$. Weights and roots of only $G_{nA}$ will be called $\rep{w}$ and $\alpha_{nA}$, respectively.
        
\item $H(\frep{R})$ is the number of hypermultiplets transforming in a representation $\frep{R}$. 
        The complete number of involved hypermultiplets is then $\dim (\frep{R}) \cdot H(\frep{R})$, where $\dim (\frep{R})$ is the dimension of the 
          representation $\cR$. One similarly defines $H(\rep{R})$.
        Let $H(q_m , q_n)$ denote the total number of hypermultiplets with $U(1)$-charges ($q_m$, $q_n$) and proceed likewise for $H(q_m , q_n , q_k , q_l)$.
        Furthermore, we write $H(\rep{R}, q_m )$ for the number of hypermultiplets transforming in the representation $\rep{R}$ and
        carrying $U(1)$-charge $q_m$. 
        An analogous statement holds for $H(\rep{R}, q_m , q_n)$. Note that when a hypermultiplet transforms in some representation $\frep{R}$ in our notation, this actually means that one complex scalar and one symplectic Majorana-Weyl fermion in the multiplet transform in $\frep{R}$, while the other complex scalar and fermion transform in the conjugate representation $\frep{R}^*$.

\item Traces with respect to the representation $\rep{R}$ are denoted by $\tr_{\rep{R}}$ and $\tr$ refers to the trace in the 
        fundamental representation.
\end{itemize}
In our conventions, the 6D $(1,0)$ anomaly polynomial is given by \cite{AlvarezGaume:1983ig}
\begin{align}
\hat{I}_8 = &-\frac{1}{360}(H-V+29T-273)[\tr\hat{\frep{R}}^4 + \frac{5}{4}(\tr\hat{\frep{R}}^2)^2]-\frac{1}{8}(9-T)(\tr\hat{\frep{R}}^2)^2 \nn \\
&-\frac{1}{6}\tr\hat{\frep{R}}^2 [\tr_{\textrm{adj}}\hat{F}^2 - \sum_{\rep{R}}  H(\rep{R}) \tr_{\rep{R}} \hat{F}^2 -\quad
\sum_{m,n,q_m , q_n}H(q_m , q_n)q_m q_n \hat{F}^m \hat{F}^n ] \nn \\
&+\frac{2}{3}[\tr_{adj}\hat{F}^4 - \sum_{\rep{R}} H(\rep{R}) \tr_{\rep{R}} \hat{F}^4] 
-\frac{8}{3}\sum_{\rep{R},m,q_m} H(\rep{R}, q_m)q_m (\tr_{\rep{R}} \hat{F}^3)\hat{F}^m \nn \\
&- 4\sum_{\rep{R},m,n,q_m , q_n} H(\rep{R},q_m , q_n) q_m q_n (\tr_{\rep{R}} \hat{F}^2) \hat{F}^m \hat{F}^n  \\
&-\frac{2}{3}\sum_{m,n,k,l,q_m , q_n , q_k , q_l}H(q_m , q_n , q_k , q_l )q_m q_n q_k q_l \hat{F}^m \hat{F}^n \hat{F}^k \hat{F}^l \ .\nn
\end{align}

As already mentioned in the previous section, under suitable conditions these 
anomalies may be canceled by a generalized Green-Schwarz mechanism 
induced by non-trivial transformations of the tensors. In fact, this is possible
if the anomaly polynomial factorizes as
\begin{align}
\hat{I}_8 = - \frac{1}{2}\Omega_{\alpha \beta}X^\alpha_4 X^\beta_4\,,
\end{align}
as can be seen by applying the descent equations to \eqref{e:GS_factor}. 
This factorization condition gives the anomaly constraints \cite{Green:1984sg,Sagnotti:1992qw}.

In the following, we list the individual anomaly conditions such that they
can be conveniently compared with the results 
of the 5D one-loop Chern-Simons terms obtained after compactifying 
the theory on a circle and integrating out all massive modes. 
The equations from purely gravitational anomalies read
 \begin{subequations} \label{e:grav_anomalies}
\begin{align}
4(12-T)=&\frac{1}{6}(H-V+5T+15) \label{anom_1}\\
\frac{1}{4} a^\alpha a^\beta \Omega_{\alpha \beta} =& \frac{1}{120}(H-V-T-3)\,. \label{anom_2}
\end{align}
\end{subequations}
The slightly unusual form of presenting these two conditions is motivated
as follows. It was shown in \cite{Bonetti:2012fn,Bonetti:2013ela} that the right-hand sides of these 
expressions are precisely the coefficients of certain 5D one-loop Chern-Simons 
terms in the circle compactified theory after integrating out massive Kaluza-Klein modes.
In fact, we recall in \autoref{sec:mixed_an} that the right-hand side of 
\eqref{anom_1} is proportional to the coefficient of the 5D Chern-Simons term 
$A^0 \wedge \tr\, \cR \wedge \cR$, while the right-hand side of 
\eqref{anom_2} is proportional to the coefficient of the 5D Chern-Simons term $A^0 \wedge F^0 \wedge F^0$.
Let us stress that this is only true as long as the hierarchy \eqref{normal_hierarchy} is obeyed
and the Kaluza-Klein modes are always heavier than the Coulomb branch modes. 
The more general case is also discussed below.

The mixed anomalies can be summarized as 
\begin{subequations} \label{e:mixed_anomalies} 
\begin{align}
\frac{1}{2}\mathcal{C}_{IJ}a^\alpha b^\beta \Omega_{\alpha \beta}  =& \frac{1}{12}\mathcal{C}_{IJ}\lambda(\mathfrak{g}) \left(A_{adj}
- \sum_{\rep{R}} H(\rep{R}) A_{\rep{R}} \right)\\
\frac{1}{2}a^\alpha b^\beta_{mn} \Omega_{\alpha \beta}  =& -\frac{1}{12}\sum_{q_m , q_n}H(q_m , q_n)q_m q_n\ .
\end{align}
\end{subequations}
Again, we have arranged the anomaly conditions in a somewhat unusual fashion and
we show in \autoref{sec:mixed_an} that the right-hand sides of these conditions precisely
match the coefficients of the Chern-Simons terms $A^0 \wedge F^I \wedge F^J$ and $A^0 \wedge F^m \wedge F^n$ 
induced after circle compactification at one-loop.

At last, the cancelation conditions for pure gauge anomalies read
\begin{align}\begin{split} \label{pure_nonAb_an}
0 & =  B_{\textrm{adj}} - \sum_{\rep{R}}  H(\rep{R}) B_{\rep{R}} \\
\frac{b^\alpha}{\lambda (\mathfrak{g})}\frac{b^\beta}{\lambda (\mathfrak{g})} \Omega_{\alpha \beta} & =  \frac{1}{3} \left(\sum_{\rep{R}}
H(\rep{R}) C_{\rep{R}} - C_{\textrm{adj}} \right) \\
0 & =  \sum_{\rep{R}, q_m}  H(\rep{R}) q_m E_{\rep{R}} \\
\frac{b^\alpha}{\lambda (\mathfrak{g})}b_{mn}^\beta \Omega_{\alpha \beta} & = \sum_{\rep{R}, q_m , q_n} H(\rep{R}, q_m , q_n) q_m  q_n A_{\rep{R}}\\
\left(b_{mn}^\alpha b_{kl}^\beta + b_{mk}^\alpha b_{nl}^\beta +b_{ml}^\alpha b_{nk}^\beta \right) \Omega_{\alpha \beta} & = \sum_{q_m , q_n , q_k , q_l}
H(q_m , q_n , q_k , q_l) q_m q_n q_k q_l\ .
\end{split}\end{align}
The constants $A_{\rep{R}}$, $B_{\rep{R}}$, $C_{\rep{R}}$, and $E_{\rep{R}}$ are defined as proportionality factors between traces in different representations as in
\begin{align}\begin{split}
& \tr_{\rep{R}} \hat{F}^2 = A_{\rep{R}} \tr \hat{F}^2 \\
& \tr_{\rep{R}} \hat{F}^3 = E_{\rep{R}} \tr \hat{F}^3 \\
& \tr_{\rep{R}} \hat{F}^4 = B_{\rep{R}} \tr \hat{F}^4 + C_{\rep{R}} (\tr \hat{F}^2 )^2 \,.
\end{split}\end{align}
Note that the anomaly cancelation conditions
\eqref{pure_nonAb_an}, too, are mapped to non-trivial identities among 5D Chern-Simons terms. However, unlike for conditions
\eqref{e:grav_anomalies} and \eqref{e:mixed_anomalies}, we are not able to show in full generality
that these are automatically satisfied for a given compactification geometry.
Nevertheless, we can determine the 6D spectrum using the 5D Chern-Simons terms 
for various examples in \autoref{s:examples}, and check that anomaly cancelation is satisfied
on a case by case basis.

\subsection{One-loop Chern-Simons terms in circle reduced theories} \label{sec:summary_1loop}

In this subsection we summarize the general formulae required to evaluate 
one-loop Chern-Simons coefficients in a 5D effective theory obtained by circle compactification.
We discuss the general situation 
of an arbitrary mass hierarchy between the Coulomb branch masses $m_{CB}(f)$ of matter fermions $f$
and the Kaluza-Klein mass $m_{KK} = 1/r$.

As was found in \cite{Witten:1996qb,Intriligator:1997pq,Bonetti:2013ela}, one can generate new Chern-Simons terms in a 5D theory 
by integrating out massive spin-1/2 fermions, spin-3/2 fermions and massive tensors. In particular, as shown 
in \cite{Bonetti:2012fn,Bonetti:2013ela}, the 5D tensors contributing 
in this loop computation have to be self-dual in the sense of \cite{Townsend:1983xs}, i.e.~the tensors must be given by complex two-forms 
$B_{\mu \nu}$ with kinetic terms $\bar{B}\wedge dB$ and mass terms $m\bar{B}\wedge \ast B$.  The induced one-loop Chern-Simons coefficients 
are \cite{Bonetti:2013ela}
\begin{align}\label{e:CS_coupling}
k_{\Lambda \Sigma \Theta} = \frac{1}{2}\bigg[&\sum_{spin\, 1/2}  (q_{1/2})_{ \Lambda} (q_{1/2})_\Sigma (q_{1/2})_\Theta \sign(m_{1/2})\\
&-5\sum_{spin\, 3/2}(q_{3/2})_\Lambda (q_{3/2})_\Sigma (q_{3/2})_\Theta \sign(m_{3/2}) -4\sum_{B}(q_B)_\Lambda (q_B)_\Sigma (q_B)_\Theta \sign(m_B)\bigg] \nn \\
\label{e:CS_coupling_hc}
k_{\Lambda }=-\frac{1}{4}\bigg[&\sum_{spin\, 1/2} (q_{1/2})_\Lambda \sign(m_{1/2})+19\! \! \! \sum_{spin\, 3/2}(q_{3/2})_\Lambda \sign(m_{3/2}) +8\sum_{B}(q_B)_\Lambda \sign(m_B)\bigg]\ ,
\end{align}
where $\Lambda,\Sigma, \Theta \neq \alpha$.
The external legs of the loops are the gauge bosons $A^\Lambda$, $A^\Sigma$, $A^\Theta$ in \eqref{e:CS_coupling} and two gravitons and a gauge boson $A^\Lambda$ in \eqref{e:CS_coupling_hc}.
The sums run over all spin-1/2 fermions, spin-3/2 fermions and self-dual tensors, respectively. 
The charge of the mode under $A^\Lambda$ is written as $(q)_\Lambda$, where the conventions are such 
that the covariant derivative reads $\partial_\mu -iqA_\mu$.  We denoted the mass by $m$ appearing in the equations of motion as:
\begin{align}
(\slashed{\partial}-m_{1/2})\psi = 0\ , \qquad
(\gamma^{\rho \mu \nu}\partial_\mu - m_{3/2} \gamma^{\rho \nu})\psi_\nu = 0\ ,\qquad
(\ast d - i m_B )B = 0
\end{align}
for a spin-1/2 fermion $\psi$, a spin-3/2 fermion $\psi_\mu$ and a self-dual tensor $B$.

In our setting we reduce 6D symplectic Majorana-Weyl spinors on a circle. The symplectic 
Majorana condition for two fermions $\psi^1$ and $\psi^2$ in six dimensions reads
\begin{align}\label{sympl_major}
 \psi^i = \varepsilon^{ij} \psi^{j\,c} \ ,
\end{align}
where $\psi^{i\,c}$ denotes the charge conjugated spinor and $\varepsilon^{ij}$ is the usual 
antisymmetric epsilon tensor in two dimensions. One can now expand the spinors in 
Fourier modes along the circle direction
\begin{align}\label{fourier}
 \psi^i (x,y) = \sum_{n= -\infty}^{+ \infty} \psi_{(n)}^i(x)e^{iny/r} \ .
\end{align}
The $\psi_{(n)}^i$ are the Kaluza-Klein modes of the fermions. To determine the fermionic 
degrees of freedom in the circle reduced theory, we apply the symplectic Majorana 
condition \eqref{sympl_major} to the expansion \eqref{fourier}
\begin{align}
 \sum_{n= -\infty}^{+ \infty} \psi_{(n)}^i(x)e^{iny/r} = \varepsilon^{ij}\sum_{n= -\infty}^{+ \infty} \psi_{(n)}^{j\,c}(x)e^{-iny/r}\ .
\end{align}
Comparing coefficients, we obtain the constraint
\begin{align}
 \psi_{(n)}^i = \varepsilon^{ij} \psi_{(-n)}^{j\,c}\ ,
\end{align}
which simply states that in 5D, the degrees of freedom of two former 6D symplectic Majorana-Weyl 
fermions are entirely comprised by the Kaluza-Klein tower of one of the fermions, e.g.~$\psi^1$.
This means that one only needs to include one fermion per multiplet when integrating out massive fermionic modes.

Put together, we have to integrate out hyperini, which gain masses on the 
Coulomb-branch, KK-modes of hyperini, massive non-Cartan gaugini, KK-modes of 
gaugini and tensorini, KK-modes of gravitini, and KK-modes of former 6D (anti-)self-dual 
tensors\footnote{KK-modes are charged under the Kaluza-Klein vector $A^0$. The covariant 
derivative reads $\partial_\mu + inA^0_\mu$.}. In general, there can be two separate contributions
to their masses. First of all, the charged hyperini and non-Cartan gaugini have
Coulomb branch masses. Secondly, there is a contribution from the KK-level 
for all KK-modes. According to \cite{Bonetti:2013ela}, the mass terms then take the form
\begin{align}\label{def-mCB}
m_{1/2} = c_{1/2} \bigg(m_{CB}  + n\, m_{KK}\bigg),\qquad m_{CB} =  (q_{1/2})_{\hat I} \zeta^{\hat I} \,,
\end{align}
where $n$ is the Kaluza-Klein level and  
$m_{CB}$ is the Coulomb branch mass of the fermion under consideration. 
The term $(q_{1/2})_{\hat I} \zeta^{\hat I}$ 
denotes the contraction of the charges $ (q_{1/2})_{\hat I} $ under the Cartan generators $\mathcal{T}_I$ in the coroot basis and 
the $U(1)$s appearing in \eqref{CB_group} 
with the $\zeta^{\hat I}$ carrying indices $\hat I$ introduced around \eqref{e:hat_notation}.
The  $\zeta^{\hat I}$ are the VEVs of the scalars 
corresponding to the $U(1)$s  in \eqref{CB_group}.
In the reductions of the 6D theories considered above, the spin-3/2 fermions and the tensors are 
neutral under the 6D gauge group. They only can admit a Kaluza-Klein mass at level $n$ of the form 
\begin{align}
m_{3/2} = -c_{3/2}\cdot n \cdot m_{KK}, \,\,\,\,
m_B  = c_B \cdot n \cdot m_{KK}.
\label{e:masses}
\end{align}
The factors $c_{1/2}$, $c_{3/2}$, $c_B$ are related to the respective representations of $SO(4)$, 
the massive little group in 5D. In the subsequent calculations, it is important that $c_{1/2}$, $c_{3/2}$ are equal to $+1$ 
for modes coming from 6D left-handed fermions and $-1$ for those coming from right-handed ones. 
Similarly, $c_B$ is $+1$ for former 6D self-dual tensors and $-1$ for anti-self-dual tensors in six dimensions.

\subsection{Gravitational and mixed anomalies via one-loop Chern-Simons terms} \label{sec:mixed_an}

In this subsection we discuss the Chern-Simons terms in \eqref{general_CS} with coefficients 
$k_{000}$, $k_{0}$, $k_{0mn}$, and $k_{0IJ}$. As we have shown in \autoref{sec:circle_reduction}, no
constant contributions to these couplings appear in a classical circle reduction of a 6D $(1,0)$ theory 
to five dimensions. The classical Chern-Simons terms $A^m \wedge F^n \wedge F^0$ and $A^I \wedge F^J \wedge F^0$ found 
in \eqref{CSnp} were shown to be 
non-gauge-invariant and arose from the anomaly canceling 6D Green-Schwarz term. They therefore have to 
disappear in the 5D one-loop effective action. This matches 
the fact that on the M-theory side only gauge-invariant terms appear.
In addition, we find that $k_{000}$, $k_{0}$, $k_{0mn}$, and $k_{0IJ}$ are generated at one-loop in the circle 
reduced action by integrating out massive Coulomb branch and Kaluza-Klein modes. We first discuss 
the general situation assuming no specific mass hierarchy for the fields. 
The simplest situation occurs if all Coulomb branch masses $m_{CB}(f)$  
are smaller than the Kaluza-Klein mass $m_{KK}$, i.e.~condition \eqref{normal_hierarchy}
is satisfied. In this case the coefficients correspond precisely to the right-hand sides of the anomaly 
equations \eqref{e:grav_anomalies} and \eqref{e:mixed_anomalies}. Anomaly cancelation 
is then more straightforward to check when matching the one-loop Chern-Simons terms 
on the circle reduction side with the geometric Chern-Simons terms in M-theory.
This special case occurs only if the zero section of the fibration is holomorphic.

First of all, let us derive the Chern-Simons term $k_{000}$ via \eqref{e:CS_coupling}. 
Since all the external legs correspond to $A^0$, we have to integrate out KK-modes of 
hyperini, gaugini, tensorini, gravitini and self-dual tensors. 
The Coulomb branch mass \eqref{def-mCB} of a hyperino with weight $w=( \rep{w},q )$ of a representation $\cR$ can 
be rewritten as 
\begin{align}
m_{CB} = \langle \alpha _I ^\vee , \rep{w} \rangle  \zeta ^I + q_m \zeta^m \ .
\end{align}
Note that on the Coulomb branch the $U(1)$-charges of a state are given by the weight of the latter. Since we have chosen vectors $A^I$ and their corresponding scalars $\zeta ^I$ to be in the coroot basis, the charge under $A^I$  of a hyperino with weight $w$ is given by $\langle \alpha _I ^\vee , \rep{w} \rangle$.
For the gaugini one uses the weights of the adjoint representation, which are the roots $\alpha$ of $G$.
For simplicity, we always use the usual basis of the Cartan generators introduced in \eqref{norm_cartan}
to calculate the Coulomb branch mass, which can therefore be written as a contraction
\begin{align}
m_{CB} = w \cdot \zeta \,,
\end{align}
where the vector $\zeta$ is taken to be in the usual basis of the Cartan generators rather than the coroot basis.
One hence finds the expression
\begin{align}\begin{split}\label{e:000}
k_{000}=&\frac{1}{120}(H-V-T-3)+\frac{1}{4}\bigg[\sum_{roots}\big\lfloor\!\mid r\alpha \cdot \zeta\mid\!\big\rfloor ^2 \Big(\big\lfloor\!\mid r\alpha \cdot \zeta\mid\!\big\rfloor +1\Big)^2 \\
&- \sum_\frep{R} H(\frep{R}) \sum_{w \in \frep{R}} \big\lfloor\!\mid r w \cdot \zeta\mid\!\big\rfloor ^2 \Big(\big\lfloor\!\mid r w \cdot \zeta\mid\!\big\rfloor +1\Big)^2 \bigg]\ ,
\end{split}\end{align}
where we have introduced the floor function $\lfloor n \rfloor$ defined as the largest integer not greater than $n$.
The explicit calculations for the results presented here and in the following are carried out in appendix \ref{app:loop}.

In the special case where $m_{CB}< m_{KK}$, one has
\begin{align}
 \big\lfloor\!\mid r\alpha \cdot \zeta\mid\!\big\rfloor = \big\lfloor\!\mid r w \cdot \zeta\mid\!\big\rfloor = 0
\end{align}
and \eqref{e:000} therefore simplifies to
\begin{align}\begin{split}
k_{000}=\frac{1}{120}(H-V-T-3) \ ,
\end{split}\end{align}
which is exactly the right-hand side of \eqref{anom_2}. By M - F-theory duality, this Chern-Simons coefficient has to be matched with the corresponding intersection number in M-theory, namely
\begin{align}
 k_{000} = \mathcal{V}_{000} \ .
\end{align}
If  the fibration has a holomorphic zero section, we can match the coefficient with the explicit 
expression for the intersection number $\mathcal{V}_{000}$ in \eqref{e:i_000}. We obtain
\begin{align} \label{match1}
\frac{1}{4} K^\alpha K^\beta \eta_{\alpha \beta} \overset{?}{=} \frac{1}{120}(H-V-T-3) \ .
\end{align}
This is precisely the anomaly condition \eqref{anom_2} of the circle reduced (1,0) theory after using the matching condition $a^\alpha = K^\alpha$ found in \eqref{e:a_matching}.
To actually show that the anomaly is canceled for the given geometry, one has to express $V,H$ and $T$ in terms of geometric data of the 
underlying elliptic fibration. While we already know how to do this for $T$ using $T=h^{1,1}(\mathcal{B})-1$, we have not yet 
discussed $H$ and $V$. The number of neutral hypermultiplets can be inferred in the reduction of the M-theory three-form. 
In \autoref{s:MtheorywithU(1)} we found it to be
\begin{align}
 H^{\textrm{neutral}}=h^{2,1}(\hat{Y}_3)+1 \ .
\end{align}
 Investigating the number of charged hypermultiplets is more involved, since they arise on the M-theory side by wrapping M2-branes on rational 
 curves in the fiber as also explained in \autoref{s:MtheorywithU(1)}. These may be determined from the topology and  intersection numbers of the
 seven-branes specified by the discriminant of the elliptic fibration. While we are able to determine them for each explicit example of section \ref{s:examples},
 finding a general formula is beyond the scope of this paper. In contrast, the number of vectors, at least for the ADE groups, 
 is given generally in terms of the dual Coxeter number $c_{G_{nA}}$ and the rank of $G_{nA}$ 
 supplemented by the number of Abelian gauge factors as
\begin{align}
 V=\dim (G) = (c_{G_{nA}} +1)\rk (G_{nA}) +n_{U(1)} \ .
\end{align}
Using the topological 
identity $K^\alpha K^\beta \eta_{\alpha \beta} = 10-h^{1,1}(\mathcal{B})$ one finds that the 
gravitational anomaly \eqref{anom_2} is canceled automatically in F-theory provided that can also find 
a relation of the type $H-V = 274-29h^{1,1}(\mathcal{B})$. Relating $H,V$ to 
topological data, one might use index theorems and an explicit expression for the 
Euler number of $\hat Y_3$ to prove such an identity (see, e.g.~\cite{Grimm:2012yq}).
Although we do not treat the general case, we show anomaly cancelation for 
concrete examples in \autoref{s:examples}.

There is an alternative interpretation for the one-loop matchings discussed above. If we trust the cancelation of gravitational 
anomalies, we can solve \eqref{match1} for $H-V$ to arrive at the expression $H-V = 274-29~h^{1,1}(\mathcal{B})$ . 
This relation can then be used to extract crucial information about e.g.~the number of charged hypermultiplets in the theory 
in terms of geometric data of the compactification space. It is also conceivable that we might be able to conjecture new 
geometric relations for the elliptic fibration in this way. 

Let us now turn to the coefficient $k_{0}$ in order to find the anomaly condition \eqref{anom_1}.
It is again generated in the circle compactified 6D action 
by integrating out massive KK-modes of hyperini, gaugini, tensorini, gravitini and self-dual tensors. Using \eqref{e:CS_coupling_hc}, we obtain
\begin{align}  
\begin{split} \label{ev_k0}
k_{0 }=&\frac{1}{6}(H-V+5T+15)-\bigg[\sum_{roots}\Big(\big\lfloor\!\mid r\alpha \cdot \zeta \mid\!\big\rfloor +1\Big)\big\lfloor\!\mid r\alpha \cdot \zeta \mid\!\big\rfloor \\
& -\sum_\frep{R} H(\frep{R}) \sum_{w \in \frep{R}} \Big(\big\lfloor\!\mid r w \cdot \zeta\mid\!\big\rfloor +1\Big)\big\lfloor\!\mid r w \cdot \zeta \mid\!\big\rfloor\bigg] \ .
\end{split}\end{align}

In the special case of $m_{CB}<m_{KK}$, this simplifies to 
\begin{align}\begin{split}
k_{0}=\frac{1}{6}(H-V+5T+15) \ ,
\end{split}\end{align}
which is the right-hand side of \eqref{anom_1}. By M - F-theory duality,
$k_0$ has to be matched with $c_0$ introduced at the end of \autoref{s:MtheorywithU(1)}.
If the investigated fibration possesses a 
holomorphic zero section, we can match the coefficient with the explicit 
expression for $c_0$ in \eqref{explicit_c0}, which gives the condition
\begin{align} \label{match2}
4(13-h^{1,1}(\mathcal{B}))\overset{?}{=}\frac{1}{6}(H-V+5T+15) \ .
\end{align}
This is the gravitational anomaly condition in \eqref{anom_1} after matching $h^{1,1}(\mathcal{B})=T+1$. 
Once again, this condition has to be checked for a given geometry, specifying $H$ and $V$.
Note, however, that \eqref{match1} already implies \eqref{match2} after inserting $K^\alpha K_\alpha = 10 - h^{1,1}(\mathcal{B})$.

\begin{figure}[h!]
\centering
\includegraphics[scale=0.8]{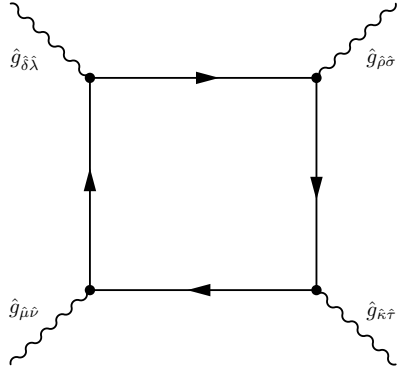}
\caption{Pure gravitational anomaly}
\label{gggg}
\end{figure}

\begin{figure}[h!]
\centering
\includegraphics[scale=0.9]{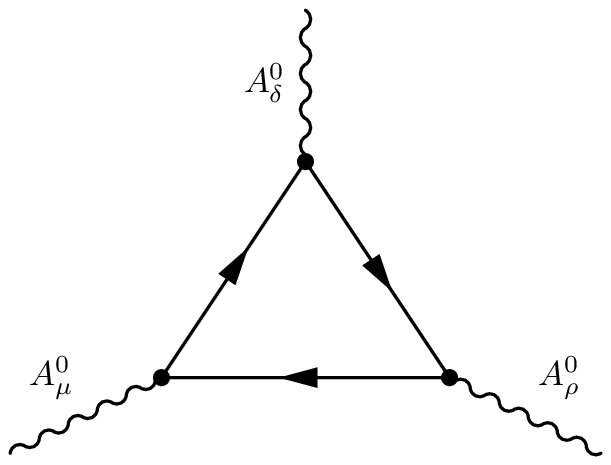}
\caption{Triangle diagram inducing $k_{000}$}
\label{000}
\end{figure}

The connection between 5D one-loop Chern-Simons terms and 6D anomalies 
becomes clearer if one considers the heuristic dimensional reduction to five dimensions 
of the 6D box-diagrams inducing the anomaly \cite{Bonetti:2012fn}. This connection 
can be described when the mass hierarchy \eqref{normal_hierarchy} is 
satisfied. In order to do that, recall that the gravitational anomaly arises from 
the diagrams displayed in figure \ref{gggg},
where all kinds of massless fermions and tensors are running in the loop.
In order to compactify this graph on a circle, we replace one of the external 
6D gravitons by the $S^1$-component of the metric $\langle r^2 \rangle$ and treat
it as a background field. After reducing the other 6D gravitons to KK-vectors $A^0$,
one obtains the triangle diagram\footnote{Note that this is not the only diagram one
finds in the reduction. As we will see in the following, there are additional triangle 
graphs arising in the reduction of this 6D box diagram.} of figure \ref{000}.

\begin{figure}[h!]
\centering
\includegraphics[scale=0.9]{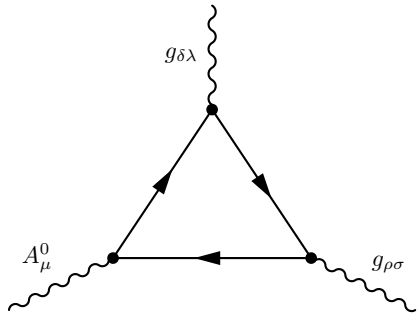}
\caption{Triangle diagram inducing $k_0$}
\label{0gg}
\end{figure}

Since only KK-modes couple to $A^0$, they are the only modes that run in the reduced 5D loop and we exactly obtain the diagram we integrated out for $k_{000}$. 
Proceeding along the same lines, the loop diagram generating $k_{0}$ can also be obtained by reducing the gravitational anomaly of figure \ref{gggg}.
While we again reduce one external 6D graviton to the background field $\langle r^2 \rangle$, this time two 6D gravitons become 5D gravitons and the
remaining 6D graviton becomes $A^0$ as displayed in figure \ref{0gg}. Since once again only KK-modes run in the reduced loop, 
this is in fact the diagram evaluated when calculating $k_0$.

Note that the reduction of anomaly diagrams only yields the complete expression 
for $k_{000}$ and $k_0$ when the Kaluza-Klein mass dominates 
in the sign-function. In other words, one has to require that the condition \eqref{normal_hierarchy} is satisfied, and the additional 
contributions in \eqref{e:000} and \eqref{ev_k0} vanish. In this case the background field method determines the relevant masses of the 
states contributing non-trivially to 5D one-loop diagrams. If \eqref{normal_hierarchy} is not satisfied,
extra states with masses dominated by the Coulomb branch mass contribute non-trivially and 
shift the Chern-Simons levels as in \eqref{e:000} and \eqref{ev_k0}. We conjecture that these arise from 6D diagrams 
where one of the external legs is a 6D vector that acquires a VEV. In contrast to a Lorentz and gauge-invariant 
6D analysis, these diagrams do not vanish in our setting, since some of the 6D symmetries 
are in fact broken. A similar logic applies to all diagrammatic reductions encountered in the following.

In order to generate $k_{0mn}$ in the circle reduced theory, we consider massive KK-modes 
coming from hyperini with $U(1)$-charges $q_m$, $q_n$ running in a loop 
with external legs $A^m, A^n, A^0$.
With the help of \eqref{e:CS_coupling}, one finds
\begin{align}
k_{0mn}=-\frac{1}{12}\sum_\frep{R} H(\frep{R}) \sum_{w \in \frep{R}}\bigg(1+6\big\lfloor\!\mid r w \cdot\zeta
\mid\!\big\rfloor\Big(\big\lfloor\!\mid r w \cdot\zeta \mid\!\big\rfloor +1\Big) \bigg)  q_m q_n \ .
\label{e:CS_mixed1_F}
\end{align}
In case that $m_{CB}<m_{KK}$, this simplifies to
\begin{align}
k_{0mn}=-\frac{1}{12}\sum_\frep{R} H(\frep{R}) \sum_{w \in \frep{R}}  q_m q_n = -\frac{1}{12}\sum_{q_m , q_n} H(q_m , q_n) q_m q_n \ ,
\end{align}
which is exactly the right-hand side of the mixed Abelian anomaly in \eqref{e:mixed_anomalies}.
We can now compare this result with the M-theory side by matching $\mathcal{V}_{0mn}=k_{0mn}$.
In the case that we have a holomorphic zero section, we can use \eqref{e:i_0gauge} and obtain
\begin{align}\label{mixed_ab_matching}
\frac{1}{2}b^\alpha_{mn}a^\beta \Omega_{\alpha \beta} \overset{?}{=} -\frac{1}{12}\sum_{q_m , q_n} H(q_m , q_n) q_m q_n 
\end{align}
after using the classical matchings $K^\alpha = a^\alpha$, $\pi (D_m \cdot D_n )^\alpha = -b^\alpha_{mn}$ and $\eta_{\alpha \beta}=\Omega_{\alpha \beta}$.
This is exactly the second mixed anomaly condition in \eqref{e:mixed_anomalies}.
Let us remark again that if we could find general geometric expressions for the right-hand side of \eqref{mixed_ab_matching}, 
we might be able to show that the mixed Abelian anomaly is canceled automatically. This is in complete analogy to the 
study of anomalies in 4D/3D F-theory reductions \cite{Cvetic:2012xn}.

\begin{figure}[h!]
\centering
\includegraphics[scale=0.8]{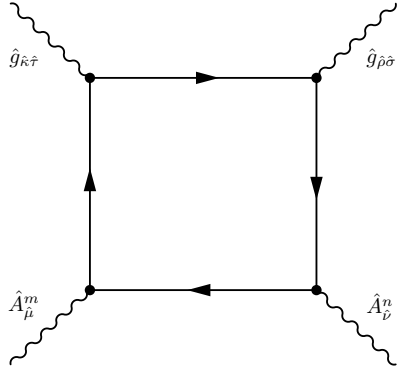}
\caption{Abelian-gravitational anomaly}
\label{mngg}
\end{figure}

\begin{figure}[h!]
\centering
\includegraphics[scale=0.9]{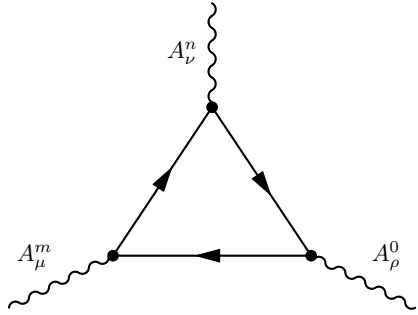}
\caption{Triangle diagram inducing $k_{0mn}$}
\label{mn0}
\end{figure}

As before, the 6D anomalous box diagram for the mixed Abelian anomaly is related by dimensional reduction to the 5D triangle 
diagram inducing the Chern-Simons terms. In the present case, we consider the diagram in figure \ref{mngg}.
Since massless fermions charged under the Abelian gauge symmetry run in the loop, this diagram describes the 
6D Abelian-gravitational anomaly.  Reducing the graph on a circle, one of the external 
gravitons becomes the background field $\langle r^2 \rangle$. The other 6D vectors reduce to the 
corresponding 5D vectors and one arrives at the triangle diagram of figure \ref{mn0}, which
is precisely the loop diagram generating $k_{0mn}$.

\begin{figure}
\centering
\includegraphics[scale=0.8]{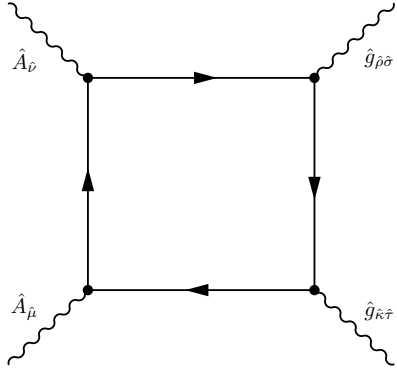}
\caption{Non-Abelian-gravitational anomaly}
\label{gg}
\end{figure}

\begin{figure}
\centering
\includegraphics[scale=0.9]{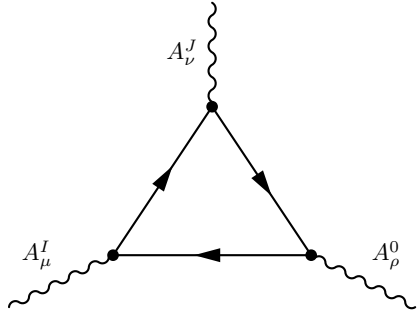}
\caption{Triangle diagram inducing $k_{0IJ}$}
\label{IJ0}
\end{figure}

Next of all, let us investigate how $k_{0IJ}$ is induced in the circle reduced theory. In order to
obtain $k_{0IJ}$, one needs to integrate out massive 5D KK-modes from hyperini charged under the corresponding Cartan 
generators, and non-Cartan gaugini running in loops with external legs $A^I$, $A^J$, $A^0$. 
After doing so, we find the Chern-Simons couplings
\begin{align}\begin{split}
k_{0IJ} =&\frac{1}{12}\bigg[\sum_{roots}\bigg(1+6\big\lfloor\!\mid r\alpha\cdot\zeta \mid\!\big\rfloor\Big(\big\lfloor\!\mid r\alpha\cdot\zeta \mid\!\big\rfloor +1\Big) \bigg) \langle  \alpha_I^\vee , \alpha_{nA} \rangle \langle \alpha_J^\vee , \alpha_{nA} \rangle \\
& - \sum_\frep{R} H(\frep{R}) \sum_{w \in \frep{R}}\bigg(1+6\big\lfloor\!\mid r w \cdot\zeta \mid\!\big\rfloor\Big(\big\lfloor\!\mid r w \cdot\zeta \mid\!\big\rfloor +1\Big) \bigg) \langle  \alpha_I^\vee , \mathbf{w} \rangle \langle \alpha_J^\vee , \mathbf{w} \rangle\bigg] \ .
\end{split}\end{align}
In the simple case that $m_{CB}<m_{KK}$, this reduces to
\begin{align}\label{k0IJ}
 k_{0IJ} =& \frac{1}{12}\bigg[\sum_{roots} \langle  \alpha_I^\vee , \alpha_{nA} \rangle \langle \alpha_J^\vee , \alpha_{nA} \rangle - \sum_{\rep{R}} H(\rep{R}) \sum_{\mathbf{w} \in \rep{R}} \langle  \alpha_I^\vee , \mathbf{w} \rangle \langle \alpha_J^\vee , \mathbf{w} \rangle\bigg] \\
 & =\frac{\mathcal{C}_{IJ}}{12} \lambda(\mathfrak{g}) \left(A_{\textrm{adj}} - \sum_{\rep{R}} H(\rep{R}) A_{\rep{R}} \right) \,, \nn 
\end{align}
where the second equality follows from a group theoretical identity proven in appendix \ref{app:group_id}. 
This is exactly the right-hand side of the mixed-non-Abelian anomaly in \eqref{e:mixed_anomalies}.
Assuming a holomorphic zero section of the fibration, matching with the M-theory intersections \eqref{e:i_0gauge} leads to
\begin{align}\label{e:exact_non_abelian_anomaly}
\frac{1}{2}\mathcal{C}_{IJ}a^\alpha b^\beta \Omega_{\alpha \beta} \overset{?}{=} \frac{\mathcal{C}_{IJ}}{12}\lambda(\mathfrak{g}) \left(A_{\textrm{adj}} - \sum_{\rep{R}} H(\rep{R}) A_{\rep{R}} \right) \ ,
\end{align}
where we have again used the classical matchings $K^\alpha = a^\alpha$, 
$S^{b,\alpha}=b^\alpha$, and $\eta_{\alpha \beta}=\Omega_{\alpha \beta}$. This is the 
mixed non-Abelian anomaly of \eqref{e:mixed_anomalies}. Once again, if we find general 
geometrical expressions for the right-hand side of \eqref{e:exact_non_abelian_anomaly}, 
we may show exact cancelation of this anomaly.
The relation to the 6D non-Abelian-gravitational anomaly can be inferred from the 
corresponding graph of figure \ref{gg} which induces the anomaly.
The modes that run in the loop are fermions charged under the non-Abelian gauge group.
Upon circle compactification, we reduce one graviton to the background field $\langle r^2 \rangle$, 
the 6D vectors to 5D vectors and the remaining graviton to $A^0$. Gauge symmetry breaking on
the Coulomb branch then takes the non Abelian vectors to Cartan vectors $A^I$ and $A^J$, and
we obtain the diagram of figure \ref{IJ0},
where fermionic KK-modes charged under the Cartan generators run in the loop. 
As before, we see that these are the loops inducing $k_{0IJ}$.

There are two more types of one-loop Chern-Simons terms in the circle reduced action that might be
related to 6D mixed anomalies. These are the terms with coefficients $k_m$ and $k_I$. 
We can generate $k_m$ by integrating out hyperini charged under $A^m$, while $k_I$ 
arises by integrating out hyperini and gaugini charged under $A^I$.
From \eqref{e:CS_coupling_hc} we obtain the Chern-Simons coefficients
\begin{align}
 k_m=&\sum_\mathcal{R} H(\mathcal{R})\sum_{w\in \mathcal{R}} \Big(2\big\lfloor\!\mid r w \cdot \zeta\mid\!\big\rfloor +1\Big)q_m \sign(w \cdot \zeta) \\
 k_I =& \sum_\mathcal{R} H(\mathcal{R})\sum_{w\in \mathcal{R}}\Big(2\big\lfloor\!\mid r w \cdot \zeta\mid\!\big\rfloor +1\Big)\langle \alpha_I^\vee , \mathbf{w} \rangle \sign(w \cdot \zeta) \\
&-\sum_{roots}\Big(2\big\lfloor\!\mid r \alpha \cdot \zeta\mid\!\big\rfloor +1\Big) \langle \alpha_I^\vee , \alpha_{nA} \rangle \sign(\alpha \cdot \zeta) \ . \nn
\end{align}
These can again be matched with the M-theory side \eqref{e:hc_coeff_M} as
\begin{align} 
 \int_{\hat Y_3 } \omega_m \wedge c_2(\hat{Y}_3) \overset{?}{=}k_m \label{kmI_match} \\ 
 \int_{\hat Y_3 } \omega_I \wedge c_2(\hat{Y}_3) \overset{?}{=}k_I \label{kmI_match2} \ .
\end{align}
The calculated loops are precisely those which appear in the circle reduction of the 6D box-diagrams inducing the 
mixed anomalies depicted in figure \ref{mngg} and \ref{gg}. Going from six dimensions 
to five dimensions, one of the vectors becomes a background field $\langle \zeta \rangle$, the other 
6D vector becomes a 5D vector and the 6D gravitons are reduced to 5D gravitons.  
Anyhow, in contrast to the matching of the one-loop coefficients 
$k_{0mn}$ and $k_{0IJ}$, the coefficients $k_m$ and $k_I$ do not admit such
simple interpretations, since they are linear in the charges.
Nevertheless, the circumstance that the loop diagrams also arise from the 6D mixed anomalies
hints to the fact that the conditions \eqref{kmI_match},\eqref{kmI_match2} are in 
fact implied by the matching for $k_{0mn}$ and $k_{0IJ}$ as long as the mass hierarchy \eqref{normal_hierarchy} is satisfied.

In summary, we have shown that the 5D Chern-Simons terms with coefficients $k_{000}$ 
and $k_0$ are related to gravitational anomalies in six dimensions, while
the terms with coefficients $k_{0mn}$, $k_{0IJ}$, $k_m$, and $k_I$ are connected to 
the mixed anomalies in the 6D theory. For fibrations admitting a holomorphic zero section 
and the special case that $m_{CB} < m_{KK}$, we can exactly reproduce the corresponding 
anomaly conditions in six dimensions from the matching
with M-theory except for $k_m$ and $k_I$. Furthermore, we have argued that by translating 
these constraints into geometry, one might be able to show exact cancelation of gravitational 
and mixed anomalies.

\subsection{Pure gauge anomalies and one-loop Chern-Simons terms} \label{sec:pure_anom}

In this subsection we work out how Chern-Simons coefficients with pure gauge 
indices are generated in the circle reduced theory by one-loop corrections.
We find that these loops can be interpreted as the dimensional reduction of 
six-dimensional box-diagrams inducing pure gauge anomalies in
the higher-dimensional theory.

Once more, we use \eqref{e:CS_coupling} to compute loop corrections to $k_{mnk}$, $k_{IJK}$, $k_{mIJ}$ and $k_{Imn}$, where the external legs of the
loop diagrams are always gauge bosons.
The explicit calculations can again be looked up in appendix \ref{app:loop}. In the general case of no particular hierarchy between $m_{CB}$ and $m_{KK}$, one finds
\begin{align}
 k_{mnk}=-\frac{1}{2}\sum_\frep{R} H(\frep{R}) \sum_{w\in \frep{R}}\Big(2\big\lfloor\!\mid r w\cdot \zeta \mid\!\big\rfloor +1\Big) q_m q_n q_k \sign(w \cdot \zeta)
\end{align}
from hyperini running in the loop charged under $A^m$, $A^n$, and $A^k$, 
\begin{align}
 k_{IJK}=&\frac{1}{2}\bigg[\sum_{roots}\Big(2\big\lfloor\!\mid r \alpha\cdot \zeta \mid\!\big\rfloor +1\Big)
 \langle \alpha_I^\vee ,\alpha_{nA} \rangle \langle \alpha_J^\vee ,\alpha_{nA} \rangle \langle \alpha_K^\vee ,\alpha_{nA} \rangle \sign(\alpha \cdot \zeta)\\
& -\sum_\frep{R} H(\frep{R})\sum_{w\in \frep{R}}\Big(2\big\lfloor\!\mid r w\cdot \zeta \mid\!\big\rfloor +1\Big)
\langle \alpha_I^\vee , \mathbf{w} \rangle \langle \alpha_J^\vee , \mathbf{w} \rangle \langle \alpha_K^\vee , \mathbf{w} \rangle \sign(w \cdot \zeta)\bigg] \nn
\end{align}
induced by non-Cartan gaugini of $G_{nA}$ and hyperini charged under the Cartan fields $A^I$, $A^J$, $A^K$,
\begin{align}
 k_{mIJ}=-\frac{1}{2}\sum_\frep{R} H(\frep{R})\sum_{w\in \frep{R}}\Big(2\big\lfloor\!\mid r w\cdot \zeta \mid\!\big\rfloor +1\Big)
 q_m  \langle \alpha_I^\vee ,\mathbf{w} \rangle \langle \alpha_J^\vee ,\mathbf{w} \rangle \sign(w \cdot \zeta)
\end{align}
generated by hyperini charged under the Cartan fields $A^I$, $A^J$ and the $A^m$, and
\begin{align}
 k_{Imn}=-\frac{1}{2}\sum_\frep{R} H(\frep{R})\sum_{w\in \frep{R}} \Big(2\big\lfloor\!\mid r w\cdot \zeta
 \mid\!\big\rfloor +1\Big) \langle \alpha_I^\vee ,\mathbf{w} \rangle q_m q_n \sign(w \cdot \zeta)
\end{align}
obtained from hyperini charged under $A^m$, $A^n$ and the Cartan generator $A^I$.

Once more, we can match these coefficients with the corresponding intersection numbers on the M-theory side.
Since we have not derived explicit general expressions for the latter, we test the matching for concrete examples in \autoref{s:examples}.

Unlike in the previous cases, matching the one-loop Chern-Simons coefficients has not allowed us to reproduce 
the anomaly cancelation conditions for pure gauge anomalies of \eqref{pure_nonAb_an}. Nevertheless, 
we suppose that all information about 6D gauge anomalies is captured 
by the coefficients $k_{mnk}$, $k_{IJK}$, $k_{mnI}$, $k_{mIJ}$.
This assumption is conceivable, since the loops we calculated in order to generate 
the coefficients can once more be obtained by dimensional
reduction of 6D anomalous box-diagrams. While the triangle diagram generating $k_{mnk}$ 
originates solely from the reduction of the pure Abelian anomaly diagram, the loop for $k_{IJK}$ 
descends from the pure non-Abelian anomaly and the Abelian-non-Abelian
anomaly with three non-Abelian and one Abelian gauge bosons as external legs. The coefficient $k_{mnI}$ is obtained
from the reduced loop of the 6D anomaly diagram with two Abelian and two non-Abelian gauge bosons as external legs. The same
6D loop reduces to the triangle diagram generating $k_{mIJ}$, which, however, can also be obtained by reducing the anomaly
with three non-Abelian and one Abelian gauge boson as external legs. Remarkably, these 5D loops  capture all possible ways
of reducing the 6D box-diagrams which induce pure gauge anomalies to five dimensions, at least when
the mass condition \eqref{normal_hierarchy} is satisfied.

\subsection{Creating new Chern-Simons terms for $m_{CB} > m_{KK}$} \label{sec:new_CS}

As already explained in \autoref{On_geometry}, certain intersection numbers of the fibered Calabi-Yau space vanish as long as the zero section is holomorphic.
After dropping the holomorphicity constraint, however, some of these intersection numbers may become non-zero.
On the other hand, since non-vanishing intersections induce Chern-Simons terms in M-theory, 
we should try to generate analogous terms in the circle reduced action as one-loop corrections. In doing so, we find out that the latter can only be non-vanishing
if there exist fields whose Coulomb-branch mass $m_{CB}$ is greater than the KK-scale $m_{KK}$.

Recall that the intersection numbers that vanished only for a holomorphic zero section were
\begin{align}
\mathcal{V}_{m00}=\mathcal{V}_{I00}=\mathcal{V}_{0Im}=0
\end{align} 
and we therefore consider the respective Chern-Simons coefficients $k_{m00}$, $k_{I00}$ and $k_{0Im}$. In the corresponding one-loop field theory
calculation in the circle reduced theory, we again use \eqref{e:CS_coupling} and obtain, as shown in \autoref{app:loop},
\begin{align}
 k_{m00}=-\frac{1}{6}\sum_\mathcal{R} H(\mathcal{R}) \sum_{w \in \mathcal{R}} &\big\lfloor\!\mid r w \cdot \zeta\mid\!\big\rfloor
 \Big(\big\lfloor\!\mid r w \cdot \zeta\mid\!\big\rfloor  +1\Big)\times \\
 &\Big(2\big\lfloor\!\mid r w \cdot \zeta\mid\!\big\rfloor +1\Big) q_m \sign(w\cdot \zeta) \nn
\end{align}
generated by KK-hyperini charged under $A^m$,
\begin{align}
 k_{I00}=\frac{1}{6} \bigg[&\sum_{roots} \big\lfloor  \mid r \alpha \cdot \zeta \mid\!\big\rfloor \Big(\big\lfloor\!\mid r \alpha \cdot
 \zeta\mid\!\big\rfloor  +1\Big)\Big(2\big\lfloor\!\mid r \alpha \cdot \zeta\mid\!\big\rfloor +1\Big)\langle \alpha_I^\vee, \alpha_{nA} \rangle \sign(\alpha \cdot \zeta) \\
& - \sum_\mathcal{R} H(\mathcal{R}) \sum_{w \in \mathcal{R}} \big\lfloor\!\mid r w \cdot \zeta\mid\!\big\rfloor \Big(\big\lfloor\!\mid r w
\cdot \zeta\mid\!\big\rfloor  +1\Big)\Big(2\big\lfloor\!\mid r w \cdot \zeta\mid\!\big\rfloor +1\Big) \times \nn \\
&\qquad \qquad \qquad \qquad \langle \alpha_I^\vee , \mathbf{w} \rangle \sign(w \cdot \zeta)\bigg] \nn
\end{align}
from KK-hyperini and KK-gaugini charged under $A^I$ and
\begin{align}
 k_{0Im}=&-\frac{1}{12}\sum_\mathcal{R} H(\mathcal{R}) \sum_{w \in \mathcal{R}}\bigg(1+6\big\lfloor\!\mid r w \cdot\zeta\mid\!\big\rfloor
 \Big(\big\lfloor\!\mid r w \cdot\zeta\mid\!\big\rfloor +1\Big) \bigg)  q_m \langle  \alpha_I^\vee , \mathbf{w} \rangle \\
=& -\frac{1}{2}\sum_\mathcal{R} H(\mathcal{R}) \sum_{w \in \mathcal{R}}\big\lfloor\!\mid r w \cdot\zeta\mid\!\big\rfloor
\Big(\big\lfloor\!\mid r w \cdot\zeta\mid\!\big\rfloor +1\Big)  q_m \langle  \alpha_I^\vee , \mathbf{w} \rangle \nn
\end{align}
by integrating out KK-hyperini charged under $A^I$ and $A^m$. In the last equality, we have used that $\sum_{\mathbf{w}}\langle
\alpha_I^\vee , \mathbf{w} \rangle =0$ , which we prove in appendix \ref{app:group_id}.

Indeed, we see that these coefficients vanish automatically if $m_{CB} < m_{KK}$. In the converse case they may still vanish,
but it would require appropriate cancelations between the different contributions.
Anyhow, in the next section we find for some examples that $m_{CB} < m_{KK}$ if and only if the fibration has a holomorphic
zero section, and that for a rational zero section the coefficients $k_{m00}$, $k_{I00}$, $k_{0Im}$ are always non-zero.

As discussed in the paragraph before \eqref{e:CS_mixed1_F}, interpreting the one-loop diagrams contributing to the 
Chern-Simons levels for a violated hierarchy \eqref{normal_hierarchy} is more involved and appears not to be
immediately related to 6D anomalies. While  $k_{m00}$ and  $k_{I00}$ could arise from 6D diagrams 
for mixed anomalies, the entire contribution to $k_{0Im}$ is expected to arise from 
diagrams whose non-vanishing is due to broken 6D symmetries. 

\section{Systematics and concrete examples of Calabi-Yau threefolds with Abelian gauge factors} \label{s:examples}

  In this section we confirm the matching of the Chern-Simons coefficients by evaluating both the geometric information
  on the M-theory side and the field theoretic quantities on the F-Theory side. Before tackling concrete examples,
  we first explain in general how to calculate the necessary geometric quantities. In doing so, we expand the prescriptions
  given in \cite{Grimm:2011fx,Cvetic:2012xn,Braun:2013yti} and find that we have to replace the relative Mori cone by the extended relative Mori
  cone in order to be able to describe Calabi-Yau manifolds with rational zero sections appropriately. The extended relative Mori cone
  has already been discussed in \cite{Grimm:2011fx}, albeit in a different context and using a different algorithm.
  
  Having explained the algorithm to calculate the extended relative Mori cone, we then turn to concrete examples. In \autoref{ss:su2xu1}
  we examine a simple example with gauge group $SU(2) \times U(1)$ and explicitly study the properties of a rational zero section
  in all glory detail. Subsections \ref{ss:su5xu1u1_1} and \ref{ss:su5xu1u1_2} contain two examples with gauge group $SU(5) \times U(1)^2$.
  While all phases of the former admit a holomorphic zero section, this is not true for the latter example and we again observe that the
  hierarchy \eqref{normal_hierarchy} is violated.
  
  To avoid confusion, let us emphasize that most of the methods discussed below readily lend themselves to an algorithmic
  implementation despite the fact that we have worked out the first example in much detail. In fact, there are only
  two obstacles that we have not yet managed to automate in a satisfactory way. The first one is to successfully
  identify divisor classes associated with an independent set of sections and the second one is to extract the set
  of all possibly occurring matter representations without over-counting too much.  The latter can of course be
  circumvented by analyzing the singularity structure of the defining equation of the Calabi-Yau manifold directly
  and we comment on doing so below. However, in order to quickly scan over many manifolds, it would be desirable
  to be able to read off all interesting quantities from combinatorial data alone. Nevertheless, both these issues seem surmountable.
  
\subsection{Generalities and an algorithm for determining $\sign{\mathbf{w}}$} 
  
Given a Calabi-Yau threefold $\hat{Y}_3$, we must calculate the following quantities in order to match the Chern-Simons coefficients
determined in the previous sections:
\begin{itemize}
   \item Intersection numbers between the divisors $D_\Lambda$
   \item The extended relative Mori cone of $\hat{Y}_3$, denoted by $\widehat{M}(\hat{Y}_3)$
\end{itemize}
Strictly speaking, one must also calculate the massless spectrum of the underlying 6D F-theory 
low-energy effective theory in order
to calculate the Chern-Simons coefficients on the circle-reduced side. However, most, if not even all, information about the
spectrum is already contained in the Mori cone and the intersection numbers. In fact, we will argue below that at least
for those examples that we have considered their knowledge was sufficient to uniquely determine the massless spectrum of the theory.
  
  We assume that it is known how to calculate intersection numbers and explain now how to extract the relevant information from
  the Mori cone of $\hat{Y}_3$. which we denote by $M(\hat{Y}_3)$\footnote{Note that most implementations of toric geometry only
  provide methods to calculate the Mori cone of the toric embedding space. To find the Mori cone of an anti-canonical hypersurface
  in a toric variety, one must take the intersection of all Mori cones pertaining to different triangulations of the embedding space
  that induce equivalent triangulations of the hypersurface. For details see for example the appendix of \cite{Grimm:2011fx}.}.
  Given $M(\hat{Y}_3)$, we construct the extended relative Mori cone $\widehat{M}(\hat{Y}_3)$ as follows:
  \begin{itemize}
   \item Take the intersection of $M(\hat{Y}_3)$ with the cone of all curves that have zero intersection with vertical divisors $D_{\alpha}$.
   \item Strictly speaking, this is all we need in order to obtain the extended relative Mori cone of $\hat{Y}_3$. However, it is useful
   to choose a different basis. Hence, for each element $m$ of this newly obtained cone do:
   \begin{itemize}
    \item Find the unique weight $\rep{w}$ of the weight space of $\fg = \mathrm{Lie}(G_{nA})$ such that
    \begin{align}
     -\langle \alpha_I, \rep{w} \rangle = D_{I} \cdot m
    \end{align}
    for all simple roots $\alpha_I$. Here, the right hand side is the intersection product between the exceptional divisor
    associated to minus the simple root $\alpha_I$ and the curve $m$.
    \item Determine the $U(1)$ charges $(q_{KK},q_n)$ of $m$ under the Kaluza-Klein vector $A^0$ and the Abelian gauge
    group factor $U(1)^{n_{U(1)}}$ by taking intersection products
    \begin{subequations}
    \begin{align}
     q_{KK} &= D_0 \cdot m \\
     q_{n}  &= D_n \cdot m \qquad n=1,\ldots, n_{U(1)}\,.
    \end{align}
    \end{subequations}
    \item The charges $(q_{KK}, q_n)$ together with the weight $\rep{w}$ determine an element 
    \begin{equation}
     \tilde{m} = (\rep{w}, q_{KK}, q_m) \in V \otimes \mathbb{Z}^{n_{U(1)}+1}\,,
    \end{equation}
    where $V$ is the weight space of $\fg$.
   \end{itemize}
   \item $\widehat{M}(Y_3)$ is the cone spanned by all elements $\tilde{m}$.
  \end{itemize}
  Note that there are
  \begin{equation}
   h^{1,1}(\hat{Y}_3) - h^{1,1}(\cB) = \rk \fg + n_{U(1)} + 1
  \end{equation}
  independent intersection numbers that an element $m$ which does not intersect vertical divisors can have.
  It is therefore crucial to include the charge under the Kaluza-Klein vector field $A^0$ to obtain a one-to-one
  map between fields on the circle reduced side and the intersection number between the curve $m$ and an arbitrary divisor
  of $\hat{Y}_3$. In previous calculations \cite{Grimm:2011fx,Morrison:2012ei,Cvetic:2012xn}, all fibral curves were
  assumed to have vanishing intersection with the zero section and therefore to carry no KK-charge. However,
  this works only as long as the Kaluza-Klein modes do not contribute to the loop-induced Chern-Simons coefficients.
  Given a weight $w = (\rep{w}, q_n)$, one can easily define its sign using the extended relative Mori cone $\widehat{M}(\hat{Y}_3)$:
  \begin{equation} \label{e:def_sign}
    \sign \left( w, n_{KK} \right) \equiv
   \left\{
    \begin{array}{ll}
      +1 \textrm{ if } (\rep{w}, n_{KK}, q_n) \in \widehat{M}(\hat{Y}_3) \\
      -1 \textrm{ otherwise}\\
    \end{array}
  \right.
  \end{equation}
  Note that the above definition gives an actual sign function, that is one satisfying $\sign(w,n_{KK}) = -\sign(-w,-n_{KK})$, only if
  either the curve associated with the weight $w$ or its conjugate, $-w$, is contained in the extended relative Mori cone. Since the Mori cone is convex,
  they can never both be contained in $\widehat{M}(\hat{Y}_3)$. However, since physical states correspond to M2-branes wrapping (anti-)holomorphic curves
  in the fiber \cite{Witten:1996qb}, one has in fact that either $w$ or $-w$ is an element of $\widehat{M}(\hat{Y}_3)$ as long as these weights belong to representations
  that actually occur in the low-energy effective theory and hence the above definition makes sense.
  
  Let us now comment on how to determine the field theory spectrum from the geometry. Obviously, the gauge algebra $\fg$ can easily be
  determined from the intersection numbers $D_{\alpha} \cdot D_I \cdot D_J$. The matter spectrum, however, is trickier to obtain and is
  usually calculated by studying the different kinds of singularity enhancement on higher-codimensional loci in the base manifold. That is,
  one searches for codimension two loci on the base, i.e.~points on $\cB$, where some of the resolution $\mathbb{P}^1$s become reducible
  and the gauge algebra enhances from $\fg$ to $\mathfrak{g}'$. At these points, matter fields are located and their non-Abelian representation
  can be obtained by decomposing the adjoint representation of $\mathfrak{g}'$ into representations of $\fg$. In the case of fibers embedded
  in arbitrary two-dimensional toric varieties, some effort might be necessary to transform the hypersurface equation into Weierstrass form,
  from which the singularity enhancement can be read off easily.
 For a general algorithm to obtain the Weierstrass form, we refer to \cite{Braun:2011ux}, in which a comprehensive scan over all Calabi-Yau threefolds was performed.
  In order to obtain matter charges under Abelian gauge group factors, one must compute the intersection number in the fiber between the pullback
  of the corresponding $U(1)$ generators and the rational fiber curve,
  see for example \cite{Morrison:2012ei,Mayrhofer:2012zy,Braun:2013yti,Borchmann:2013jwa,Cvetic:2013nia}.
  
  However, simply requiring that \eqref{e:def_sign} behaves indeed as a well-defined sign function as discussed above already puts considerably
  strong constraints on the allowed representations of $\fg$. Note, however,
  that this does not allow us to determine the different singlet representations, since the single weight of the representation $\mathbf{1}_{\lambda q_n}$
  satisfies this criterion for any value of $\lambda$ as long as $\mathbf{1}_{q_n}$ does.
  
Having determined the set of all occurring representations, their respective multiplicities can then be determined by finding solutions
to the Chern-Simons matchings. In all our examples, these were sufficiently restrictive to give a unique solution.
It would of course be desirable to be able to show in general that the Chern-Simons matching always determine the matter multiplicities uniquely.
  
With the necessary generalities discussed, we now turn to specific examples and match the Chern-Simons coefficients for each one of them.

    \subsection{First example: Gauge group $SU(2) \times U(1)$} \label{ss:su2xu1}
    Let us begin by discussing a simple example, which has both a phase with holomorphic zero section as well as a phase
    in which the zero section is non-holomorphic. Specifically, we take our Calabi-Yau threefold to be embedded in the
    toric ambient space whose rays are listed in \autoref{t:vb5_su2}.
    \begin{table}[h] 
	 \centering 
	 \begin{tabular}{rrrr|cc} 
 	 \multicolumn{4}{c|}{Point $n_z \in \nabla \cap N$}& Coordinate $z$ & Divisor class $[V(z)]$ \\ 
	 \hline 
	 	$-1$&$-1$&$-1$&$-1$& $h_{0}$ & $H$\\ 
	 	$0$&$0$&$0$&$1$& $h_{1}$ & $H$\\ 
	 	$-2$&$-1$&$1$&$0$& $d_{0}$ & $H-D_1$\\ 
	 	$-1$&$0$&$1$&$0$& $d_{1}$ & $D_1$\\ 
	 	$-1$&$0$&$0$&$0$& $f_{0}$ & $F_0$\\ 
	 	$0$&$1$&$0$&$0$& $f_{1}$ & $F_1$\\ 
	 	$1$&$0$&$0$&$0$& $f_{2}$ & $H + F_0 + F_1$\\ 
	 	$-1$&$-1$&$0$&$0$& $f_{3}$ & $-2 H + D_1 + F_1$
 	 \end{tabular} 
 	 \caption{The toric data of the ambient space $W_I$ of the smooth Calabi-Yau threefold $Y_I$ with Hodge numbers are $h^{1,1}(Y_I)=4$  and $h^{2,1}(Y_I)=84$.}
 	 \label{t:vb5_su2}
    \end{table}
    Since the projection onto the last two lattice coordinates is a well-defined fan morphism, it induces a toric morphism $\pi': W \rightarrow \mathbb{P}^2$
    from the toric ambient space $W$ to the base manifold $\cB = \mathbb{P}^2$. The kernel of the fan morphism is a two-dimensional reflexive polytope
    and therefore an anti-canonical hypersurface will in fact cut out an elliptic curve inside the generic fiber\footnote{The generic fiber is defined
    to be the fiber over all points of the open torus inside $\cB$. It is given by the toric variety obtained from the kernel of the corresponding fan morphism.
    In this specific case, the open torus is given by all points $[u:v:w] \in \mathbb{P}^2$ such that $u v w \neq 0$ and one can see that the generic fiber
    is a $\mathbb{F}_1$.}  of $\pi'$.
    Hence, the anti-canonical hypersurface inside $W_I$ indeed defines an elliptically fibered Calabi-Yau threefold with its projection map given
    by $\pi = \pi' \rvert_{Y_I}$.
    
     Next of all, one can confirm that there exists a total of four fine star triangulations. To see that these descend to only two inequivalent
     triangulations of the hypersurface, we examine their Stanley-Reisner ideals. All four of them share the common elements
     \begin{equation}
      d_1 f_3, f_1 f_3, f_0 f_2, h_0 h_1 d_1, d_0 f_1 f_2, h_0 h_1 d_0 f_2, h_0 h_1 d_0 f_1\,. \label{e:vb5_su2_SRI_common}
     \end{equation}
     The additional elements depend on the choice of triangulation and the four possible combinations are
     \begin{equation}
      \left\{
    \begin{array}{ll}
      d_1 f_0 \\
      d_0 f_1 \\
    \end{array}
    \right\} \times \left\{
    \begin{array}{ll}
     h_0 h_1 d_0 \\
     f_0 f_3 \\
    \end{array} \right\}\,. \label{e:vb5_su2_SRI}
     \end{equation}
     However, by writing down the equation $p=0$ for a generic anti-canonical hypersurface inside this toric ambient space, one can confirm that
     \begin{align}
      p \rvert_{f_0=f_3=0} \sim f_1 d_1 f_2^2 \qquad \textrm{and} \qquad p \rvert_{h_0=h_1=d_0=0} \sim f_1 d_1 f_2^2\,.
     \end{align}
     In both cases the common elements of the Stanley-Reisner ideals make it impossible to find solutions to $p = 0$ and hence there are no
     points on the Calabi-Yau threefold for which $f_0=f_3=0$ or $h_0=h_1=d_0=0$. We therefore find that the second factor of \eqref{e:vb5_su2_SRI}
     is irrelevant and there are only two inequivalent triangulations of the Calabi-Yau threefold -- one corresponding to including $d_1 f_0$ in
     the Stanley-Reisner ideal and the other corresponding to choosing $d_0 f_1$ instead. Their respective fans are given in  \eqref{e:vb5_su2_fan_hol}
     and \eqref{e:vb5_su2_fan_rat}.

     To proceed further, we define a basis of divisors. Since $h^{1,1}(\cB)=1$, there is precisely one independent vertical divisor, namely
      \begin{equation}
      H = \pi^{-1}([1:1:0])\,.
      \end{equation}
     There is only a single exceptional blow-up divisor $D_1$ and therefore the gauge group of the resulting low-energy effective theory will be $SU(2)$. 
     
     In this example, the most interesting part are the sections. From the Hodge numbers of $Y_I$ and the fact that the gauge group is $SU(2)$,
     we see that the Mordell-Weil group must have rank $1$. First, however, we concentrate on the zero section $D_{\hat{0}}$, which is realized
     as the toric divisor $F_0$. In order to understand the impact of the two different triangulations, we try to find an explicit form for
     the section by using the equation defining $Y_I$ inside the toric ambient space $W_I$. Since $f_0 f_2$ is contained in both Stanley-Reisner
     ideals, we set $f_0=0$ and $f_2=1$ to find
     \begin{equation}
     \begin{split}
      p(f_0=0, f_2=1) &= f_3 \left( \alpha_1 d_{0}^{2} d_{1}^{2} +\alpha_2 h_{0} d_{0} d_{1} + \alpha_3 h_{1} d_{0} d_{1} +\alpha_4 h_{0}^{2}
                         + \alpha_5 h_{0} h_{1} + \alpha_6 h_{1}^{2} \right) \\
			& \quad - \beta d_{1} f_{1}\,, \label{e:vb5_su2_zero_section}
      \end{split}
     \end{equation}
     where $\alpha_i$ and $\beta$ are generic constants. We can now see the crucial difference between the two inequivalent triangulations:
     \begin{enumerate}
      \item Let us first assume that $d_1 f_0$ is an element of the Stanley-Reisner ideal. In that case we can safely scale $d_1$ to $1$.
      Furthermore, for generic $\beta$, $f_3=0$ would imply that $f_1=0$, too, which is excluded by \eqref{e:vb5_su2_SRI_common}.
      Hence we can assume that $f_3 \neq 0$ and scale it to $1$ as well. One thus obtains the explicit form for the section
      \begin{equation}
       D_{\hat{0}} : [h_0 :h_1 : d_0] \mapsto [h_0 : h_1 :d_0 : 1 : 0 : f_1(h_0,h_1,d_0): 1 : 1]
       \end{equation}
      where
      \begin{equation}
       f_1(h_0,h_1,d_0) = \frac{1}{\beta} \left(\alpha_1 d_{0}^{2} d_{1}^{2} +\alpha_2 h_{0} d_{0} d_{1} + \alpha_3 h_{1} d_{0} d_{1} +\alpha_4 h_{0}^{2}
                         + \alpha_5 h_{0} h_{1} + \alpha_6 h_{1}^{2} \right)\,.
      \end{equation}
      In particular, one sees that the zero section is \emph{holomorphic} and we call the corresponding Calabi-Yau threefold $Y_{I,\textrm{hol.}}$.
      
      \item Alternatively, we can take $d_0 f_1$ to be contained in the Stanley-Reisner ideal. In this case there is nothing 
      that prevents $d_1$ from becoming $0$ and therefore we cannot simply scale it to $1$ anymore. As a consequence, we
      cannot find a \emph{holomorphic} expression for $f_1$ in terms of the base coordinates. With this triangulation, $D_{\hat{0}}$ defines a \emph{rational} zero section and we denote the corresponding threefold by $Y_{I,\textrm{rat.}}$.
      
      Furthermore, note the following. Setting $f_0=d_1=0$, we can scale $f_2$ and $f_3$ to $1$ and find
      \begin{equation}
       p(d_1=0, f_0=0, f_2=1, f_3=1) = \alpha_4 h_{0}^{2} + \alpha_5 h_{0} h_{1} + \alpha_6 h_{1}^{2}
      \end{equation}
      with $f_1$ left unconstrained. Since $h_0$ and $h_1$ cannot both be zero at the same time and the above equation implies
      that $h_0=0 \leftrightarrow h_1=0$ for generic $\alpha_i$, we can set $h_1=1$. This leaves us with the quadratic constraint 
      \begin{equation}
       0 = \alpha_4 h_0^2 + \alpha_5 h_0 + \alpha_6 \label{e:vb5_su2_non_flat_points}
      \end{equation}
      on $h_0$ and two unconstrained coordinates $d_0$ and $f_1$. So far we have used three out of four scaling relations and
      therefore the intersection between $D_{\hat{0}}$ and $D_1$ has complex dimension $1$ and, in particular,
      \begin{equation}
       D_{\hat{0}} \cdot D_1 \neq 0
      \end{equation}
      in the Chow ring of the Calabi-Yau threefold. This is exactly what we expect from \eqref{e:i_0I} for a \emph{non-holomorphic} zero section.
     \end{enumerate}
     
     Let us therefore quickly summarize the content of the Stanley-Reisner ideal and its relation to the properties of the zero section:
     \begin{equation}
      d_1 f_3, f_1 f_3, f_0 f_2, f_0 f_3, h_0 h_1 d_0, h_0 h_1 d_1, d_0 f_1 f_2, \times  \left\{
    \begin{array}{ll}
      d_1 f_0: D_{\hat{0}} \textrm{ holomorphic}\\
      d_0 f_1: D_{\hat{0}} \textrm{ rational} \\
    \end{array}
    \right.
     \end{equation}
     Unfortunately, we cannot repeat the same discussion for the second section, the generator of the Mordell-Weil group,
     since only one section is realized torically. Nevertheless, one can still determine its homology class, namely
     \begin{equation}
      [\sigma_1] = [F_1] - [F_0]\,,
     \end{equation}
     which can be shown to have the correct intersection numbers with the remaining divisors and contains a unique global
     section over the Calabi-Yau threefold. A more detailed discussion of this issue for a slightly different example is
     contained in \cite{Braun:2013yti}. Lastly, plugging in the defining equations, the shifted base divisor $D_0$ and the $U(1)$ divisor $D_{U(1)}$ are
    \begin{align}
    D_0 &= D_{\hat{0}} + \frac{3}{2} H \\
    D_{U(1)} &= 2 \sigma_1 - 2 D_{\hat{0}} - 16 H + 2 D_1 \,,
    \end{align}
    where we have taken the freedom to re-scale the $U(1)$ divisor by a factor of $2$ in order to obtain integer charges.

     Going through the algorithm outlined at the beginning of this section, one can determine the cones $\widehat{M}$ for
     both triangulations of the reflexive polytope and finds
     \begin{subequations}
     \begin{align}
      \widehat{M}\left(Y_{I,\textrm{hol.}}\right) &= \mathrm{Cone} \left( e_2 + 4 e_{U(1)} + e_{KK}, -6 e_{U(1)} - e_{KK}, e_1 \right) \label{e:vb5_su2_mori_holomorphic} \\
      \widehat{M}\left(Y_{I,\textrm{rat.}}\right) &= \mathrm{Cone} \left(-e_2 - 4 e_{U(1)} - e_{KK}, 4 e_{U(1)} + e_{KK}, -e_1 - 2 e_{U(1)} \right)\,.
     \end{align}
     \end{subequations}
     Here, we have picked $e_i$, $i=1,2$ to be the generators of the $\mathfrak{su}(2)$ weight lattice and imposed the equivalence relation $\sum_i e_i \sim 0$.
     Clearly, the curve corresponding to the weight $\tilde{m} = e_2 + 4 e_{U(1)} + e_{KK}$ is flopped in the transition from one triangulation to another.
     In the Calabi-Yau threefold with holomorphic zero section $\sign(\tilde{m}) = 1$, while convexity of the Mori cone implies that $\sign(\tilde{m}) = -1$
     for the threefold with rational zero section.
     
Next of all, we wish to determine the matter spectrum. As mentioned above, one can either try to extract this data from $\widehat{M}(\hat{Y}_{I})$,
or examine the singularity enhancements by studying the explicit hypersurface equation. In this particular case, the charged matter spectrum can
be found to consist of the representations
\begin{equation}
	\rep{2}_0,\ \rep{2}_2,\ \rep{2}_4,\ \rep{1}_2,\ \rep{1}_4\, , \label{e:vb5_su2_spectrum}
\end{equation}
where the subscript indicates the $U(1)$ charge of the state.
Note that even though there \emph{is} matter transforming under the antisymmetric representation $\Lambda^2 (\rep{2}) = \rep{1}$ of $SU(2)$,
it carries no charge under any of the Cartan generators and can therefore be neglected in the following analysis. Given this set of representations,
we now wish to determine whether or not there exist multiplicities $H(\frep{R})$ such that all Chern-Simons coefficients can be matched.
Before doing so, we remark on the crucial difference between the two triangulations. In the case of the \emph{holomorphic} zero section,
one can use \eqref{e:vb5_su2_mori_holomorphic} to confirm that
\begin{equation}
   \sign( w, n_{KK} ) = 1 \ \textrm{for} \ n_{KK}  \geq 1
\end{equation}
and
\begin{equation}
  \sign( w, n_{KK} ) = -1 \ \textrm{for} \ n_{KK} \leq -1\,
\end{equation}
for all weights $w$ of the representations $\frep{R}$ in \eqref{e:vb5_su2_spectrum}. As a 
consequence, all contributions from Kaluza-Klein modes running in the loops either cancel 
among each other perfectly or add up in a simply summable way discussed 
in section \ref{sec:one-loop_CS}.
For the \emph{non-holomorphic} zero section this is no longer true. 
As noted above, there is a single curve which undergoes
a flop transition from one triangulation to another and therefore 
\begin{equation}
	e_2 + 4 e_{U(1)} + e_{KK}
\end{equation}
is no longer contained in $\widehat{M}(Y_{I,\textrm{rat.}})$. No curve with negative Kaluza-Klein charge lies in
$\widehat{M}(Y_{I,\textrm{rat.}})$. As a consequence, there are two Kaluza-Klein modes 
whose contributions to the Chern-Simons terms
has to be treated differently in the calculation. This was encountered for in section \ref{sec:one-loop_CS}
by considering the general cases with violated condition \eqref{normal_hierarchy}.
     
Taking this into account, one can calculate the induced Chern-Simons terms on the field theoretic side 
for generic matter multiplicities $H(\frep{R})$.
Matching them with the intersection numbers on the M-theory side gives a system of linear equations whose unique solution is
\begin{align} \label{number_ex1}
    H(\rep{2}_0) = 12 \qquad H(\rep{2}_2) = 8 \qquad H(\rep{2}_4) = 2 \nonumber \\
    H(\rep{1}_2) = 112 \qquad H(\rep{1}_4) = 36\,.
\end{align}
To check anomaly cancelation for this spectrum one also needs to read off the anomaly coefficients. 
For the base $\cB=\bbP^2$ one has 
\beq \label{example_eta_a}
   \Omega_{11} = H\cdot H =1\ , \qquad a^1 = -3 \ ,
\eeq
where the basis element generating $H^{1,1}(\cB)$ is $H$.
In this example the location of the seven-branes are specified by
\beq
  b^1_{SU(2)} = 1\ , \qquad b^1_{U(1)} = 64\ . 
\eeq
Given these explicit expressions and the spectrum \eqref{number_ex1}, it
is straightforward to check that all 6D anomalies are canceled.

\noindent\textbf{An intriguing observation}
     
\noindent Before finishing with this example, we would like to make one further observation. First of all, let us make contact with the analysis
of phase transitions in \cite{Witten:1996qb}. As we have just noted, there are exactly two points in the base manifold $\cB$ over which matter in the
$\rep{1}_4$ representation is located. To each of these matter points belong two isolated fibral curves, represented by the weights $e_1 + 4 e_{U(1)}$ 
and $e_2 + 4 e_{U(1)}$, plus the whole tower of Kaluza-Klein states for each weight.
Flopping $\cC \equiv e_2 + 4 e_{U(1)} + e_{KK}$ in the transition from one triangulation to another, one therefore flops \emph{two} curves in the manifold,
      one associated to each matter point. According to Witten's analysis, we therefore expect all intersection numbers
      \begin{equation}
       D_\Lambda \cdot D_\Sigma \cdot D_\Theta 
      \end{equation}
      to jump by
      \begin{equation}
       2 (D_\Lambda \cdot \cC) (D_\Sigma \cdot \cC) (D_\Theta \cdot \cC)\ ,
      \end{equation}
      which is precisely what we find.
      
      However, in the triangulation with a rational zero section, there is one more intriguing fact. In the previous analysis, we observed that
      there are precisely two points in the base manifold over which the zero section wraps an entire fiber component instead of marking a single point,
      namely those for which \eqref{e:vb5_su2_non_flat_points} was fulfilled. However, these are precisely the points over which matter in the $\rep{1}_4$
      representation is located. We therefore believe that in addition to the field theoretic arguments presented in the previous section, there must also
      be a clear geometric interpretation of why non-holomorphic zero sections cause Kaluza-Klein modes to be light enough to contribute corrections to the
      Chern-Simons coupling.

  \subsection{Second example: Gauge group $SU(5) \times U(1)^2$} \label{ss:su5xu1u1_1}
  Next of all, we consider a Calabi-Yau threefold giving rise to a $U(1)^2$ Abelian gauge factor. Its defining reflexive polytope is given in \autoref{t:e2}.
  As before, we choose the base manifold to be $\cB = \mathbb{P}^2$.
\begin{table}[h] 
	 \centering 
	 \begin{tabular}{rrrr|cc} 
 	 \multicolumn{4}{c|}{Point $n_z \in \nabla \cap N$}& Coordinate $z$ & Divisor class $[V(z)]$ \\ 
	 \hline 
	 	$3$&$2$&$1$&$1$& $h_{0}$ & $H$\\ 
	 	$3$&$2$&$0$&$-1$& $h_{1}$ & $H$\\ 
	 	$3$&$2$&$-1$&$0$& $d_{0}$ & $H - D_1 - D_2 - D_3 - D_4$\\ 
	 	$2$&$1$&$-1$&$0$& $d_{1}$ & $D_1$\\ 
	 	$1$&$0$&$-1$&$0$& $d_{2}$ & $D_2$\\ 
	 	$0$&$0$&$-1$&$0$& $d_{3}$ & $D_3$\\ 
	 	$1$&$1$&$-1$&$0$& $d_{4}$ & $D_4$\\ 
	 	$3$&$2$&$0$&$0$& $f_{0}$ & $D_{\hat{0}}$\\ 
	 	$-1$&$-1$&$0$&$0$& $f_{1}$ & $\sigma_1$\\ 
	 	$-1$&$0$&$0$&$0$& $f_{2}$ & $\sigma_2$\\ 
	 	$1$&$0$&$0$&$0$& $f_{3}$ & $3 H - D_1 - 2 D_2 - D_3 + D_{\hat{0}} - \sigma_1 + \sigma_2$ \\ 
	 	$-2$&$-1$&$0$&$0$& $f_{4}$ & $6 H - D_1 - 2 D_2 -2 D_3 - D_4 + 2 D_{\hat{0}} - \sigma_1$
 	 \end{tabular} 
 	 \caption{The toric data of the ambient space $W_{II}$ of the smooth Calabi-Yau threefold $Y_{II}$ with Hodge numbers are $h^{1,1}(Y_{II})=8$
 	 and $h^{2,1}(Y_{II})=75$.}
 	 \label{t:e2}
 \end{table}
  The $216$ different fine star triangulations of the toric ambient space give rise to twelve inequivalent triangulations of the embedded hypersurface $Y_3$.
  Since all of these triangulations have a holomorphic zero section, we limit ourselves to studying the particular triangulation whose fan is given by \eqref{e:e2_fan}.
  Compared to the previous example, the main difference lies in the rational sections. There are now two independent Mordell-Weil group generators and, conveniently,
  they are both realized as toric divisors $f_1=0$ and $f_2=0$, respectively. Furthermore, the rational sections $\sigma_1$ and $\sigma_2$ do not intersect
  the zero section, i.e.~$D_{\hat{0}} \cdot \sigma_i = 0$.
  
  Since the base manifold is again a $\mathbb{P}^2$, the shifted base divisor reads $D_0 = D_{\hat{0}} + \frac{3}{2} H$ as before. Applying the Shioda map and rescaling by a factor of five yields the $U(1)$ generators
  \begin{subequations}
  \begin{align}
    D_5 &= 5 \sigma_1 - 5 D_{\hat{0}} - 15 H + 3 D_1 + 6 D_2 + 4 D_3 + 2 D_4 \\
    D_6 &= 5 \sigma_2 - 5 D_{\hat{0}} - 15 H + 1 D_1 + 2 D_2 + 3 D_3 + 4 D_4\,.
  \end{align}
  \end{subequations}
  By the same logic as before, one calculates that
  \begin{equation}
   \begin{split}
   \widehat{M}(Y_{II}) & = \mathrm{Cone} \Bigl( -e_4 - 3 e_{U(1)_1} - e_{U(1)_2}, e_4 - 2 e_{U(1)_1} - 4 e_{U(1)_2}, e_3 +3 3 e_{U(1)_1} + e_{U(1)_2},\\
      & \qquad \qquad \quad e_1+e_5 + e_{U(1)_1} - 3 e_{U(1)_2}, e_2+e_4 + e_{U(1)_1} +2 e_{U(1)_2}, \\
      & \qquad \qquad \quad e_1 - e_2. -e_1+e_5 + e_{KK}, -5e_{U(1)_1} +5 e_{U(1)_2} + e_{KK} \Bigr).
   \end{split}
  \end{equation}
  The matter spectrum turns out to be
  \begin{equation}
   \rep{5}_{-2,-4},\ \rep{5}_{-2,1},\ \rep{5}_{3,1},\ \rep{10}_{1,2},\ \rep{1}_{5,0},\ \rep{1}_{0,5},\ \rep{1}_{5,5}\,. \label{e:e2_spectrum}
  \end{equation}
  As before, the non-Abelian sector can be determined directly from demanding that the sign function on the weight space is well-defined. Having determined the set of all
  possible representations, we search for a solution for the match of the 5-dimensional Chern-Simons coefficients in order to determine the number 
  of representations the low-energy effective theory contains. Again, a unique solution exists and it reads
  \begin{align} \label{number_ex2}
   &H(\rep{5}_{-2,-4}) = 5\quad H(\rep{5}_{-2,1}) = 7\quad H(\rep{5}_{3,1}) =  7\nonumber \\
   &H(\rep{10}_{1,2}) = 3\quad H(\rep{1}_{5,0}) = 28 \quad H(\rep{1}_{0,5}) = 35 \quad H(\rep{1}_{5,5}) = 35\,.
  \end{align}
To conclude, we check that all 6D anomalies are canceled 
for this example. Since the base is again $\bbP^2$ we use \eqref{example_eta_a} and 
the brane locations specified by
\beq
   b^1_{SU(5)} = 1 \ , \qquad b^{1}_{U(1)\, 11} = 120 \ , \ \ b^{1}_{U(1)\, 12} = 65 \ , \ \ b^{1}_{U(1)\, 22} = 130 \ ,
\eeq
to show anomaly cancelation for the spectrum \eqref{number_ex2}.
  
\subsection{Third example: Gauge group $SU(5) \times U(1)^2$} \label{ss:su5xu1u1_2}
  
Lastly, we present an example with gauge group $SU(5) \times U(1)^2$, which, unlike the previous one, has triangulations in which the zero section is non-holomorphic.
\begin{table}[h] 
	 \centering 
	 \begin{tabular}{rrrr|cc} 
 	 \multicolumn{4}{c|}{Point $n_z \in \nabla \cap N$}& Coordinate
$z$ & Divisor class $[V(z)]$ \\ 
	 \hline 
	 	$3$&$1$&$-1$&$-1$& $h_{0}$ & $H$\\ 
	 	$0$&$-3$&$0$&$1$& $h_{1}$ & $H$\\ 
	 	$-1$&$-1$&$1$&$0$& $d_{0}$ & $H - D_1 - D_2 - D_3 - D_4$\\ 
	 	$-1$&$0$&$1$&$0$& $d_{1}$ & $D_1$\\ 
	 	$0$&$1$&$1$&$0$& $d_{2}$ & $D_2$\\ 
	 	$0$&$0$&$1$&$0$& $d_{3}$ & $D_3$\\ 
	 	$0$&$-1$&$1$&$0$& $d_{4}$ & $D_4$\\ 
	 	$-1$&$-1$&$0$&$0$& $f_{0}$ & $D_{\hat{0}}$\\ 
	 	$1$&$2$&$0$&$0$& $f_{1}$ & $\sigma_1$\\ 
	 	$-1$&$0$&$0$&$0$& $f_{2}$ & $\sigma_2$\\ 
	 	$0$&$1$&$0$&$0$& $f_{3}$ & $H-D_1-3 D_2 - 2 D_3 - D_4 + 2 D_{\hat{0}} - 3 \sigma_1 + \sigma_2$\\ 
	 	$1$&$-1$&$0$&$0$& $f_{4}$ &$-2 H- D_2 -  D_3 - D_4 + D_{\hat{0}} - \sigma_1 + \sigma_2$
 	 \end{tabular} 
 	 \caption{The toric data of the ambient space $W_{III}$ of the smooth Calabi-Yau threefold $Y_{III}$ with
                      Hodge numbers are $h^{1,1}(Y_{III})=8$  and $h^{2,1}(Y_{III})=75$.}
\end{table}
Of the $324$ different triangulations admitted by the toric ambient space, only $18$ descend to inequivalent triangulations of the
anti-canonical hypersurface. Half of these possess a holomorphic zero section. Apart from the holomorphicity of the zero section, the only
other difference between the different phases is the sub-wedge of the Weyl chamber that the VEV of the adjoint scalar lies
in \cite{Intriligator:1997pq,Grimm:2011fx,Hayashi:2013lra}. We therefore concentrate on one triangulation with a holomorphic zero section
and another one in which the zero section is non-holomorphic. Their respective fans are given by \eqref{e:f1_hol} and \eqref{e:f1_rat}.
  
  Choosing an appropriate basis of divisors is fairly straightforward, since both Mordell-Weil group generators are realized torically. 
  After rescaling by a factor of five in order to avoid fractional charges, we therefore find that the shifted divisors are
  \begin{subequations}
   \begin{align}
    D_0 &= D_{\hat{0}} + \frac{3}{2} H \\
    D_{U(1)_1} &= 5 \sigma_1 - 5 D_{\hat{0}} - 15 H + 3 D_1 +6 D_2 + 4 D_3 + 2 D_4\\
    D_{U(1)_2} &= 5 \sigma_2 - 5 D_{\hat{0}} - 40 H + 4 D_1 + 3 D_2 + 2 D_3 + D_4\,.
   \end{align}
  \end{subequations}
  Next of all, one calculates that the cones are given by
  \begin{align}
   \widehat{M}(Y_{III,\textrm{hol.}}) &= \textrm{Cone} \Bigl( e_5 - 2 e_{U(1)_1} - 6 e_{U(1)_2},  e_2 + 3 e_{U(1)_1} + 4 e_{U(1)_2},\nonumber \\
   & \qquad \qquad \ - e_1 + 2 e_{U(1)_1} + 6 e_{U(1)_2} + e_{KK}, -5 e_{U(1)_1} - 15 e_{U(1)_2} - e_{KK},  \nonumber \\
    & \qquad \qquad \  5 e_{U(1)_1}, -5 e_{U(1)_1} - 5 e_{U(1)_2}, e_3 -e_4, \nonumber \\
    & \qquad \qquad \ -e_1 - e_5 - e_{U(1)_1} - 3 e_{U(1)_2}, e_1 + e_4 + e_{U(1)_1} + 3 e_{U(1)_2} \Bigr)
  \end{align}
  and
  \begin{align}
   \widehat{M}(Y_{III,\textrm{rat.}}) &= \textrm{Cone} \Bigl( e_2 + 3 e_{U(1)_1} + 4 e_{U(1)_2}, e_1 - 2 e_{U(1)_1} - 6 e_{U(1)_2} - e_{KK}, \nonumber \\
   & \qquad \qquad \ -e_1 + e_5 + e_{KK}, 5 e_{U(1)_1}, - 5 e_{U(1)_1} -5 e_{U(1)_2}, e_3-e_4,  \nonumber \\
   & \qquad \qquad \ -e_1-e_5 -e_{U(1)_1} - 3 e_{U(1)_2}, e_1 + e_4 + e_{U(1)_1} + 3 e_{U(1)_2} \Bigr)\,.
  \end{align}
  Comparing these two cones, one finds a number of differences corresponding to changing the sub-wedge of the Weyl chamber \cite{Hayashi:2013lra}. However, there is one additional flop
  \begin{equation}
   -e_1 + 2 e_{U(1)_1} + 6 e_{U(1)_2} + e_{KK} \leftrightarrow e_1 - 2 e_{U(1)_1} - 6 e_{U(1)_2} - e_{KK}
  \end{equation}
  which has the effect that the two weights $e_1 - 2 e_{U(1)_1} - 6 e_{U(1)_2} \pm e_{KK}$ do not have opposite signs anymore. Therefore the
  contributions of the corresponding Kaluza-Klein modes do not cancel and must be taken into account when matching their Chern-Simons terms.
  
  The matter spectrum can be determined to be
  \begin{equation}
    \rep{5}_{-2,-6},\ \rep{5}_{-2,-1},\ \rep{5}_{3,4},\ \rep{10}_{1,3},\ \rep{1}_{0,5},\ \rep{1}_{5,5},\ \rep{1}_{5,10}\,.
  \end{equation}
  Taking the Kaluza-Klein modes into account, one can match the Chern-Simons coefficients obtained from integrating out matter on the field
  theory with those given by intersection numbers of the M-theory geometry. Once again, there is a unique solution and the multiplicities one obtains are
  \begin{align}
   H(\rep{1}_{0,5}) = 35, \quad H(\rep{1}_{5,5}) = 28, \quad H(\rep{1}_{5,10}) = 35 \nonumber \\
   H(\rep{5}_{-2,-6}) = 5, \quad H(\rep{5}_{-2,-1}) = 7, \quad H(\rep{5}_{3,4}) = 7 \nonumber \\
   H(\rep{10}_{1,3}) = 3\,.
  \end{align}
Once again, one straightforwardly checks that all 6D anomalies are canceled 
for this example. To do so, we use \eqref{example_eta_a} and 
the brane locations specified by
\beq
   b^1_{SU(5)} = 1 \ , \qquad b^{1}_{U(1)\, 11} = 120 \ , \ \ b^{1}_{U(1)\, 12} = 185 \ , \ \ b^{1}_{U(1)\, 22} = 380 \ .
\eeq

\section{Conclusions}

In this paper we calculated the six-dimensional effective F-theory
action of a theory with $\cN = (1,0)$ supersymmetry and Abelian as well as non-Abelian gauge group factors.
In doing so, we exploited the duality between M-theory on an elliptically fibered manifold
and Type IIB string theory on the corresponding base manifold with varying dilaton
and seven-branes. We dimensionally reduced M-theory on a resolved elliptically fibered
Calabi-Yau threefold and determined the resulting 5D theory. To implement the M-theory to
F-theory limit we also performed a circle reduction of a generic 6D $\cN= (1,0)$ 
supergravity. Comparing the two 5D theories we were able to connect the characteristic
data specifying the $(1,0)$ theory to the geometry of the Calabi-Yau threefold.

Having performed both reductions and compared the resulting actions,
we found that the circle reduced action contained additional fields not present on the M-theory
side. First of all, there are Kaluza-Klein modes originating from the 6D/5D circle reduction.
Second of all, M-theory compactification on a \emph{resolved} Calabi-Yau corresponds to
choosing a point in the Coulomb branch of the resulting gauge theory,
or, equivalently, giving masses to the W-bosons. Hence, a second set of fields, namely 
the W-bosons, their superpartners, and fields charged under the gauge group, becomes massive.
Neither one of these sets of fields appears in the M-theory low-energy
effective action. In order to compare our two reductions, we therefore had to manually integrate out these
fields on the circle reduction side. In contrast, the M-theory reduction yielded a number of additional
Chern-Simons terms and their
supersymmetric completions that are absent in the classical circle reduction. We showed that all such terms
are in fact induced at one loop when integrating out massive fields.

By doing so, we encountered new and interesting connections of the physics of
5D field theories to the geometry of elliptically fibered Calabi-Yau threefolds.
As has long been known, Abelian gauge group factors in the
effective theory are obtained from Calabi-Yau manifolds admitting multiple sections.
In particular, considering non-holomorphic sections as $U(1)$ generators
can lead to a richer spectrum of Abelian charges. In this work, we
furthermore explored the implications of having a non-holomorphic zero section. While a
holomorphic zero section appears to always enforce a hierarchy between Kaluza-Klein masses and
Coulomb branch masses, this is no longer true for a non-holomorphic zero section. In fact, we
studied concrete examples and found cases for which the Kaluza-Klein scale was smaller than some
of the Coulomb branch masses. Geometrically, this means that in performing the F-theory limit and
sending the fiber volume to zero, one must pay close attention to the relative volumes of the generic
torus fiber and the exceptional resolution divisors. In this context, we gave an improved and extended
algorithm for determining the relevant subset of the Mori cone known as the \emph{extended relative} Mori cone.

Taking these insights into account, we described the matching of the one-loop Chern-Simons
terms on the circle reduction side with the M-theory results even
in the presence of a non-holomorphic zero section. Due to their
origin, the loop-induced Chern-Simons terms encode information about the charged spectrum of
the 6D theory. However, since 6D theories can potentially be anomalous, every anomaly free
spectrum in six dimensions must obey certain constraints, which are
thus translated into constraints on the Chern-Simons coefficients. We analyzed these constraints for
purely gravitational, mixed and pure gauge anomalies. In the case of the former two, we were
able to show in full generality that the one-loop Chern-Simons coefficients take the
form of the one-loop 6D (1,0) anomalies induced by the chiral spectrum if the mass hierarchy
\eqref{normal_hierarchy} is satisfied. For a given geometry it is then straightforward to check
anomaly cancelation. To do this in general requires more insights on the geometrical relations
for elliptically fibered Calabi-Yau threefolds.
For the pure gauge anomalies or a violated condition \eqref{normal_hierarchy}
the situation becomes even more involved, since in that case none of the one-loop Chern-Simons
terms captures the 6D anomaly contributions directly.
Nevertheless, all information appears to be encoded in the Chern-Simons terms
and we worked out the correct matchings for specific examples.
To complete this picture, we included higher curvature corrections in
the M-theory effective action. Matching these with loop-induced contributions on the circle reduction
side gave a further consistency check of our reduction. 
    
There are various interesting future directions that should be explored.
It would be desirable to be able to generally derive all
properties of the 6D $(1,0)$ theory from the
5D effective action in the Coulomb branch. For the charged spectrum
this appears always to be possible for a given example, but
closed expressions for the number of fields and appearing
representations are so far unknown. Combined with geometric
relations for elliptic fibrations one should then be able to
generally prove anomaly cancelation in F-theory.
Moreover, one can extend the strategy suggested in \cite{Bonetti:2013cza}
and ask if a given low-energy 5D theory can ever
arise from an anomaly free 6D $(1,0)$ theory with $U(1)$ factors.
This again requires to check non-trivial relations among a given set
of Chern-Simons terms. The much harder question is to
ask if every consistent set of Chern-Simons terms arising from
an anomaly free circle reduced 6D theory can arise in M-theory.
To answer this question, it would be desirable to
find geometric bounds on their values and rule out geometrically impossible patterns.
Finally, let us also stress that more detailed information is
also needed to recover the complete 6D theory. In particular, the metric for the matter 
multiplets was never determined in our analysis. One-loop Chern-Simons terms turned out to be independent
of its precise form and it is therefore necessary to look at further
corrections in five dimensions that might capture this additional information.

\subsubsection*{Acknowledgments}

We gratefully acknowledge interesting discussions with Federico Bonetti, Volker Braun, Stefan Hohenegger,
Denis Klevers, Eran Palti, Daniel Park, Tom Pugh, Raffaele Savelli, Wati Taylor, and Timo Weigand. We would particularly like 
to thank Volker Braun for sharing his knowledge and his SAGE
implementations for toric geometry with us.
This research was supported by a grant of the Max Planck Society.

\appendix

\section{Conventions and group theory identities} 

\subsection{Conventions} \label{a:conventions}

For all spacetime dimensions $d$, let us adopt the mostly plus convention for the metric $g_{\mu\nu}$,
and the $(+++)$ conventions of \cite{Misner:1974qy} for the Riemann tensor.
Furthermore, we denote the Levi-Civita tensor by $\epsilon_{\mu_1 \dots \mu_d}$ 
and use the above metric to raise its indices. With this definition we have in any coordinate system $(x^0, x^1, \dots ,x^{d-1})$ that
\begin{equation}
\epsilon_{01 \dots (d-1)} = +\sqrt{-\det g_{\mu\nu}} \;.
\end{equation}
Then the following identity is satisfied for arbitrary $k=0,...,d$:
\begin{align}
\epsilon_{\mu_1 \dots \mu_k \lambda_{k+1} \dots \lambda_d}  \epsilon^{\nu_1 \dots \nu_k \lambda_{k+1} \dots \lambda_{d}} 
= - k!(d-k!) \delta^{\nu_1}_{[\mu_1} \dots \delta^{\nu_k}_{\mu_k]} \;.
\end{align}
We expand differential $p$-forms as
\begin{equation}
\lambda = \tfrac{1}{p!} \lambda_{\mu_1 \dots \mu_p} \; dx^{\mu_1} \wedge \dots \wedge dx^{\mu_p} \;,
\end{equation}
such that the wedge product of a $p$- and a $q$-form satisfies
\begin{equation}
(\alpha \wedge \beta)_{\mu_1 \dots \mu_{p+q}} 
= \tfrac{(p+q)!}{p!q!} \alpha_{[\mu_1 \dots \mu_p} \beta_{\mu_{p+1} \dots \mu_{p+q}]} \;.
\end{equation}
Next of all, exterior differentiation of a $p$-form yields
\begin{equation}
(d\alpha)_{\mu_0 \dots \mu_p} = (p+1) \partial_{[\mu_0} \alpha_{\mu_1 \dots \mu_p]} \; .
\end{equation}
In real coordinates and arbitrary spacetime dimension $d$, we take the Hodge dual of a $p$-form to be defined by the following expression:
\begin{equation}
(*\alpha)_{\mu_1 \dots \mu_{d-p}} = \tfrac{1}{p!} \alpha^{\nu_1 \dots \nu_p} 
\epsilon_{\nu_1 \dots \nu_p \mu_1 \dots \mu_{d-p}} \;.
\end{equation}
As a consequence,
\begin{equation}
\alpha \wedge *\beta = \tfrac{1}{p!} \alpha_{\mu_1 \dots \mu_p} \beta^{\mu_1 \dots \mu_p} \; *1
\end{equation}
is satisfied identically for arbitrary $p$-forms $\alpha, \beta$.

\subsection{Group theory identities} \label{app:group_id} 
In this section, we briefly state the group theory conventions used in this paper and then proceed to prove three identities used to match one-loop Chern-Simons terms from 5D F-theory with intersection numbers on the M-theory side in \autoref{sec:one-loop_CS}. For the sake of brevity, we denote the roots of the non-Abelian group by $\alpha$ instead of $\alpha_{nA}$. For an introduction to the theory of Lie algebras and the representations, we refer for example to \cite{Fuchs:1997jv}.

Let us begin by defining the coroot intersection matrix as
\begin{align} \label{def-cCIJ}
   \cC_{I J} = \frac{1}{\lambda(\fg)} \langle \alpha^\vee_{I}, \alpha^\vee_{J} \rangle 
                 = \frac{1}{\lambda(\fg)}\frac{2}{\langle \alpha_{J}, \alpha_{J} \rangle} C_{I J}\,,
  \end{align}
  where $\langle \alpha^\vee_{I}, \alpha^\vee_{J} \rangle$ denotes the inner product between two coroots of the Lie algebra $\fg$ 
  and $\alpha_I$ are the simple roots of $\fg$. We also define
  \begin{align}\label{lambda}
   \lambda(\fg)=\frac{2}{\langle \alpha_{max} , \alpha_{max} \rangle} \ ,
  \end{align}
  where $\alpha_{max}$ is the root of the Lie algebra $\fg$ with maximal length. The Cartan matrix is refered to as $C_{IJ}$.
  Note that for the simply-laced groups of ADE-type, $\cC_{I J}$ and the 
  Cartan matrix $C_{I J}$ coincide.
  Throughout this work the conventions for the normalization of the Cartan generators $T^M$ are chosen such that
  \begin{align}\label{norm_cartan}
   \tr \, (T^M T^N ) = \delta^{MN} \ ,
  \end{align}
  where the trace is taken in the fundamental representation of $\fg$.
  Note that this also fixes the normalization of the roots and weights.

 Having fixed all notation, we proceed by proving the second equality in \eqref{k0IJ}.
To do so, we show that
\begin{align}\label{e:adj_id}
A_{\textrm{adj}}\lambda (\mathfrak{g})\mathcal{C}_{IJ} = \sum_{roots}\langle \alpha^\vee_I , \alpha \rangle \langle \alpha^\vee_J , \alpha \rangle
\end{align}
\begin{align}\label{e:R_id}
A_{\rep{R}} \lambda (\mathfrak{g}) \mathcal{C}_{IJ} = \sum_{w \in \rep{R}}\langle  \alpha_I^\vee , w \rangle \langle \alpha_J^\vee , w \rangle\ ,
\end{align}
where the second equation is a generalization of the first. These hold for any simple Lie algebra $\mathfrak{g}$ and for all non-trivial, finite-dimensional irreducible representations $\rep{R}$. 

Following \cite{Fuchs:1997jv} we first define an inner product on the Lie algebra $\mathfrak{g}$
\begin{align}\begin{split}\label{e:killing}
\kappa : \mathfrak{g} \times \mathfrak{g} &\rightarrow \mathbb{C} \\
x,y &\mapsto \tr ( \ad_x \circ \ad_y )\ , 
\end{split}\end{align}
where the trace is taken in the adjoint representation of the Lie algebra. The above product is called the Killing form and it is bilinear and symmetric. It was proven by Cartan that for finite-dimensional semi-simple Lie algebras the Killing form $\kappa$ is non-degenerate and, hence, so is its restriction to  any Cartan sub-algebra $\fg_\circ \subset \fg$. We can therefore use the Killing form to identify the Cartan sub-algebra $\mathfrak{g}_\circ$ with the dual space $\mathfrak{g}_\circ^\star$, the space spanned by the roots. In particular, we identify $\alpha \in \mathfrak{g}_\circ^\star$ with $T^\alpha \in \mathfrak{g}_\circ$ such that
\begin{align}
\alpha (T) = c_\alpha \, \kappa (T^\alpha , T )\quad \forall \, T \in \mathfrak{g}_\circ\ ,
\end{align}
where $c_\alpha$ is some normalization constant. If one then chooses a basis of the Cartan sub-algebra $\lbrace T^M \rbrace_{M=1, \dots, \dim (\mathfrak{g}_\circ )}$ generating the non-Abelian gauge group, one can expand every $T^\alpha$ as
\begin{align}
T^\alpha = a_M^\alpha T^M\ ,
\end{align}
In accordance with \eqref{norm_cartan} we have normalized the Cartan generators as
\begin{align}
\kappa (T^M T^N ) = A_{\textrm{adj}} \delta^{MN}\ .
\end{align}
Identifying $\mathfrak{g}_\circ$ and $\mathfrak{g}_\circ^\star$ enables us to define a non-degenerate product on $\mathfrak{g}_\circ^\star$ via the Killing form by setting
\begin{align}\label{e:inner_product}
(\alpha , \beta ) := c_\alpha c_\beta \kappa (T^\alpha T^\beta ) = c_\beta \alpha (T^\beta )\ .
\end{align}
 for any two roots $\alpha, \beta \in \mathfrak{g}_\circ^\star$. By bilinearity, this extends to all of $\mathfrak{g}_\circ^\star$.

Let us now use the following identity from {\cite{Fuchs:1997jv} for any $\lambda, \mu \in \mathfrak{g}_\circ^\star$:
\begin{align}
(\lambda , \mu ) = \sum_{roots} (\alpha , \lambda ) (\alpha , \mu )\ .
\end{align} 
The right hand side of this equation can be expanded as
\begin{align}\begin{split}
\sum_{roots}(\alpha , \lambda)(\alpha , \mu) & = \sum_{roots}c_\alpha c_\lambda \kappa (T^\alpha , T^\lambda)c_\alpha c_\mu \kappa (T^\alpha , T^\mu)\\
& = \sum_{roots}c_\alpha c_\lambda a_M^\alpha a_N^\lambda \kappa (T^M , T^N)c_\alpha c_\mu a^\alpha_K a^\mu_L \kappa (T^K , T^L)\\
& = \sum_{roots}c_\alpha c_\lambda \frac{1}{c_\alpha A_{\textrm{adj}}} \alpha (T^M) \frac{1}{c_\lambda A_{\textrm{adj}}} \lambda (T^N) A_{\textrm{adj}} \delta^{MN}\times \\
&\,\,\,\,\,\,\,\,\, c_\alpha c_\mu \frac{1}{c_\alpha A_{\textrm{adj}}} \alpha (T^K) \frac{1}{c_\mu A_{\textrm{adj}}} \mu (T^L) A_{\textrm{adj}} \delta^{KL}\\
& =\sum_{roots} \frac{1}{A_{\textrm{adj}}^2}\alpha (T^M) \lambda (T^M) \alpha (T^K) \mu (T^K)\\
&= \sum_{roots} \frac{1}{A^2_{\textrm{adj}}} \langle \alpha , \lambda \rangle \langle \alpha , \mu \rangle\ .
\end{split}\end{align}
Similarly, the left hand side can be rewritten as
\begin{align}\begin{split}
(\lambda , \mu ) & = c_\lambda c_\mu \kappa (T^\lambda , T^\mu) = c_\lambda c_\mu a_M^\lambda a_N^\mu \kappa (T^M , T^N) = c_\lambda c_\mu \frac{1}{c_\lambda A_{\textrm{adj}}} \lambda (T^M) \frac{1}{c_\mu A_{\textrm{adj}}} \mu (T^N) A_{\textrm{adj}} \delta^{MN}\\
& = \frac{1}{A_{\textrm{adj}}}\lambda (T^M) \mu (T^M) = \frac{1}{A_{\textrm{adj}}} \langle \lambda , \mu \rangle\ .
\label{e:adj_lhs}
\end{split}\end{align}
Combining the two equations then yields
\begin{align}\label{e:group_id_1}
A_{\textrm{adj}} \langle \lambda , \mu \rangle = \sum_{roots} \langle \alpha , \lambda \rangle \langle \alpha , \mu \rangle\ .
\end{align}
Now note that
\begin{align}
\langle \alpha_I^\vee , \alpha_J^\vee \rangle & = \frac{4\langle \alpha_I , \alpha_J \rangle }{\langle \alpha_I , \alpha _I \rangle \langle \alpha_J , \alpha_J \rangle} = \lambda (\mathfrak{g}) \mathcal{C}_{IJ}
\end{align}
and insert the coroots $\alpha_I^\vee$ and $\alpha_J^\vee$ for $\lambda$ and $\mu$ to obtain
\begin{align}
A_{\textrm{adj}} \lambda (\mathfrak{g}) \mathcal{C}_{IJ}= \sum_{roots} \langle \alpha , \alpha_I^\vee \rangle \langle \alpha , \alpha_J^\vee \rangle\ ,
\end{align}
which is exactly \eqref{e:adj_id}.

Let us now proceed and prove \eqref{e:R_id}. As shown in \cite{Fuchs:1997jv}, for any simple Lie algebra $\mathfrak{g}$ and any finite-dimensional, non-trivial irreducible representation $\rep{R}$, the trace over $\rep{R}$ is proportional to the trace in the adjoint representation. Hence,
\begin{align}
\kappa_{\rep{R}} (x,y) := \tr (\rep{R}(x)\rep{R}(y)) = K_{\rep{R}} \kappa (x,y)
\end{align}
for all $x,y \in \mathfrak{g}$ with the proportionality factor $K_{\rep{R}}$ depending of course on the representation $\rep{R}$. Using the definition of the inner product \eqref{e:inner_product}, we then have for $\lambda , \mu \in \mathfrak{g}_\circ^\star$ that
\begin{align}\begin{split}
(\lambda , \mu ) & = c_\lambda c_\mu \kappa_{\textrm{adj}}(T^\lambda , T^\mu) = c_\lambda c_\mu a_M^\lambda a_N^\mu \frac{1}{K_{\rep{R}}}\kappa_{\rep{R}} (T^M , T^N)\\
& = c_\lambda c_\mu \frac{1}{c_\lambda A_{\textrm{adj}}} \lambda (T^M) \frac{1}{c_\mu A_{\textrm{adj}}} \mu (T^N) \frac{1}{K_{\rep{R}}} \sum_{\mathbf{w} \in \rep{R}} \mathbf{w}(T^M)\mathbf{w}(T^N)\\
& = \frac{1}{A_{\textrm{adj}}^2}\lambda (T^M) \mu (T^N) \frac{1}{K_{\rep{R}}} \sum_{\mathbf{w} \in \rep{R}} \mathbf{w}(T^M) \mathbf{w}(T^N)  = \frac{1}{A_{\textrm{adj}}^2} \frac{1}{K_{\rep{R}}} \sum_{\mathbf{w} \in \rep{R}} \langle \lambda , \mathbf{w} \rangle \langle \mu , \mathbf{w} \rangle\ .
\end{split}\end{align}
In the third equality we used that the weights can be chosen to form an orthonormal basis of the representation space. Inserting \eqref{e:adj_lhs}, one then finds
\begin{align}\begin{split}
&K_{\rep{R}} A_{\textrm{adj}} \langle \lambda , \mu \rangle = \sum_{\mathbf{w} \in \rep{R}} \langle \lambda , \mathbf{w} \rangle \langle \mu , \mathbf{w} \rangle \\
\Rightarrow \, & A_{\rep{R}} \langle \lambda , \mu \rangle = \sum_{\mathbf{w} \in \rep{R}} \langle \lambda , \mathbf{w} \rangle \langle \mu , \mathbf{w} \rangle\ ,
\end{split}\end{align}
which, after plugging in the coroots, finally yields \eqref{e:R_id}:
\begin{align}
A_{\rep{R}} \lambda(\mathfrak{g}) \mathcal{C}_{IJ} = \sum_{\mathbf{w} \in \rep{R}} \langle \alpha_I^\vee , \mathbf{w} \rangle \langle \alpha_J^\vee , \mathbf{w} \rangle\ .
\end{align}

Last of all, we prove the identity
\begin{equation}
 \sum_{\mathbf{w} \in \rep{R}} \langle\alpha, \mathbf{w} \rangle \label{e:sum_w_alpha} = 0
\end{equation}
for any root $\alpha$ and any highest weight representation $\rep{R}$.

Given a representation $\rep{R}$ of a Lie algebra $\fg$
and a simple root $\alpha$, $\fg$ always contains an $\mathfrak{sl}(2,\mathbb{C})$ subalgebra defined as
\begin{equation}
 s_{\alpha} = \fg_{\alpha} \oplus \fg_{-\alpha} \oplus \left[\fg_{\alpha},\fg_{-\alpha}\right]\,.
\end{equation}
Here, $\fg_{\alpha}$ is the linear subspace of $\fg$ spanned by elements $\mathfrak{l} \in \fg$ such that $[T^M,\mathfrak{l}] = \alpha^M$, where $T^M$ form the basis of the Cartan subalgebra of $\fg$. Now, the idea is to decompose $\rep{R}$ into chains of representations of $s_{\alpha}$ in order to reduce the problem to dealing with $\mathfrak{sl}(2,\mathbb{C})$ representations. And in fact, this can easily be accomplished as follows. 
Given any weight $\mathbf{w}$ of $\rep{R}$, acting with $\fg_{\pm \alpha}$ either annihilates
$\mathbf{w}$ or gives another weight $\mathbf{w}' = \mathbf{w} \pm \alpha$ of $\rep{R}$, since $s_{\alpha}$ is a subalgebra of $\fg$.
The different orbits under the action of $s_{\alpha}$ therefore form a partition of the weigths $\mathbf{w} \in \rep{R}$.
For each such orbit, we pick the highest weight $\mathbf{v}$ with of the $\mathfrak{sl}(2,\mathbb{C})$ representation associated
with $s_{\alpha}$ and denote its dimension by $d_{\mathbf{v}}$. Then $\rep{R}$ decomposes as
\begin{equation}
 \rep{R} = \bigoplus_{\mathbf{v}} \left( V_{\mathbf{v}} \oplus V_{\mathbf{v} - \alpha}  \ldots \oplus V_{\mathbf{v} - (d_{\mathbf{v}}-1) \alpha} \right)\,,
\end{equation}
where $\mathbf{v}$ ranges over highest weights of $s_{\alpha}$ orbits and $V_{\mathbf{w}}$ is the subspace of $\rep{R}$ spanned by $\mathbf{w}$. One can now rearrange \eqref{e:sum_w_alpha} into sums over $\mathfrak{sl}(2,\mathbb{C})$ representations and take advantage of the fact that the representation theory of highest weight representations of $\mathfrak{sl}(2,\mathbb{C})$ is very simple. Since the weights of a such a representation with dimension $d$ are just integer numbers given by
\begin{equation}
 d-1, d-3, \dots, -(d-1), -(d-3)\,,
\end{equation}
one can evaluate
\begin{equation}
 \sum_{\mathbf{w} \in \rep{R}} \langle \mathbf{w},\alpha \rangle = \sum_{\mathbf{v}} \sum_{i=0}^{d_{\mathbf{v}}-1} \langle \mathbf{v} - i \alpha, \alpha \rangle
 = \sum_{\mathbf{v}} \sum_{i=0}^{d_{\mathbf{v}}-1} \left(d-1-2 i \right) = 0\,.
\end{equation}

\section{Circle reduction of the 6D action}
\label{app:circle_reduction} 
In this part of the appendix we explicitly carry out the circle reduction of 6D $\mathcal{N} =(1,0)$ supergravity as sketched in \ref{sec:circle_reduction}. We closely follow \cite{Bonetti:2011mw} in the following.

Upon compactification on a circle of radius $r$ the 6D metric is reduced to
\begin{align}
d\hat{s}^2=\tilde{g}_{\mu \nu}dx^\mu dx^\nu +r^2 Dy^2 \,,
\end{align}
where
\begin{align}
 Dy = dy  - A^0 , \qquad A^0 = A^0_\mu dx^\mu \qquad F^0 = dA^0 \,.
\end{align}
Here $\tilde{g}_{\mu \nu}$ is the 5D metric and the tilde indicates that one still has to perform a Weyl rescaling to obtain the Einstein-Hilbert term in the canonical form. Recall that 6D quantities and indices are denoted by a hat and that 5D fields do not depend on the circle coordinate $y$. The Kaluza-Klein vector $A^0$ enjoys a $U(1)$ gauge symmetry from $S^1$-diffeomorphisms and has the usual Abelian field strength $F^0$. The reduction of the Vielbeine is found to be
\begin{align}
 \hat{e}^a=\tilde{e}^a_\mu dx^\mu\ , \qquad \hat{e}^5=rDy \,.
\end{align}
The spin connection reduces to
\begin{align}
 \hat{\omega}_{ab}=\tilde{\omega}_{ab}+\tilde{\mathfrak{a}}^{(0)}_{ab}Dy\ , \qquad \hat{\omega}_{a5}=\tilde{\mathfrak{b}}^{(1)}_a + \tilde{\mathfrak{c}}^{(0)}_a Dy\,, 
\end{align}
where we have introduced the functions $\tilde{\mathfrak{a}}^{(0)}_{ab}$, $\tilde{\mathfrak{c}}^{(0)}_a$ and the one-form $\tilde{\mathfrak{b}}^{(1)}_a$ given by
\begin{align}
 \tilde{\mathfrak{a}}^{(0)}_{ab}=\frac{1}{2}r^2 \tilde{e}^\mu_a \tilde{e}^\nu_b F^0_{\mu \nu}\ , \qquad \tilde{\mathfrak{b}}^{(1)}_a=\frac{1}{2}r \tilde{e}^\lambda_a F^0_{\lambda \mu} dx^\mu\ , \qquad \tilde{\mathfrak{c}}^{(0)}_a=-\tilde{e}^\lambda_a \tilde{\nabla}_\lambda r \,.
\end{align}
At leading order, the reduction of the Ricci-scalar is 
\begin{align}
 \hat{R} = \tilde{R} + \dots \ , 
\end{align}
where we neglect higher curvature contributions.\footnote{We stress that for the moment we approach only a two-derivative reduction. Therefore higher curvature contributions are omitted in the following. This affects the Green-Schwarz term, the tensor kinetic terms and the Einstein-Hilbert term. See also \autoref{sec:circle_reduction}.} The vectors are reduced according to
\begin{align}
 \hat{A}= A + \zeta Dy \ , \qquad \hat{A}^m= A^m + \zeta^m Dy \,,
\end{align}
where $A$, $A^m$ are 5D vectors and $\zeta$, $\zeta^m$ are 5D scalars. The reduction of the tensors reads 
\begin{align}
 \hat{B}^\alpha = B^\alpha - [ A^\alpha - \frac{1}{2}a^\alpha \tr (\tilde{\mathfrak{a}}^{(0)}\tilde{\omega}) - 2 \frac{b^\alpha}{\lambda (\mathfrak{g})} \tr (\zeta A) -2b^\alpha _{mn} \zeta ^m A^n)]\wedge Dy
\end{align}
with a 5D tensor $B^\alpha$ and a 5D vector $A^\alpha$. While the Abelian vector $A^\alpha$ has the usual field strength $F^\alpha = dA^\alpha$, the gauge invariant field strength for $B^\alpha$ turns out to be
\begin{align}
 G^\alpha = dB^\alpha -A^\alpha \wedge F^0 + \frac{1}{2}a^\alpha \tilde{\omega}^{CS}_{grav} +2\frac{b^\alpha}{\lambda (\mathfrak{g})} \omega ^{CS} + 2 b^\alpha _{mn} \omega ^{CS,mn} \,.
\end{align}

As already mentioned in \autoref{sec:circle_reduction}, the 6D scalars reduce trivially to 5D scalars.

One can now insert these reductions into the 6D action \eqref{e:6D_action}. We show the results for the different terms separately. The Einstein-Hilbert term is reduced to
\begin{align}
 \hat{S}^{(6)}_{EH}  = \int _{M_6}  \frac{1}{2} \hat{R} \hat{\ast} 1 = \int _{M_6}  \frac{1}{2} r \tilde{R} \tilde{\ast} 1 \wedge Dy \,.
\end{align}
To obtain the corresponding term in the 5D effective action, one has to integrate over the circle direction, which is just a trivial integration of $Dy$. Now the reduction of the Green-Schwarz terms takes the form\footnote{In the following we omit terms without a $Dy$-factor, since these forms are integrated to zero along the circle direction.}
\begin{align}
 S_{GS}^{(6)} & = \int_{M_6} - \Omega_{\alpha \beta} \frac{b^\alpha}{\lambda (\mathfrak{g})} \hat{B}_\beta \wedge tr \hat{F}\wedge\hat{F} - \Omega_{\alpha \beta} b^{\alpha}_{mn} \hat{B}_\beta \wedge tr \hat{F}^{m} \wedge\hat{F}^{n}  \\
 & = \int_{M_6} -\frac{1}{2} \Omega_{\alpha \beta} G^\alpha \wedge (\mathcal{F}^\beta - F^\beta) \wedge Dy +  \Omega_{\alpha \beta}\frac{b^\alpha}{\lambda (\mathfrak{g})} A^\beta \wedge tr (F \wedge F) \wedge Dy \nn \\
& \qquad \quad + \Omega_{\alpha \beta} b^\alpha _{mn} A^\beta \wedge F^m \wedge F^n \wedge Dy - 2  \Omega_{\alpha \beta} \frac{b^\alpha}{\lambda (\mathfrak{g})} \omega ^{CS}\wedge \big[2 \frac{b^\beta}{\lambda (\mathfrak{g})} tr (\zeta F)\nn \\
& \qquad \quad - \frac{b^\beta}{\lambda (\mathfrak{g})} tr(\zeta \zeta)F^0 + 2 b^\beta _{mn} \zeta ^m F^n - b^\beta _{mn} \zeta ^m \zeta ^n F^0\big] \wedge Dy - 2  \Omega_{\alpha \beta} b^\alpha _{kl} \omega ^{CS,kl}\wedge \nn \\ 
& \qquad \qquad \big[2 \frac{b^\beta}{\lambda (\mathfrak{g})} tr (\zeta F) - \frac{b^\beta}{\lambda (\mathfrak{g})} tr(\zeta \zeta)F^0 + 2 b^\beta _{mn} \zeta ^m F^n - b^\beta _{mn} \zeta ^m \zeta ^n F^0\big] \wedge Dy\nn \\
& \qquad \quad - 2\Omega_{\alpha \beta} \frac{b^\alpha}{\lambda (\mathfrak{g})} \frac{b^\beta}{\lambda (\mathfrak{g})} tr \zeta A \wedge \big[tr F \wedge F + tr \zeta \zeta F^0 \wedge F^0 -2 tr \zeta F \wedge F^0\big] \wedge Dy\nn \\
& \qquad \quad -2 \Omega_{\alpha \beta} \frac{b^\alpha}{\lambda (\mathfrak{g})} b^{\beta}_{mn}\zeta ^m A^n \wedge \big[tr F \wedge F + tr \zeta \zeta F^0 \wedge F^0 -2 tr \zeta F \wedge F^0\big] \wedge Dy\nn \\
& \qquad \quad -2 \Omega_{\alpha \beta} b^{\alpha}_{mn} \frac{b^\beta}{\lambda (\mathfrak{g})} tr \zeta A \wedge \big[ F^m \wedge F^n +  \zeta ^m \zeta ^n F^0 \wedge F^0 -2  \zeta ^m F^n \wedge F^0\big] \wedge Dy\nn \\
& \qquad \quad -2 \Omega_{\alpha \beta} b^{\alpha}_{mn} b^\beta _{kl}  \zeta ^k A^l \wedge \big[ F^m \wedge F^n +  \zeta ^m \zeta ^n F^0 \wedge F^0 -2  \zeta ^m F^n \wedge F^0\big] \wedge Dy\nn \,.
\end{align}
The kinetic terms for the Abelian vectors are reduced to
\begin{align}
  \int _{M_6}& -2 \Omega _{\alpha \beta} j^\alpha b^\beta _{mn} \hat{F}^m \wedge \hat{\ast} \hat{F}^n\\
& = \int _{M_6} -2 r \Omega _{\alpha \beta} j^\alpha b^\beta _{mn} (F^m - \zeta ^m F^0) \wedge \tilde{\ast}(F^n - \zeta ^n F^0) \wedge Dy \nn \\
& \qquad \quad -2 r^{-1} \Omega _{\alpha \beta} j^\alpha b^\beta _{mn} d \zeta ^m \wedge \tilde{\ast}d \zeta ^n \wedge Dy \,, \nn
\end{align}
while the reduction for the non-Abelian vectors was found in \cite{Bonetti:2011mw} to be
\begin{align}
  \int _{M_6}& -2 \Omega _{\alpha \beta} j^\alpha b^\beta \tr \hat{F} \wedge \hat{\ast} \hat{F}\\
& = \int _{M_6} -2 r \Omega _{\alpha \beta} j^\alpha b^\beta \tr (F - \zeta  F^0) \wedge \tilde{\ast}(F - \zeta  F^0) \wedge Dy \nn \\
& \qquad \quad -2 r^{-1} \Omega _{\alpha \beta} j^\alpha b^\beta \tr D \zeta  \wedge \tilde{\ast}D \zeta  \wedge Dy \,,\nn
\end{align}
where we have introduced the covariant derivative for the adjoint scalars in the vector multiplets as
\begin{align}
 D\zeta = d \zeta + [A, \zeta ] \ .
\end{align}
The kinetic terms of the 6D tensors are found to reduce to
\begin{align}
 \int _{M_6}& -\frac{1}{4} g_{\alpha \beta} \hat{G}^\alpha \wedge \hat{\ast} \hat{G}^\beta \\
 & = \int _{M_6} -\frac{1}{4} r g_{\alpha \beta} G^\alpha \wedge \tilde{\ast} G^\beta \wedge Dy - \frac{1}{4} r^{-1} g_{\alpha \beta} \mathcal{F}^\alpha \wedge \tilde{\ast} \mathcal{F}^\beta \wedge Dy \ , \nn
\end{align}
where $\mathcal{F}^\alpha$ was defined in \eqref{modified_cF}. 
While terms involving neutral 6D scalars reduce trivially to five dimensions, this is not true for terms with charged scalars. One computes
\begin{align}\label{e:scalar_red}
 \int _{M_6}& -h_{UV} \hat{D} q^U \wedge \hat{\ast} \hat{D} q^V\\
 & = \int _{M_6} -rh_{UV} Dq^U \wedge \tilde{\ast} Dq^V \wedge Dy\nn \\
& \qquad \quad  - r^{-1} h_{UV} (\zeta ^{\rep{R}_U} q^U + \zeta ^m q_m^{(U)} q^U)(\zeta ^{\rep{R}_V} q^V  + \zeta ^m q_m^{(V)} q^V) \tilde{\ast}1 \wedge Dy\,.\nn 
\end{align}
The expression $Dq^U$ encodes the 5D covariant derivative
\begin{align}
 Dq^U=dq^U + A^{\rep{R}_U} q^U - i q_m^{(U)} A^m q^U
\end{align}
and the $\zeta^{\rep{R}_U}$ denote the scalars from the 5D vector multiplet in the representation $\rep{R}_U$ of the Lie-algebra, where $\rep{R}_U$ is the representation $q^U$ transforms in. The last line in \eqref{e:scalar_red} only contributes to the 5D scalar potential. It is completed by reducing the 6D scalar potential, which we did not carry out. Finally, the combination of all of these terms makes up the full circle reduced classical bosonic two-derivative pseudo-action.

As in six dimensions, there is still some redundancy in this 5D pseudo-action. In contrast to the 6D case, we are nevertheless able to write down a proper action without any additional duality constraints. This works by dualizing the action, in particular replacing all tensors $G^\alpha$ by the vectors $F^\alpha$. The connection between the vectors and tensors can be seen by reducing the duality constraint \eqref{self_duality} to
\begin{align}
\label{self_duality_reduced}
 r g_{\alpha \beta} \tilde{\ast} G^\beta = -\Omega_{\alpha \beta}\mathcal{F}^\beta \,.
\end{align}
We can safely modify the Lagrangian by adding a total derivative
\begin{align}
 \Delta S^{(5)F} & =\int _{M_5} -\frac{1}{2} \Omega _{\alpha \beta} dB^\alpha \wedge F^\beta \\
& = \int _{M_5} -\frac{1}{2} \Omega _{\alpha \beta} G^\alpha \wedge F^\beta + \frac{1}{2} \Omega _{\alpha \beta} (-A^\alpha \wedge F^0 + 2 \frac{b^\alpha}{\lambda (\mathfrak{g})} \omega ^{CS} + 2b^\alpha _{mn} \omega ^{CS,mn})\wedge F^\beta \,. \nn
\end{align}
Varying the new action with respect to $G^\alpha$ gives precisely the reduced duality constraint \eqref{self_duality_reduced}. The terms in the 5D action that change in the dualization procedure are
\begin{align}
 \int _{M_5}& -\frac{1}{4} r g_{\alpha \beta} G^\alpha \wedge \tilde{\ast} G^\beta - \frac{1}{4} r^{-1} g_{\alpha \beta} \mathcal{F}^\alpha \wedge \tilde{\ast} \mathcal{F}^\beta \\
&  -\frac{1}{2} \Omega_{\alpha \beta} G^\alpha \wedge (\mathcal{F}^\beta - F^\beta) +  \Omega_{\alpha \beta} \frac{b^\alpha}{\lambda (\mathfrak{g})} A^\beta \wedge tr (F \wedge F)\nn \\
&  + \Omega_{\alpha \beta} b^\alpha _{mn} A^\beta \wedge F^m \wedge F^n -\frac{1}{2} \Omega _{\alpha \beta} G^\alpha \wedge F^\beta \nn \\
&  - \frac{1}{2} \Omega _{\alpha \beta} A^\alpha \wedge F^0 \wedge F^\beta + \Omega _{\alpha \beta} \frac{b^\alpha}{\lambda (G)} \omega ^{CS} \wedge F^\beta + \Omega _{\alpha \beta} b^\alpha _{mn} \omega ^{CS,mn} \wedge F^\beta\nn  \\
 =& \int _{M_5} -\frac{1}{2} r^{-1} g_{\alpha \beta} \mathcal{F}^\alpha \wedge \tilde{\ast}\mathcal{F}^\beta  + 2 \Omega _{\alpha \beta} \frac{b^\alpha}{\lambda (\mathfrak{g})} A^\beta \wedge tr F \wedge F \nn \\
& + 2 \Omega _{\alpha \beta} b^\alpha _{mn} A^\beta \wedge F^m \wedge F^n - \frac{1}{2} \Omega _{\alpha \beta} A^0 \wedge F^\alpha \wedge F^\beta \nn \ ,
\end{align}
where we inserted the reduced duality constraint \eqref{self_duality_reduced}.

The Einstein-Hilbert term is not in its canonical form yet. Performing the Weyl rescaling $\tilde{g}_{\mu \nu} = r^{-2/3}g_{\mu \nu}$  turns out to give the right result
\begin{align}
 S^{(5)F}_{EH}  = \int _{M_5} & \frac{1}{2} R \ast 1 \,.
\end{align}
Note that the Hodge star operator scales as $\tilde{\ast}\alpha = r^{-5/3}(r^{2/3})^p \ast \alpha$~, where $\alpha$ is a p-form.

The final step is to push the theory onto the Coulomb branch, which means that we give a VEV to the scalars in the 5D vector multiplets.
The W-bosons get massive and break the gauge group to its maximal torus. Additionally, the charged hypermultiplets acquire a mass and do not appear in the effective action. Including only massless modes, one obtains the final form \eqref{circle_action} for the classical 5D action on the Coulomb branch.
\begin{align}
S^{(5)F}  = \int _{M_5} &+ \frac{1}{2} R \ast 1 -\frac{2}{3}r^{-2} dr \wedge \ast dr 
         - \frac{1}{2} g_{\alpha \beta} dj^\alpha \wedge \ast dj^\beta -h_{uv} dq^u \wedge \ast dq^v \\
&  -2 r^{-2} \Omega_{\alpha \beta} j^\alpha b^\beta_{\hat I \hat J} \, d \zeta ^{\hat I} \wedge \ast d \zeta ^{\hat J}
    - \frac{1}{4} r^{8/3} F^0 \wedge \ast F^0   - \frac{1}{2} r^{-4/3} g_{\alpha \beta}\, \mathcal{F}^\alpha \wedge \ast \mathcal{F}^\beta  \nn  \\
&  
     -2 r^{2/3}  \Omega_{\alpha \beta} j^\alpha  b^\beta_{\hat I \hat J} \, (F^{\hat I} - \zeta ^{\hat I} F^0) \wedge \ast(F^{\hat J} - \zeta ^{\hat J} F^0)\nn \\
& - \frac{1}{2} \Omega _{\alpha \beta} \, A^0 \wedge F^\alpha \wedge F^\beta  
                                  +  2 \Omega_{\alpha \beta} b^\alpha_{\hat I \hat J} \, A^\beta \wedge F^{\hat I} \wedge F^{\hat J} \nn \\
& - 2\Omega_{\alpha \beta} b^\alpha_{\hat I \hat J} b^\beta_{\hat I \hat J} \zeta ^{\hat K} \zeta ^{\hat L} \zeta ^{\hat I} A^{\hat J} \wedge F^0 \wedge F^0 \nn \\
&  + 2  \Omega_{\alpha \beta} (b^\alpha_{\hat I \hat J} b^\beta_{\hat K \hat L} + 2 b^\alpha_{\hat I \hat K} b^\beta_{\hat J \hat L}) \zeta ^{\hat K} \zeta ^{\hat L} A^{\hat I} \wedge F^{\hat J} \wedge F^0 \nn \\
&  - 2\Omega_{\alpha \beta} (2 b^\alpha_{\hat I \hat J} b^\beta_{\hat K \hat L} + b^\alpha_{\hat I \hat L} b^\beta_{\hat J \hat K}) \zeta ^{\hat L} A^{\hat I} \wedge F^{\hat J} \wedge F^{\hat K} \ , \nn
\end{align}
where we have chosen the Cartan generators to be in the coroot basis and used the notation introduced around \eqref{e:hat_notation}.
In order to obtain the full quantum effective action one has to integrate out the massive modes. This is partly done in \autoref{sec:one-loop_CS}
and induces new Chern-Simons couplings.

\section{Loop calculations}
\label{app:loop}
In this section we explicitly derive the one-loop induced Chern-Simons coefficients in \autoref{sec:one-loop_CS}. For completeness, we once more write down the basic formulae \eqref{e:CS_coupling} found in \cite{Bonetti:2013ela}, namely
\begin{align}
k_{\Lambda \Sigma \Theta} = \frac{1}{2}\bigg[&\sum_{spin\, 1/2}  (q_{1/2})_{ \Lambda} (q_{1/2})_\Sigma (q_{1/2})_\Theta \sign(m_{1/2})\\
&-5\sum_{spin\, 3/2}(q_{3/2})_\Lambda (q_{3/2})_\Sigma (q_{3/2})_\Theta \sign(m_{3/2}) -4\sum_{B}(q_B)_\Lambda (q_B)_\Sigma (q_B)_\Theta \sign(m_B)\bigg] \nn \\
k_{\Lambda }=-\frac{1}{4}\bigg[&\sum_{spin\, 1/2} (q_{1/2})_\Lambda \sign(m_{1/2})+19\! \! \! \sum_{spin\, 3/2}(q_{3/2})_\Lambda \sign(m_{3/2}) +8\sum_{B}(q_B)_\Lambda \sign(m_B)\bigg]\ .
\end{align}
The masses are given by
\begin{align}\label{masses_app}
m_{1/2} = c_{1/2} \bigg( q_{1/2} \cdot \zeta + \frac{n}{r}\bigg)\ ,\qquad
m_{3/2} = -c_{3/2}\, \frac{n}{r}\ , \qquad
m_B  = c_B \, \frac{n}{r}\ ,
\end{align}
where the coefficients $c$ refer to the respective representations of $SO(4)$, as explained in \autoref{sec:one-loop_CS}.

In the following, most of our effort will go into summing up contributions from Kaluza-Klein modes in these equations.
Since the $n$'th KK-mode carries charge $n$ under the Kaluza-Klein vector $A^0$, the infinite sum over the KK-modes can
in principle take one of the following four different forms 
\begin{align}\begin{split}
\sum_{n = -\infty}^{+\infty}\sign (x+n)\qquad &\sum_{n = -\infty}^{+\infty}n\sign (x+n)\\
\sum_{n = -\infty}^{+\infty}n^2 \sign (x+n)\qquad &\sum_{n = -\infty}^{+\infty}n^3 \sign (x+n).
\label{e:KK_sums}
\end{split}\end{align}
Here, the parameter $x$ takes the values
\begin{align}
x=
\begin{cases} 
r \alpha \cdot \zeta \\
r w \cdot \zeta\ ,
\end{cases}
\end{align}
as can be seen by looking at the expressions for the masses in \eqref{masses_app} and by noting that
the $U(1)$-charge on the Coulomb branch of the circle reduced action is encoded by the weights $w$ and roots $\alpha$ of $G$.
In previous papers, it was mostly assumed that $\left| x \right| < 1 $, implying that the Kaluza-Klein scale always exceeds
the Coulomb branch masses. As we found in the examples in this work, this need not always be correct and we therefore discuss general $x$. Let us define
\begin{align}
k:= \big \lfloor \left| x \right| \big \rfloor\ ,
\end{align}
then the first equation in \eqref{e:KK_sums} reads 
\begin{align}\begin{split}
\sum_{n = -\infty}^{+\infty}\sign (x+n)&=\sum_{n = -k}^{+k}\sign (x+n) + \sum_{n = k+1}^{+\infty}\sign (x+n)+ \sum_{n = -\infty}^{-k-1}\sign (x+n)\\
&=\sum_{n = -k}^{+k}\sign (x) + \sum_{n = k+1}^{+\infty}\sign (n)+ \sum_{n = -\infty}^{-k-1}\sign (n)=(2k+1)\sign(x)\ .
\end{split}\end{align}
Next of all, we calculate
\begin{align}\begin{split}
\sum_{n = -\infty}^{+\infty}n^2 \sign (x+n)&=\sum_{n = -k}^{+k}n^2 \sign (x) + \sum_{n = k+1}^{+\infty}n^2 \sign (n)+ \sum_{n = -\infty}^{-k-1}n^2 \sign (n)\\
&= 2 \sum_{n=1}^{k}n^2 \sign (x) =\frac{k(k+1)(2k+1)}{3} \sign (x)\ , 
\end{split}\end{align}
where we performed the sum in the last step. The remaining two sums require zeta function regularization. Using
\begin{align}
\zeta (-1) = - \frac{1}{12} \qquad \zeta (-3) = \frac{1}{120}\,,
\end{align}
we compute that
\begin{align}\begin{split}
\sum_{n = -\infty}^{+\infty}n \sign (x+n)&=\sum_{n = -k}^{+k}n \sign (x) + \sum_{n = k+1}^{+\infty}n \sign (n)+ \sum_{n = -\infty}^{-k-1}n \sign (n)\\
&= \sum_{n = k+1}^{+\infty}n + \sum_{n=1}^k n - \sum_{n=1}^k n + \sum_{n = -\infty}^{-k-1}(-n) +\sum_{n=-k}^{-1} (-n) - \sum_{n=-k}^{-1} (-n)\\
&= 2\zeta (-1) -2 \sum_{n=1}^k n = -\frac{1}{6} - (k+1)k
\end{split}\end{align}
and
\begin{align}\begin{split}
\sum_{n = -\infty}^{+\infty}n^3 \sign (x+n)&=\sum_{n = -k}^{+k}n^3 \sign (x) + \sum_{n = k+1}^{+\infty}n^3 \sign (n)+ \sum_{n = -\infty}^{-k-1}n^3 \sign (n)\\
&= \sum_{n = k+1}^{+\infty}n^3 + \sum_{n=1}^k n^3 - \sum_{n=1}^k n^3 + \sum_{n = -\infty}^{-k-1}(-n^3) +\sum_{n=-k}^{-1} (-n^3) - \sum_{n=-k}^{-1} (-n^3)\\
&= 2\zeta (-3) -2 \sum_{n=1}^k n^3 = \frac{1}{60} - \frac{k^2 (k+1)^2}{2}.
\end{split}\end{align}
Apart from the following caveat, it is now straightforward to calculate the Chern-Simons coefficients induced at one-loop. When we broke the non-Abelian gauge group to its maximal torus, we used the Cartan generators in the coroot basis of the remaining gauge fields. As already pointed out in \autoref{sec:mixed_an}, the charge $q_I$ of a field of weight $\mathbf{w}$ under the Cartan generator $\mathcal{T}_I$ therefore reads
\begin{align}
q_I = \langle \alpha_I^\vee , \mathbf{w} \rangle
\end{align}
and is what has to be inserted in the formulae for the one-loop induced Chern-Simons coefficient.    

\section{Some geometry}

  \subsection{Exact identities for the second Chern class} \label{a:c_2}
    Let us show explicitly that an elliptically fibered Calabi-Yau threefold $\hat{Y}_3$ obeys
    \begin{equation}
     \int_{\hat{Y}_3} \omega_{\alpha} \wedge c_2(\hat{Y}_3) = - 12 K_{\alpha}\,,
    \end{equation}
    where as before $\omega_{\alpha} = \pi^*(\omega_{\alpha}^b)$ is obtained by pulling back the $(1,1)$-form $\omega_{\alpha}^b \in H^{1,1}(\cB)$. To do so, we first note that the adjunction formula implies that
    \begin{equation}
    \begin{split}
     c_2(D_{\alpha}) &= c_2(\hat{Y}_3) \rvert_{D_{\alpha}} + \omega_{\alpha} \wedge \omega_{\alpha} - \omega_{\alpha} \wedge c_1(Y_3) \rvert_{D_{\alpha}} \\
     &= c_2(\hat{Y}_3) \rvert_{D_{\alpha}} + \omega_{\alpha} \wedge \omega_{\alpha}\,, \label{e:adjunction_app}
     \end{split}
    \end{equation}
    since $\hat{Y}_3$ is Calabi-Yau. Recalling that triple intersections of vertical divisors vanish, we can therefore rewrite the above integral as
    \begin{equation}
    \int_{\hat{Y}_3} \omega_{\alpha} \wedge c_2(\hat{Y}_3) = \int_{\hat{Y}_3} \omega_{\alpha} \wedge c_2(D_{\alpha}) = \int_{D_{\alpha}} c_2(D_{\alpha}) = \chi(D_{\alpha})\,.
    \end{equation}
    We are left with calculating the Euler characteristic of the vertical divisor $D_{\alpha}$. Fortunately, we can exploit that $D_{\alpha}$ is obtained by smoothly fibering the generic fiber manifold over $D_{\alpha}^b$. In particular, $D_{\alpha}^b$ is a smooth manifold of complex dimension $1$ and we have rid ourselves of the reducible fiber components that $\hat{Y}_3$ has. Hence, we can use Theorem 4.3 of \cite{Aluffi:2009tm} and reduce the integral over $D_{\alpha}$ to an integral over only the base of the fibration. In fact, for one-dimensional base manifolds one finds that
    \begin{equation}
     \chi(D_{\alpha}) = 12 \int_{D_{\alpha}^b} c_1(\cB)\rvert_{D_{\alpha}^b} = - 12 K_{\alpha}\,,
    \end{equation}
    no matter whether the elliptic fiber is embedded in an $E_6$, $E_7$ or $E_8$ model, which concludes our short proof.
    
    For completeness, let us briefly show how to calculate $c_{0}$ assuming now that the zero section is \emph{holomorphic}. Note that this merely reproduces the calculation in \cite{Bonetti:2011mw}. Using \eqref{e:adjunction_app} for $D_{\hat{0}}$ instead of $D_{\alpha}$, one finds that
    \begin{equation}
    \begin{split}
     \int_{\hat{Y}_3} \omega_{\hat{0}} \wedge c_2(\hat{Y}_3) &= \int_{\hat{Y}_3} \omega_{\hat{0}} \wedge \left( c_2(\cB) - \omega_{\hat{0}} \right)
     = \int_{\cB} c_2(\cB) - c_1(\cB) \wedge c_1(\cB) \\
     &= -8 + 2 h^{1,1}(\cB)\,,
     \end{split}
    \end{equation}
    where we have used adjunction for a second time in order to obtain $\omega_{\hat{0}}^2 = -\omega_{\hat{0}} \wedge c_1(\cB)$. Inserting \eqref{e:base_shift},
    one finds that $c_0 = c_{\hat{0}} - \frac{1}{2} K^{\alpha} K_{\alpha}$ and computes
    \begin{equation}
     K^{\alpha} K_{\alpha} = \int_{\cB} c_1(\cB) \wedge c_1(\cB) = 10 - h^{1,1}(\cB)\,.
    \end{equation}
    Putting everything together, one finally ends up with
    \begin{equation}
     c_0 = 52 - 4 h^{1,1}(\cB) \qquad \textrm{if } D_{\hat{0}} \textrm{ is holomorphic.}
    \end{equation}

    \subsection{Toric fans}
  In this subsection we list the complete data of the fans used for the construction of the Calabi-Yau threefolds examined in this paper.

  \begin{equation}
   \begin{split}
    \Sigma_{I,\textrm{hol.}} &= \Bigl\{\langle h_{0} h_{1} f_{0} f_{1} \rangle , \langle
		    h_{0} h_{1} f_{0} f_{3} \rangle , \langle h_{0} h_{1} f_{1} f_{2}
		    \rangle , \langle h_{0} h_{1} f_{2} f_{3} \rangle , \langle h_{0} d_{0}
		    d_{1} f_{1} \rangle , \\ 
 	 	 & \qquad \langle h_{0} d_{0} d_{1} f_{2} \rangle , \langle
		h_{0} d_{0} f_{0} f_{1} \rangle , \langle h_{0} d_{0} f_{0} f_{3}
		\rangle , \langle h_{0} d_{0} f_{2} f_{3} \rangle , \langle h_{0} d_{1}
		f_{1} f_{2} \rangle , \\ 
 	 	 & \qquad \langle h_{1} d_{0} d_{1} f_{1} \rangle , \langle
		h_{1} d_{0} d_{1} f_{2} \rangle , \langle h_{1} d_{0} f_{0} f_{1}
		\rangle , \langle h_{1} d_{0} f_{0} f_{3} \rangle , \langle h_{1} d_{0}
		f_{2} f_{3} \rangle , \\ 
 	 	 & \qquad \langle h_{1} d_{1} f_{1} f_{2} \rangle \Bigr\}
   \end{split} \label{e:vb5_su2_fan_hol}
  \end{equation}
  \begin{equation}
   \begin{split}
    \Sigma_{I,\textrm{rat.}} &= \Bigl\{\langle h_{0} h_{1} f_{0} f_{1} \rangle , \langle
h_{0} h_{1} f_{0} f_{3} \rangle , \langle h_{0} h_{1} f_{1} f_{2}
\rangle , \langle h_{0} h_{1} f_{2} f_{3} \rangle , \langle h_{0} d_{0}
d_{1} f_{0} \rangle , \\ 
 	 	 & \qquad \langle h_{0} d_{0} d_{1} f_{2} \rangle , \langle
h_{0} d_{0} f_{0} f_{3} \rangle , \langle h_{0} d_{0} f_{2} f_{3}
\rangle , \langle h_{0} d_{1} f_{0} f_{1} \rangle , \langle h_{0} d_{1}
f_{1} f_{2} \rangle , \\ 
 	 	 & \qquad \langle h_{1} d_{0} d_{1} f_{0} \rangle , \langle
h_{1} d_{0} d_{1} f_{2} \rangle , \langle h_{1} d_{0} f_{0} f_{3}
\rangle , \langle h_{1} d_{0} f_{2} f_{3} \rangle , \langle h_{1} d_{1}
f_{0} f_{1} \rangle , \\ 
 	 	 & \qquad \langle h_{1} d_{1} f_{1} f_{2} \rangle \Bigr\}
   \end{split} \label{e:vb5_su2_fan_rat}
  \end{equation}
    \begin{equation}
   \begin{split}
      \Sigma_{II} &= \Bigl\{\langle h_{0} h_{1} f_{0} f_{2} \rangle , \langle
      h_{0} h_{1} f_{0} f_{3} \rangle , \langle h_{0} h_{1} f_{1} f_{3}
      \rangle , \langle h_{0} h_{1} f_{1} f_{4} \rangle , \langle h_{0} h_{1}
      f_{2} f_{4} \rangle , \\ 
		      & \qquad \langle h_{0} d_{0} d_{1} d_{4} \rangle , \langle
      h_{0} d_{0} d_{1} f_{0} \rangle , \langle h_{0} d_{0} d_{4} f_{0}
      \rangle , \langle h_{0} d_{1} d_{2} d_{4} \rangle , \langle h_{0} d_{1}
      d_{2} f_{0} \rangle , \\ 
		      & \qquad \langle h_{0} d_{2} d_{3} d_{4} \rangle , \langle
      h_{0} d_{2} d_{3} f_{1} \rangle , \langle h_{0} d_{2} f_{0} f_{3}
      \rangle , \langle h_{0} d_{2} f_{1} f_{3} \rangle , \langle h_{0} d_{3}
      d_{4} f_{4} \rangle , \\ 
		      & \qquad \langle h_{0} d_{3} f_{1} f_{4} \rangle , \langle
      h_{0} d_{4} f_{0} f_{2} \rangle , \langle h_{0} d_{4} f_{2} f_{4}
      \rangle , \langle h_{1} d_{0} d_{1} d_{4} \rangle , \langle h_{1} d_{0}
      d_{1} f_{0} \rangle , \\ 
		      & \qquad \langle h_{1} d_{0} d_{4} f_{0} \rangle , \langle
      h_{1} d_{1} d_{2} d_{4} \rangle , \langle h_{1} d_{1} d_{2} f_{0}
      \rangle , \langle h_{1} d_{2} d_{3} d_{4} \rangle , \langle h_{1} d_{2}
      d_{3} f_{1} \rangle , \\ 
		      & \qquad \langle h_{1} d_{2} f_{0} f_{3} \rangle , \langle
      h_{1} d_{2} f_{1} f_{3} \rangle , \langle h_{1} d_{3} d_{4} f_{4}
      \rangle , \langle h_{1} d_{3} f_{1} f_{4} \rangle , \langle h_{1} d_{4}
      f_{0} f_{2} \rangle , \\ 
		      & \qquad \langle h_{1} d_{4} f_{2} f_{4} \rangle \Bigr\}
   \end{split} \label{e:e2_fan}
  \end{equation}
  \begin{equation}
   \begin{split}
\Sigma_{III,\textrm{hol.}} &= \Bigl\{\langle h_{0} h_{1} f_{0} f_{2} \rangle , \langle
h_{0} h_{1} f_{0} f_{4} \rangle , \langle h_{0} h_{1} f_{1} f_{3}
\rangle , \langle h_{0} h_{1} f_{1} f_{4} \rangle , \langle h_{0} h_{1}
f_{2} f_{3} \rangle , \\ 
 	 	 & \qquad \langle h_{0} d_{0} d_{1} d_{4} \rangle , \langle
h_{0} d_{0} d_{1} f_{2} \rangle , \langle h_{0} d_{0} d_{4} f_{0}
\rangle , \langle h_{0} d_{0} f_{0} f_{2} \rangle , \langle h_{0} d_{1}
d_{2} d_{3} \rangle , \\ 
 	 	 & \qquad \langle h_{0} d_{1} d_{2} f_{2} \rangle , \langle
h_{0} d_{1} d_{3} d_{4} \rangle , \langle h_{0} d_{2} d_{3} f_{1}
\rangle , \langle h_{0} d_{2} f_{1} f_{3} \rangle , \langle h_{0} d_{2}
f_{2} f_{3} \rangle , \\ 
 	 	 & \qquad \langle h_{0} d_{3} d_{4} f_{4} \rangle , \langle
h_{0} d_{3} f_{1} f_{4} \rangle , \langle h_{0} d_{4} f_{0} f_{4}
\rangle , \langle h_{1} d_{0} d_{1} d_{4} \rangle , \langle h_{1} d_{0}
d_{1} f_{2} \rangle , \\ 
 	 	 & \qquad \langle h_{1} d_{0} d_{4} f_{0} \rangle , \langle
h_{1} d_{0} f_{0} f_{2} \rangle , \langle h_{1} d_{1} d_{2} d_{3}
\rangle , \langle h_{1} d_{1} d_{2} f_{2} \rangle , \langle h_{1} d_{1}
d_{3} d_{4} \rangle , \\ 
 	 	 & \qquad \langle h_{1} d_{2} d_{3} f_{1} \rangle , \langle
h_{1} d_{2} f_{1} f_{3} \rangle , \langle h_{1} d_{2} f_{2} f_{3}
\rangle , \langle h_{1} d_{3} d_{4} f_{4} \rangle , \langle h_{1} d_{3}
f_{1} f_{4} \rangle , \\ 
 	 	 & \qquad \langle h_{1} d_{4} f_{0} f_{4} \rangle \Bigr\} \label{e:f1_hol}
   \end{split}
  \end{equation}
  \begin{equation}
   \begin{split}
    \Sigma_{III,\textrm{rat.}} &= \Bigl\{\langle h_{0} h_{1} f_{0} f_{2} \rangle , \langle
h_{0} h_{1} f_{0} f_{4} \rangle , \langle h_{0} h_{1} f_{1} f_{3}
\rangle , \langle h_{0} h_{1} f_{1} f_{4} \rangle , \langle h_{0} h_{1}
f_{2} f_{3} \rangle , \\ 
 	 	 & \qquad \langle h_{0} d_{0} d_{1} d_{4} \rangle , \langle
h_{0} d_{0} d_{1} f_{0} \rangle , \langle h_{0} d_{0} d_{4} f_{0}
\rangle , \langle h_{0} d_{1} d_{2} d_{3} \rangle , \langle h_{0} d_{1}
d_{2} f_{2} \rangle , \\ 
 	 	 & \qquad \langle h_{0} d_{1} d_{3} d_{4} \rangle , \langle
h_{0} d_{1} f_{0} f_{2} \rangle , \langle h_{0} d_{2} d_{3} f_{1}
\rangle , \langle h_{0} d_{2} f_{1} f_{3} \rangle , \langle h_{0} d_{2}
f_{2} f_{3} \rangle , \\ 
 	 	 & \qquad \langle h_{0} d_{3} d_{4} f_{1} \rangle , \langle
h_{0} d_{4} f_{0} f_{4} \rangle , \langle h_{0} d_{4} f_{1} f_{4}
\rangle , \langle h_{1} d_{0} d_{1} d_{4} \rangle , \langle h_{1} d_{0}
d_{1} f_{0} \rangle , \\ 
 	 	 & \qquad \langle h_{1} d_{0} d_{4} f_{0} \rangle , \langle
h_{1} d_{1} d_{2} d_{3} \rangle , \langle h_{1} d_{1} d_{2} f_{2}
\rangle , \langle h_{1} d_{1} d_{3} d_{4} \rangle , \langle h_{1} d_{1}
f_{0} f_{2} \rangle , \\ 
 	 	 & \qquad \langle h_{1} d_{2} d_{3} f_{1} \rangle , \langle
h_{1} d_{2} f_{1} f_{3} \rangle , \langle h_{1} d_{2} f_{2} f_{3}
\rangle , \langle h_{1} d_{3} d_{4} f_{1} \rangle , \langle h_{1} d_{4}
f_{0} f_{4} \rangle , \\ 
 	 	 & \qquad \langle h_{1} d_{4} f_{1} f_{4} \rangle \Bigr\} \label{e:f1_rat}
   \end{split}
  \end{equation}


\bibliography{Abelian_6D}
\bibliographystyle{utcaps}

\end{document}